\documentclass[fleqn,usenatbib]{mnras}

\usepackage{newtxtext,newtxmath}

\usepackage[T1]{fontenc}

\usepackage[detect-all]{siunitx}

\DeclareRobustCommand{\VAN}[3]{#2}
\let\VANthebibliography\thebibliography
\def\thebibliography{\DeclareRobustCommand{\VAN}[3]{##3}\VANthebibliography}

\usepackage{graphicx}	
\usepackage{amsmath}	

\usepackage{amssymb}	
\usepackage{threeparttable}
\usepackage{subcaption}
\usepackage{textcase} 
\usepackage{hyperref} 

\title[Shear biases from undetected galaxies]{
Controlling weak-lensing shear biases from undetected galaxies in the era of Stage IV Surveys
}

\author[L. M. Voigt]{
L. M. Voigt$^{1}$\thanks{E-mail: lv18675@essex.ac.uk}
\\
$^{1}$School of Mathematics, Statistics and Actuarial Science, University of Essex, Wivenhoe Park, Colchester, CO4 3SQ
}

\date{Accepted XXX. Received YYY; in original form ZZZ}

\pubyear{2023}

\begin{document}
\label{firstpage}
\pagerange{\pageref{firstpage}--\pageref{lastpage}}
\maketitle

\begin{abstract}
Gravitational lensing of background galaxies by intervening matter is a powerful probe of the cosmological model. In the era of Stage IV surveys, contamination from galaxies below the detection threshold has emerged as a significant source of bias. Adopting a noise-bias-free machine-learning method to estimate shear, we quantify the impact of faint galaxies for a Euclid-like survey. In our baseline simulations, faint blends induce a multiplicative shear bias of -0.008, well above Euclid's requirement. Similar to previous studies, we find that calibration simulations must include neighbouring galaxies to AB apparent magnitudes as faint as 27.0 (+2.1, -0.9) and within approximately 1.0 (+0.2, -0.2) arcsec of each bright sample galaxy (BSG; the galaxy for which shear is measured). By varying faint galaxy properties, we identify which ones significantly affect shear biases and quantify how well they must be constrained. Crucially, we find that biases not only depend on the mean projected faint-galaxy density and apparent-magnitude distribution across the sample, but also on how these quantities vary with the observed brightness of the BSG. Furthermore, biases are sensitive to radial and tangential alignments and positional anisotropy of faint galaxies relative to BSGs. 
By contrast, shear coherence between BSGs and faint galaxies, parallel orientation alignments, and variations in the faint galaxy size–magnitude relation have negligible impact within the parameter ranges explored.
Our results guide calibration simulations and highlight the critical role of deep observations in measuring the properties of faint galaxies.
\end{abstract}






\section{Introduction}
\label{sec:int}
Weak gravitational lensing provides a powerful probe of cosmic structure, making it one of the main science drivers for large-scale cosmological surveys. Stage IV missions such as \emph{Euclid}\footnote{https://www.euclid-ec.org/}\citep{Laureijs2011,2013LRR....16....6A}, the Vera C. Rubin Observatory's Legacy Survey of Space and Time\footnote{https://rubinobservatory.org/} \citep[LSST;][]{2012arXiv1211.0310L}, and NASA's \textit{Nancy Grace Roman} Space Telescope\footnote{https://roman.gsfc.nasa.gov/} \citep{2015arXiv150303757S} aim to deliver sub-percent level constraints on cosmological parameters, including the dark energy equation-of-state parameter $w_0$. Achieving this level of precision requires stringent control of systematic biases in weak lensing measurements.

Sources of systematic bias include galaxy shape measurement \citep[][]{step1_2006,step2_2007,great10,great3}, photometric redshifts \citep[][]{Hildebrandt2010}, intrinsic alignments \citep[e.g.][]{PhysRevD.70.063526,2010A&A...523A...1J,Heymans_2013}, and the modelling of non-linear structure and baryonic effects \citep[see also][for a review]{Mandelbaum2018review}. Shape measurement systematics include model-fitting bias \citep{voigt_modelBias_2010}, noise bias \citep[][]{kacprzak_noiseBias_2012,refregier_noiseBias_2012}, and modelling of the point spread function \citep[PSF;][]{paulin_psf_2008}, including wavelength-dependent effects \citep{Cypriano2010,voigt_colDep}. A bias that has emerged more recently as a significant concern is blending -- flux contamination from neighbouring or overlapping galaxies that affects both the detection and shape measurement of sources \citep[e.g.][]{2018MNRAS.475.4524S_samuroff}. These galaxies may be physically associated with the source (e.g., within groups or clusters) or unrelated galaxies projected at small angular separation.  

While some bright blends—i.e., those above the detection threshold—can be mitigated using deblending algorithms \citep[e.g.][]{Arcelin_2020,2024PASA...41...35Z} or catalogue-level flagging based on \textsc{SExtractor}'s output parameters \citep{2011ASPC..442..435B,Zuntz2018}, galaxies below the detection threshold cannot be directly removed yet still contribute flux to the images of galaxies used for shear estimation. We refer to the galaxies used for shear estimation as bright sample galaxies (BSGs), and their undetected neighbours as faint galaxies. These faint blends, despite being undetected, can introduce significant shear biases if their presence is ignored i.e., if the bright galaxy used for shear estimation is assumed to be isolated. Stage III surveys, for example the Dark Energy Survey (DES)\footnote{https://www.darkenergysurvey.org/} Year 3 analysis, have incorporated deep image simulations to model these effects, extending the simulated galaxy population to magnitudes as faint as $m_{\rm AB} \sim 27.5$ \citep{2022MNRAS.509.3371M}. Although some shear measurement techniques aim to reduce dependence on simulations, such as \textsc{metacalibration} and \textsc{metadetection} \citep{2017arXiv170202600H, Sheldon_2017_metacalibration, Sheldon_2023}, simulations nevertheless remain a key component of weak lensing analyses—particularly for those using model-fitting methods such as \textsc{lens}fit \citep[][]{Miller2007,Li2023}. In such cases, assumptions about the faint galaxy population can affect shear calibration and shift cosmological constraints. However, the extent to which the properties of the faint galaxy population in these simulations must match those in the data still requires further investigation.

In this paper, we carry out a systematic study of the biases arising from faint blends, using \emph{Euclid} as the reference survey. 
We quantify the sensitivity of the shear bias to several faint galaxy properties, such as the limiting magnitude to which galaxies must be simulated and the maximum separation from the BSG at which faint blends still contribute significantly---parameters that have been explored in previous studies using separate galaxy simulation and shear measurement methods \citep[][hereafter H17, M19 and H21, respectively]{hoekstra2017,martinet_2019,hoekstra2021}. We extend this work to also include additional properties such as the angular distribution and orientation of faint galaxies relative to the BSG, the shear applied to faint galaxies, the slope of the faint galaxy apparent magnitude distribution, and the form of the size--magnitude relation. Crucially, we also investigate the impact of correlations between the properties of faint blends and those of the BSG—specifically, dependencies of the local faint galaxy density and magnitude distribution slope on BSG brightness. To our knowledge, this is the first time such correlations have been explicitly quantified in the context of shear calibration. These effects are particularly relevant for Stage IV surveys, where even sub-percent level biases can impact cosmological inferences.

We simulate galaxy images with a fixed PSF using simplified models, with the aim of isolating the impact of the faint galaxy population on shear bias. To measure shear, we use the convolutional neural network (CNN) method developed in \citet{voigt2024}, which employs a committee of shallow CNNs trained to recover unbiased shear estimates in the presence of noise. This supervised learning approach relies on a training set of simulated galaxies with known shears, from which the model learns to predict shear for new data. Model-fitting and PSF-related biases are avoided by using consistent galaxy and PSF models in both the training and test simulations. To isolate biases arising from the faint galaxy population, we deliberately exclude faint galaxies from the training simulations while including them in the test simulations, which are intended to mimic real survey data. By comparing shear measurements across simulations with varying faint galaxy properties, we quantify the sensitivity of 
biases to the faint population. These results inform the level of realism required in calibration simulations to meet the stringent systematic error budgets of \emph{Euclid}-like Stage IV weak lensing surveys.

The paper is organised as follows.  We describe the analytic galaxy and PSF models in Section~\ref{sec:gal_psf_models} and provide the ellipticity and shear definitions in Section~\ref{sec:ellip_shear}.
Section~\ref{sect:pops} summarises the apparent magnitude, ellipticity, size, morphology and signal-to-noise ($S/N$) distributions adopted for the BSG and faint galaxy populations.
We detail the simulation setup and shape measurement methodology in Sections~\ref{sec:sims} and \ref{sec:shear_measurement_method}, respectively, followed by a discussion in Section~\ref{sec:biases_noSat} of the shear biases obtained for isolated galaxies. In addition, in Section~\ref{sec:biases_noSat} we present a novel method for obtaining signal-to-noise estimates from noisy images. In Section~\ref{sec:faint}, we present results for the biases arising from faint galaxy contamination. These biases are then further explored in Section~\ref{sec:clustering}, which investigates the impact of correlations between faint galaxy properties and the apparent magnitude of the BSG.
Finally, we summarise and discuss the implications of our results in Section~\ref{sec:discussion}.

\section{The galaxy and PSF models}
\label{sec:gal_psf_models}
We simulate a population of single-component disc and elliptical galaxies with constant ellipticity isophotes. 
The projected intensity distributions are modelled using S\'ersic profiles \citep{1968adga.book.....S} with intensity $I(\textbf{\textit{x}})$ at position $\textbf{\textit{x}}$ given by
\begin{equation}
I(\textbf{\textit{x}})=I_\mathrm{0}\exp\left\{-k\left[(\textbf{\textit{x}}-\textbf{\textit{x}}_{\mathrm{0}})^T\textbf{\textit{C}}(\textbf{\textit{x}}-\textbf{\textit{x}}_\mathrm{0})\right]^{\frac{1}{2n_{\mathrm{s}}}}\right\},
\label{eqn:gal_intensity}
\end{equation}
where $I_\mathrm{0}$ is the peak intensity, $\textbf{\textit{x}}_{\mathrm{0}}$ the position of the galaxy's centre and $n_{\mathrm{s}}$ the S\'ersic index. The matrix $\textbf{\textit{C}}$ encodes the axis lengths and orientation of the elliptical isophotes, and is given by
\begin{equation}
\textbf{\textit{C}}=
\begin{pmatrix}
C_{11} & C_{12} \\ 
C_{21} & C_{22}
\end{pmatrix},
\end{equation}
with
\begin{equation}
C_{11}=\frac{\cos^2(\phi)}{a^2}+\frac{\sin^2(\phi)}{b^2},
\end{equation}
\begin{equation}
C_{12}=C_{21}=\frac{1}{2}\left(\frac{1}{a^2}-\frac{1}{b^2}\right)\sin{(2\phi)},
\end{equation}
and
\begin{equation}
C_{22}=\frac{\sin^2(\phi)}{a^2}+\frac{\cos^2(\phi)}{b^2}.
\end{equation}
Here, $a$, $b$, and $\phi$ are the semi-major and semi-minor axis lengths ($a \geq b$) and the orientation (measured counter-clockwise from the $x$-axis) of the galaxy, respectively. For $k=1.9992 n_{\mathrm{s}} - 0.3271$ and a circular profile, $a(=b)$ is the radius enclosing half the total flux \citep[][]{1997MNRAS.287..221G}.

We model the PSF as an elliptical Gaussian with ellipticity components 0.01 and 0.02 along and at $45^{\circ}$ to the $x$-axis, respectively (see Section \ref{sec:ellip_shear} for definitions).
The full width at half maximum is 0.17\,arcsec (corresponding to a half-light radius of 0.084\,arcsec), sampled on a 0.1\,arcsec pixel grid.
The PSF size and pixel scale are chosen to be representative of the \emph{Euclid} VISible (VIS) instrument. While the true VIS PSF follows an Airy pattern modified by optical and detector effects, a Gaussian approximation captures its overall size and shape
sufficiently well for studying blending-induced biases. The PSF model is fixed throughout and assumed to be precisely known.

\section{Ellipticity and shear}
\label{sec:ellip_shear}
A galaxy with elliptical isophotes can be described by a complex ellipticity
\begin{equation}
    e^{\mathrm{u}} = e^{\mathrm{u}}_1+ie^{\mathrm{u}}_2 = \left( \frac{a - b}{a + b} \right) e^{2i\phi},
\label{eqn:eu}
\end{equation}
where $a$, $b$ and $\phi$ are defined in Section~\ref{sec:gal_psf_models}. 

Gravitational lensing transforms image-plane positions according to a Jacobian matrix. The mapping from lensed coordinates \( (x^{\mathrm{l}}, y^{\mathrm{l}}) \) to unlensed coordinates \( (x^{\mathrm{u}}, y^{\mathrm{u}}) \) is given by
\begin{equation}
\begin{pmatrix}
x^{\mathrm{u}} \\ 
y^{\mathrm{u}}
\end{pmatrix}
=
\begin{pmatrix}
1 - \kappa - \gamma_1 & -\gamma_2 \\
-\gamma_2 & 1 -\kappa + \gamma_1
\end{pmatrix}
\begin{pmatrix}
x^{\mathrm{l}} \\ 
y^{\mathrm{l}}
\end{pmatrix},
\label{eqn:coords}
\end{equation}
where \( \gamma = \gamma_1 + i \gamma_2 \) is the gravitational shear and \( \kappa \) is the dimensionless surface mass density \citep[see, e.g.,][]{BARTELMANN2001291}. 
This transformation shears and magnifies the unlensed image such that the observed (lensed) complex ellipticity is
\begin{equation}
    e^{\mathrm{l}} = \frac{e^{\mathrm{u}} + g}{1 + g^* e^{\mathrm{u}}},
\end{equation}
where the reduced shear, \( g = g_1 + i g_2 \), is related to \(\gamma\) and \(\kappa\)
via
\begin{equation}
g = \frac{\gamma}{1 - \kappa}.
\end{equation}

In the weak lensing regime, where \( |\gamma| \ll 1 \) and \( \kappa \ll 1 \), the reduced shear approximates the true shear and the lensed ellipticity simplifies to
\begin{equation}
    e^{\mathrm{l}} \approx e^{\mathrm{u}} + \gamma.
\end{equation}

The Jacobian in Equation~\ref{eqn:coords} also changes the overall image size. In general the projected area transforms as
\begin{equation}
    ab \;\longrightarrow\; \frac{ab}{(1-\kappa)^2 - |\gamma|^2}.
\end{equation}
For the case of \(\kappa \approx 0\) this reduces to
\begin{equation}
    ab \;\longrightarrow\; \frac{ab}{1 - |\gamma|^2}.
\end{equation}

Assuming galaxies have randomly oriented intrinsic shapes, the mean unlensed ellipticity averages to zero: \( \langle e^{\mathrm{u}} \rangle = 0 \). A shear estimator can therefore be defined as the average observed ellipticity:
\begin{equation}
    \gamma^{\mathrm{est}} = \langle e^{\mathrm{l}} \rangle = \gamma \pm \frac{\sigma_{\mathrm{e}}}{\sqrt{n_{\mathrm{gal}}}},
\end{equation}
where \( \sigma_{\mathrm{e}} \) \citep[$\approx0.26$;][]{Gatti2021_DESY3ShapeCat} is the dispersion of each ellipticity component, and \( n_{\mathrm{gal}} \) is the number of galaxies in the ensemble.

In practice, the measured shear, \( \gamma^{\mathrm{est,b}}_i \), is a biased estimate of the true shear due to various systematics, including blending (see Section~\ref{sec:int}). This bias is commonly parameterised as
\begin{equation}
\gamma^{\mathrm{est,b}}_i = (1 + m_i)\gamma_i + c_i,
\label{eqn:shear_est}
\end{equation}
where \( m_i \) and \( c_i \) are the multiplicative and additive biases on the \( i \)-th component of the shear. For \emph{Euclid}, these biases must satisfy \( |m_i| < 2 \times 10^{-3} \) and \( |c_i| < 3 \times 10^{-4} \) in order for them to be sub-dominant to the expected statistical uncertainties \citep{amara&refregier2008,cropper2013}.

\section{The bright and faint galaxy populations}
\label{sect:pops}
We divide galaxies into two populations: BSGs, with apparent magnitudes between 20 and 24.5, and faint galaxies, which fall below the \emph{Euclid} VIS band detection threshold. 
BSGs are detected galaxies used for shear estimation after standard selection cuts\footnote{Detected galaxies must also meet specific selection criteria to be included in the shear catalogue, typically based on signal-to-noise ratio and size.}, whereas faint galaxies are not detected but may blend with the brighter sources used for shear estimation. 

\subsection{Apparent magnitude distribution}
We sample galaxies from a cumulative distribution function, where the mean projected number density per arcmin$^2$ of galaxies with AB apparent magnitude less than or equal to $m_{\mathrm{AB}}$ is given by:
\begin{equation}
\left< N(m_{\mathrm{AB}}) \right> = \frac{A_{\mathrm{m}}}{\alpha_{\mathrm{m}}\ln(10)} 10^{\alpha_{\mathrm{m}}(m_{\mathrm{AB}})^{\beta_{\mathrm{m}}}},
\label{eqn:num_cum}
\end{equation}
where $A_{\mathrm{m}}$, $\alpha_{\mathrm{m}}$ and $\beta_{\mathrm{m}}$ are population-dependent parameters that control the normalisation, slope and curvature of the distribution. This functional form provides flexibility to match observed number counts across both the bright and faint galaxy populations.

We first consider the parameter values used for the BSGs.
We adopt a slope of $\alpha_{\mathrm{m}} = 0.36$, consistent with values used in other studies (e.g. H17), and 
set the amplitude \( A_{\mathrm{m}} \) so that the number density of bright galaxies matches the expected $\sim30$ galaxies per arcmin$^2$ in the \emph{Euclid} VIS band \citep[e.g.,][]{Laureijs2011}. The parameter values adopted for the bright galaxy distribution, provided in Table~\ref{tab:slopes}, are fixed throughout the paper.

\begin{table*}
\caption{Parameter values used for the galaxy apparent magnitude distribution (Equations~\ref{eqn:num_cum} and \ref{eqn:nm}) and the size--magnitude relation (Equations~\ref{eqn:size-mag_bright} and \ref{eqn:size-mag_faint}) for the bright and faint populations. For the faint population, we list the fiducial values for both the full and simplified linear forms of the apparent magnitude distribution. The linear setup is used in Sections~\ref{subsec:alphaf} and \ref{sec:clustering}.
$m_{\mathrm{AB,lw}}$ and $m_{\mathrm{AB,up}}$ denote the lower and upper apparent magnitudes for each population. For the size--magnitude relation, the dispersion parameters $\alpha_{\mathrm{\sigma}} = -0.0166$ and $\beta_{\mathrm{\sigma}} = 0.5633$ are fixed for both populations.}
\centering
\begin{tabular}{ccccccccc}
\hline
Population Type & Model & $m_{\mathrm{AB,lw}}$ & $m_{\mathrm{AB,up}}$ & $A_{\mathrm{m}}$ &$\alpha_{\mathrm{m}}$ & $\beta_{\mathrm{m}}$ & $\alpha_{\mathrm{r}}$ & $\beta_{\mathrm{r}}$\\
\hline
Bright & -- & 20 & 24.5 & $3.8564\times10^{-8}$ & 0.36 & 1 & $-0.1324$ & 2.65 \\
Faint & Fiducial Full  & 24.5 & 29 & $-2.1095\times10^{9}$ &  $-5.25970\times10^{5}$& $-3.9427$ & $-0.0330$ & 1.06  \\
Faint & Fiducial Linear & 24.5 & 27 & $8.2451\times10^{-3}$ & 0.139 & 1 & $-0.0330$ & 1.06  \\
\hline
\end{tabular}
\label{tab:slopes}
\end{table*}

\begin{figure}
\centering
\includegraphics[width=0.45\textwidth]{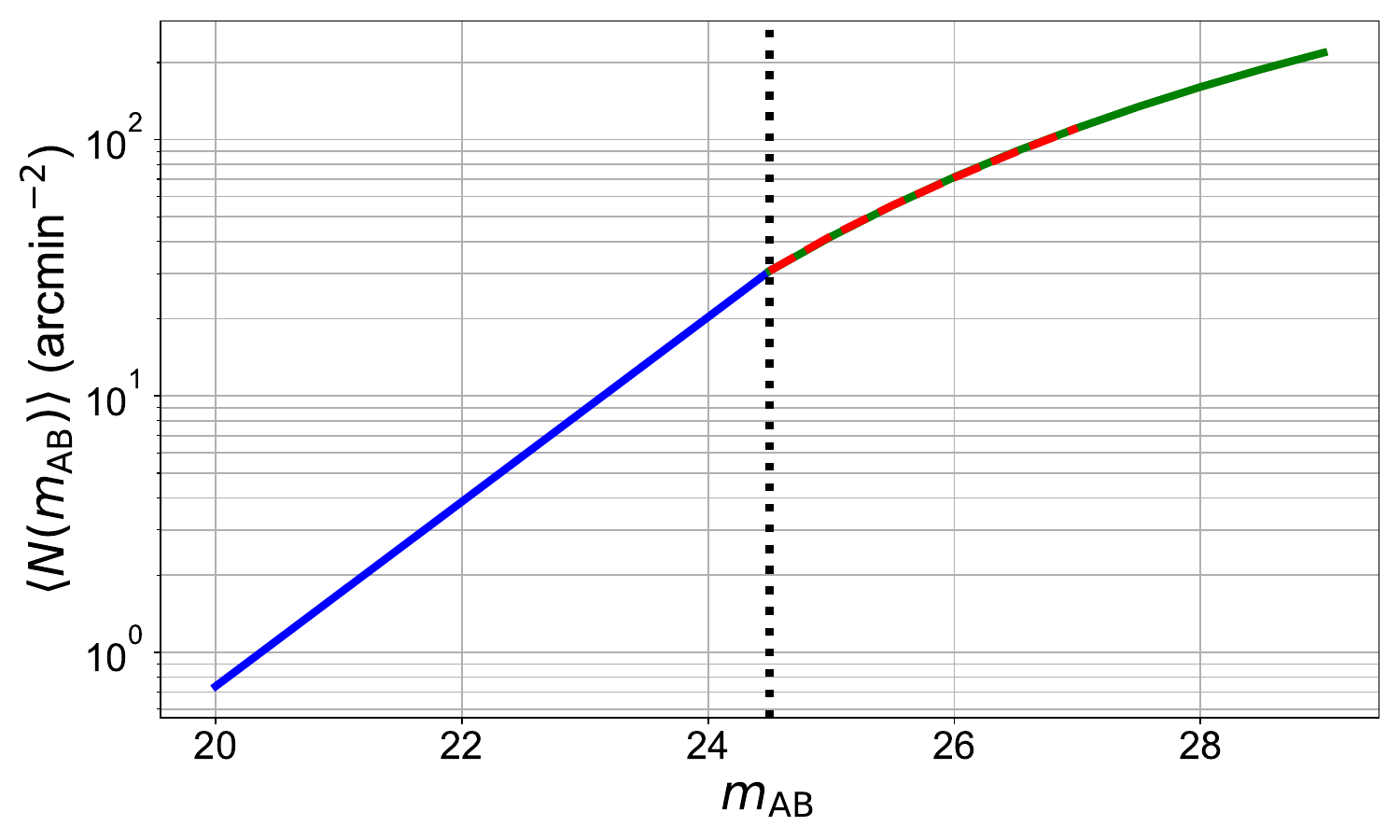}
\caption{Mean cumulative number density of galaxies per arcmin$^2$, $\left<N(m_{\mathrm{AB}})\right>$, for the bright (blue) and faint (green) populations as a function of apparent magnitude. The functional form is given in Equation~\ref{eqn:num_cum} and parameter values in Table~\ref{tab:slopes}. The linear approximation to the faint galaxy distribution over the range $24.5 < m_{\mathrm{AB}} \leq 27$ is also shown (red dashed), offset to match the green curve at $m_{\mathrm{AB}} = 24.5$ (see Section~\ref{subsec:alphaf}). The black vertical dotted line shows the division between the bright and faint populations at $m_{\mathrm{AB}}=24.5$.
} 
\label{fig:pm}
\end{figure}

In this study, faint galaxies are defined as those with apparent magnitudes fainter than the \emph{Euclid} detection limit of 24.5 and brighter than 29.
The parameters for the faint population are chosen to reproduce the projected number densities reported in \citet[][]{2024A&A...691A.319E}, specifically 250 and 90 galaxies per arcmin$^2$ for apparent magnitudes below 29.5 and 26.5, respectively.
The cumulative number density distributions are shown for the bright and faint galaxy populations in Figure~\ref{fig:pm}. In addition, Table~\ref{tab:Nsat} lists the mean number density of faint galaxies for different limiting magnitudes. 

To enable analysis of the sensitivity of shear biases to the faint-end slope of the apparent magnitude distribution, we also define a simplified “linear” model with $\beta_{\mathrm{m}} = 1$ over the range $24.5 < m_{\mathrm{AB}} < 27$ (see Figure~\ref{fig:pm}). This approximation is used in Sections~\ref{subsec:alphaf} and \ref{sec:clustering} to assess the impact of varying the slope parameter, $\alpha_{\mathrm{m,f}}$, of the faint population, including potential correlations with the apparent magnitude of the BSG. The fiducial parameters adopted for this model are listed in Table~\ref{tab:slopes}, with further discussion in Section~\ref{subsec:alphaf}.  

\begin{figure*}
\centering
\includegraphics[width=0.7\textwidth]{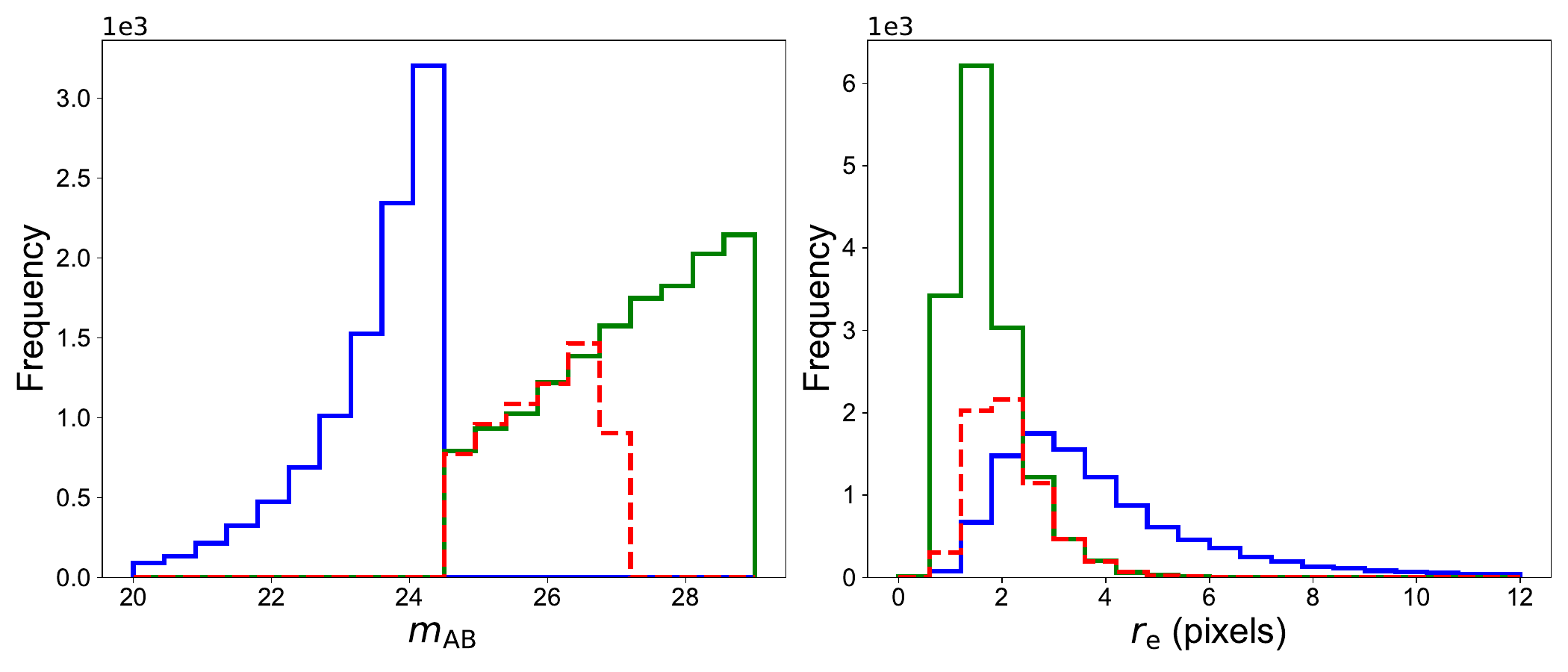}
\caption{Distributions of apparent magnitudes ($m_{\mathrm{AB}}$, left) and effective radii ($r_{\mathrm{e}}$, right) for bright (blue solid) and faint galaxies. Faint galaxy distributions are shown for both the full (green solid) and simplified ``linear" (red dashed) models for the apparent magnitude distribution. Parameter values are listed in Table~\ref{tab:slopes}; see also Figure~\ref{fig:bias_alphamf}. Bright galaxies are simulated over the range $20 < m_{\mathrm{AB}} < 24.5$, while faint galaxies span $24.5 < m_{\mathrm{AB}} < 29$ (full model) or $24.5 < m_{\mathrm{AB}} < 27$ (simplified model). Histograms use equal bin widths and identical $x$-axis ranges in each panel. Distributions are based on $10^{4}$ BSGs and associated faint galaxies simulated within 3\,arcsec of each BSG (see Table~\ref{tab:Nsat}).
} 
\label{fig:histograms}
\end{figure*}

Figure~\ref{fig:histograms} shows the apparent magnitude distributions of the bright and faint populations, with the fiducial linear approximation overlaid for the faint end.
This framework provides a flexible model for exploring how the undetected galaxy population contributes to biases in shear estimation.

\begin{table}
\caption{Mean number density of faint galaxies (per arcmin$^2$) as a function of limiting magnitude, $m_{\mathrm{lim}}$, using the full model for the faint apparent magnitude distribution (see Table~\ref{tab:slopes}). $\left<N_{3}\right>$ is the mean number of faint galaxies within 3\,arcsec of a bright galaxy, assuming a uniform random spatial distribution. 
}
\centering
\begin{tabular}{cccc}
\hline
$m_{\mathrm{lim}}$  &  $\left< N(m_{\mathrm{lim}}) \right>-\left<N(24.5)\right>$  & $\left<N_{3}\right>$\\
     & arcmin$^{-2}$  &   & \\
\hline
25 & 11.13 & 0.09\\
\\
26  & 40.69  & 0.32      \\
\\
27  & 80.34    & 0.63  \\
           \\
28  & 129.70    & 1.02  \\
           \\ 
29  & 187.59   & 1.47   \\        
\hline
\end{tabular}
\label{tab:Nsat}
\end{table}

\subsection{Galaxy size, ellipticity and morphology distributions}
Galaxy sizes are assigned based on distinct size–magnitude relations for the bright and faint populations. We define the effective radius, $r_{\mathrm{e}} = \sqrt{ab}$ (where $a$ and $b$ are the semi-major and semi-minor axes defined in Section~\ref{sec:gal_psf_models}), such that for a circular galaxy ($a = b$), the effective radius equals the half-light radius. For bright galaxies, the logarithm of the effective radius is drawn from a normal distribution with mean
\begin{equation}
\langle\log_{10}{r_{\mathrm{e}}}\rangle = \alpha_{\mathrm{r}} m_{\mathrm{AB}} + \beta_{\mathrm{r}},
\label{eqn:size-mag_bright}
\end{equation}
and a magnitude-dependent dispersion given by
\begin{equation}
\sigma_{\log_{10}{r_{\mathrm{e}}}} = \alpha_{\sigma} m_{\mathrm{AB}} + \beta_{\sigma}.
\label{eqn:size-mag_dispersion}
\end{equation}
For faint galaxies, $r_{\mathrm{e}}$ is drawn from a normal distribution with mean
\begin{equation}
\langle r_{\mathrm{e}} \rangle = \alpha_{\mathrm{r}} m_{\mathrm{AB}} + \beta_{\mathrm{r}},
\label{eqn:size-mag_faint}
\end{equation}
and the same form of magnitude-dependent dispersion. The population-dependent parameters $\alpha_{\mathrm{r}}$ and $\beta_{\mathrm{r}}$ are listed in Table~\ref{tab:slopes}, while the dispersion parameters are fixed at $\alpha_{\sigma} = -0.0166$ and $\beta_{\sigma} = 0.5633$ for both populations. For all galaxies, we set the maximum effective radius in simulations to 1.2 arcsec and adopt a minium $r_{\mathrm{e}} = 0$. 
The bright-end relation is motivated by observational results from H17, and the shallower faint-end trend by (M19; see their Fig.~1).

We model the intrinsic (unlensed) ellipticities of both bright and faint galaxies using a Rayleigh distribution, with probability density function
\begin{equation}
f(e) = \frac{e}{\sigma_{\mathrm{e}}^2} \exp\left(-\frac{e^2}{2\sigma_{\mathrm{e}}^2}\right),
\end{equation}
where the ellipticity magnitude is \( e = \sqrt{e_1^2 + e_2^2} \). The mode of the Rayleigh distribution—which corresponds to the dispersion of each underlying normally distributed ellipticity component—is fixed at \(\sigma_{\mathrm{e}} = 0.25\), consistent with values adopted in previous lensing simulation studies 
(e.g., H21) and supported by recent observational results \citep{Gatti2021_DESY3ShapeCat}.  
The distribution is commonly truncated at a maximum ellipticity between 0.7 and 0.9 \citep[e.g.,][]{great08_results,tewes2019}; here, we adopt a maximum intrinsic ellipticity of 0.8, corresponding to a maximum post-sheared galaxy ellipticity of approximately $0.87$.
The unlensed galaxy position angle \(\phi\) is uniformly distributed in the range \([0, \pi)\).

For the morphology distribution, we simulate a two-type population comprising single-component disc ($n_{\mathrm{s}}=1$) and elliptical ($n_{\mathrm{s}}=4$) galaxies, with discs comprising 80\% of the total. This simplification is justified because model-fitting biases are not addressed in this work.

\subsection{Signal-to-noise}
\label{subsec:snr}
\begin{figure*}
\centering
\includegraphics[width=0.9\textwidth]{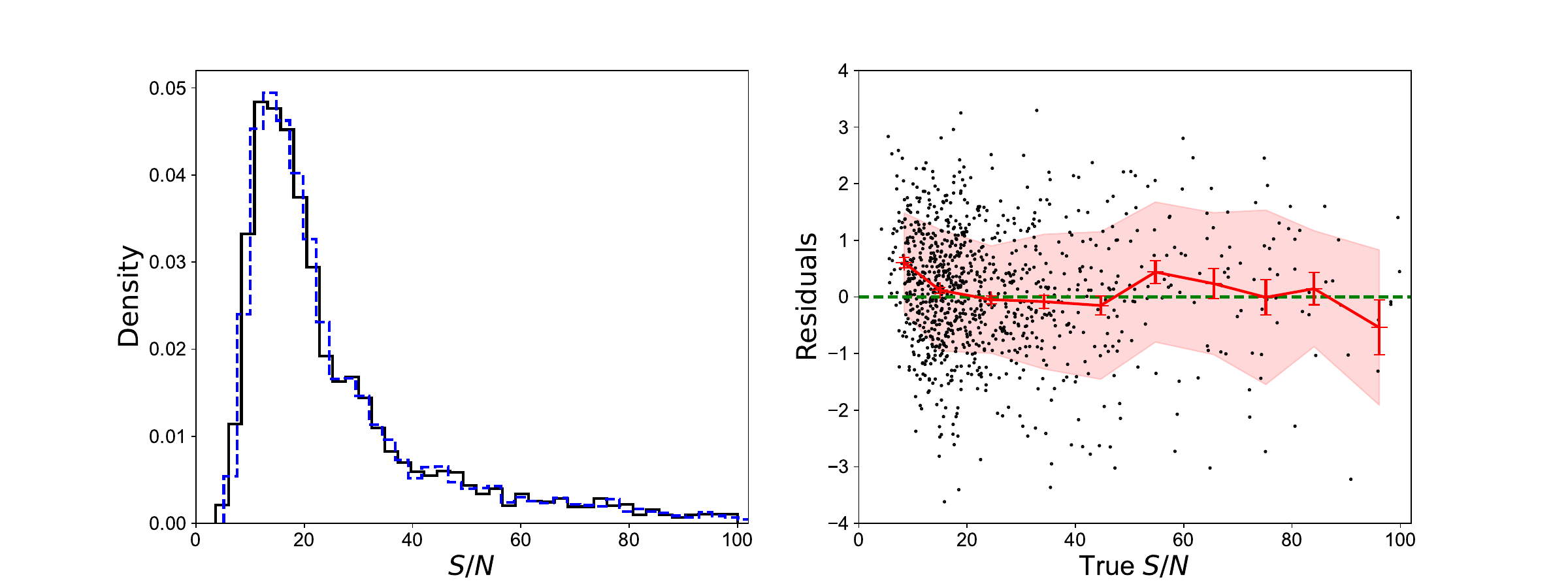}
\caption{Signal-to-noise distribution (left) for an independent set of $10^4$ galaxies, where the flux is obtained from (i) noise-free postage stamps (black solid) and (ii) predicted from noisy stamps using CNN$_{\mathrm{snr}}$ (blue dashed). Residuals (right) show predicted minus true $S/N$ for a random subset of $10^3$ galaxies (black dots). The red points show the binned mean residuals with error bars indicating the standard error on the mean, and the light red shaded band indicates the sample standard deviation within each bin. The green dashed line marks zero residuals. Results are shown for zero shear and test set fiducial galaxy parameter values.}
\label{fig:snr_hist_res}
\end{figure*}

The signal-to-noise ratio is defined as
\begin{equation}
    S/N = \frac{\sqrt{\sum I^2}}{\sigma_{\mathrm{n}}},
\label{eqn:snr}
\end{equation}
where the sum is taken over all pixels in the postage stamp and $\sigma_{\mathrm{n}}^2$ is the variance of the constant Gaussian noise added to each pixel. The peak intensity, $I_0$ (see Equation~\ref{eqn:gal_intensity}), is related to the flux, $F$, through the equation
\begin{equation}
    I_0=\frac{F}{2\pi n_{\mathrm{s}}k^{-2n_{\mathrm{s}}}r_{\mathrm{h}}^2\Gamma(2n_{\mathrm{s}})},
\end{equation}
where $F=F_0 10^{-0.4m_{\mathrm{AB}}}$ and $\Gamma$ is the gamma function. 
We set $F_0/\sigma_{\mathrm{n}}$ such that the resulting signal-to-noise distribution peaks at 
$S/N \sim 10$--20 (see Figure~\ref{fig:snr_hist_res}), consistent with the expected \emph{Euclid} galaxy populations \citep[e.g.][see their Fig.~21]{2025A&A...697A...2E}.

In both the training and test sets, we exclude galaxies with $S/N > 100$. We impose no lower limit on the signal-to-noise in the training stage. We note that imposing a lower limit on the signal-to-noise ratio during training—even if lower than the test set cut—results in larger biases.

\section{Simulating the images}
\label{sec:sims}
In this section, we describe the simulations used to generate PSF-convolved galaxy images on postage stamps with 0.1\,arcsec pixels.
We follow a similar procedure to that described in previous work \citep{voigt_modelBias_2010,voigt2024} and adopted for the shape measurement pipeline \texttt{IM3SHAPE} \citep{2013MNRAS.434.1604Z}. 

Galaxy and PSF profiles are simulated separately on convolution grids\footnote{The grids used for numerically convolving the galaxy with the PSF are larger than the final postage stamps to avoid edge effects.} with $n_{\mathrm{pix}}$ image pixels per side, where each pixel is subdivided into $n_{\mathrm{conv}}^2$ subpixels. For exponential profiles, the flux in each subpixel is computed assuming the intensity is constant across the subpixel and equal to the value at its center. For de Vaucouleurs profiles, to accurately capture the steep central peak, subpixels within the central $3 \times 3$ image pixels are further subdivided into $n_{\mathrm{int}}^2$ sub-subpixels, and the intensity is integrated over these. We use $n_{\mathrm{conv}}=3$ and $n_{\mathrm{int}}=9$; increasing these values does not significantly change the intensity profiles. 
The galaxy intensity is set to zero for $(\mathbf{x}-\mathbf{x}_0)^T\mathbf{C}(\mathbf{x}-\mathbf{x}_0) > R_{\rm cut}^2$ (see Equation~\ref{eqn:gal_intensity}), with $R_{\rm cut}=4$ so that the truncation occurs at semi-axes $4a$ and $4b$.

The BSG is positioned randomly within the central pixel of the postage stamp, with its centre coordinates drawn independently as 
\begin{equation}
    x_{0,\mathrm{b}}, y_{0,\mathrm{b}} \sim \mathcal{U}(-0.05, 0.05) \,\mathrm{arcsec}.
\end{equation}
If a faint neighbouring galaxy is present, its position is sampled randomly within a square region centred on the BSG, with side length $2\theta_{r}$, such that
\begin{equation}
    x_{0,\mathrm{f}} = x_{0,\mathrm{b}} + \mathcal{U}(-\theta_{r}, \theta_{r}), \quad
    y_{0,\mathrm{f}} = y_{0,\mathrm{b}} + \mathcal{U}(-\theta_{r}, \theta_{r})
\end{equation}
and is simulated only if its centre lies within a circular region of radius \(\theta_{r}\) around the BSG, i.e., if
\begin{equation}
    (x_{0,\mathrm{f}} - x_{0,\mathrm{b}})^2 + (y_{0,\mathrm{f}} - y_{0,\mathrm{b}})^2 \leq \theta_{r}^2.
\end{equation}
The fiducial value for $\theta_{r}$ adopted in this paper is 3\,arcsec, and the positions described above correspond to the lensed coordinates of the galaxies.

In practice, BSG and faint galaxy intensity profiles are simulated separately on convolution grids, summed, and then convolved with the PSF. The convolution is performed using the \texttt{convolve2d} function from the \texttt{signal} module in SciPy \citep{2020SciPy-NMeth}. Following convolution, the images are binned and cropped to produce postage stamps of size $n_{\mathrm{stamp}} \times n_{\mathrm{stamp}}$ pixels. In this work, we set $n_{\mathrm{stamp}} = 17$ and $n_{\mathrm{pix}} = 19$.

\section{The shear measurement method}
\label{sec:shear_measurement_method}
A wide range of techniques have been developed to infer weak gravitational lensing shear from galaxy shapes \citep[see review article][and references therein]{Mandelbaum2018review}. These include moment-based approaches, such as KSB \citep{1995ApJ...449..460K}, which were used in early detections of cosmic shear \citep{Bacon_2000, Kaiser2000, VanWaerbeke2000, Wittman_2000}, and model-fitting methods, such as \textsc{im3shape} \citep{2013MNRAS.434.1604Z} and \textsc{Lensfit} \citep{Miller2007}, adopted in later surveys \citep[e.g.,][]{Miller2013, Troxel2018, wright2025kidslegacycosmologicalconstraintscosmic}.
As survey data improve and systematic requirements tighten, increasingly sophisticated techniques have been developed to estimate shear with reduced biases. These include simulation-calibrated model-fitting methods \citep[e.g.,][]{2017MNRAS.465.1454H, Zuntz2018, Li2023} and self-calibrating approaches such as \textsc{metacalibration} \citep{2017arXiv170202600H}, which was employed in the DES Year 3 analyses \citep{Amon2022_DESY3, Secco2022_DESY3}.
In recent years, machine learning techniques have also been explored for shear estimation \citep[e.g.][]{tewes2019, Ribli_2019}. 
In this work, we employ a CNN-based shear measurement method introduced by \citet{voigt2024}. This approach avoids noise bias without relying on external calibration, making it particularly well-suited for isolating the effects of galaxy blending on shear estimation. We describe the method in detail below.

\subsection{The CNN model architecture}
\label{subsec:arch}
The shape measurement method employs two committees of shallow CNNs; one for estimating the first component of the lensed ellipticity, $e^{\mathrm{l}}_1$, and another for the second component, $e^{\mathrm{l}}_2$, for each galaxy in the catalogue. These are the shear estimators, $\gamma^{\mathrm{est,b}}_1$ and $\gamma^{\mathrm{est,b}}_2$, given in Equation~\ref{eqn:shear_est}. 
We refer to each CNN model within a committee as CNN$_{e_i}$. 
The model architecture 
is summarised in Table~\ref{tab:arch} and described in detail in \citet{voigt2024}. 
In brief, for each CNN$_{e_i}$, PSF-convolved postage stamps are fed into the first convolutional layer \footnote{tensorflow.keras.layers.Conv2D}, consisting of $n_{\mathrm{fil}}$ filters 3 by 3 pixels across. We use a stride of one and do not include any padding, resulting in $n_{\mathrm{fil}}$ feature maps on grids with width $(n_{\mathrm{stamp}}-2)$, where $n_{\mathrm{stamp}}$ is the width of the postage stamp in image pixels. The activation function adopted for this layer is a Rectified Linear Unit \citep[ReLU;][]{nair2010rectified}. The output from the first layer is flattened\footnote{tensorflow.keras.layers.Flatten} and passed through a dense layer\footnote{tensorflow.keras.layers.Dense} with a hyperbolic tangent activation function, ensuring the output lies within the valid ellipticity range $[-1, 1]$. 
The total number of trainable parameters in each CNN$_{e_i}$ is $9,401$. Each committee consists of 31\footnote{We initially trained a larger ensemble; the committee size reflects the number of models that converged successfully during training.} independently trained CNN$_{e_i}$ models, and the per-component ellipticity estimate for each galaxy is computed as the mean over the predictions of the committee members (see Section~\ref{subsec:shear_est_procedure}).

\begin{table*}
\caption{Architecture of the CNN models used to estimate shear (CNN$_{e_i}$) and $S/N$ (CNN$_{\mathrm{snr}}$). $n_{\mathrm{stamp}}$ is the width (in pixels) of the square postage stamp image; $n_{\mathrm{fil}}$ is the number of convolutional filters; $n_{\mathrm{batch}}$ is the number of samples per training batch, set equal to the number of noise realisations $n_{\mathrm{real}}$. The model does not include pooling or dropout layers. The convolutional layer uses a Rectified Linear Unit (ReLU) activation function, and in the dense layer, CNN$_{e_i}$ uses a hyperbolic tangent activation while CNN$_{\mathrm{snr}}$ uses a linear activation.}
\centering
\begin{tabular}{cccc}
\hline
Layer & Layer type & Output shape & Trainable parameters \\
\hline
1 & Convolution (2D) & $\left(n_{\mathrm{batch}},\, n_{\mathrm{stamp}}{-}2,\, n_{\mathrm{stamp}}{-}2,\, n_{\mathrm{fil}}\right)$ 
  & $(3 \times 3 + 1) \times n_{\mathrm{fil}}$ \\
2 & Flatten & $\left(n_{\mathrm{batch}},\, (n_{\mathrm{stamp}}{-}2)^2 \times n_{\mathrm{fil}}\right)$ 
  & 0 \\
3 & Dense (fully connected) & $\left(n_{\mathrm{batch}},\, 1\right)$ 
  & $(n_{\mathrm{inputs}} + 1)$\textsuperscript{a} \\
\hline
\end{tabular}
\label{tab:arch}
\vspace{1ex}
\begin{flushleft}
\textsuperscript{a}Where $n_{\mathrm{inputs}} = (n_{\mathrm{stamp}}{-}2)^2 \times n_{\mathrm{fil}}$ is the number of flattened input features; the +1 accounts for the bias term.
\end{flushleft}
\end{table*}

\subsection{Training the CNN models}

The trainable parameters (i.e., the weights and biases) in each CNN$_{e_i}$ are optimised by minimising the difference between the measured (biased) lensed ellipticity, $e_i^{\mathrm{l,b}}$, and the true lensed ellipticity, $e_i^{\mathrm{l}}$, using a mean-square-bias (MSB) loss function \citep{2010ApJ...720..639G,tewes2019}, given by:
\begin{equation}
 \mathrm{MSB} = \frac{1}{n_{\mathrm{gal}}}\sum_{n=1}^{n_{\mathrm{gal}}}\left[\frac{1}{n_{\mathrm{real}}}\sum_{m=1}^{n_{\mathrm{real}}}\left(e_{i;n,m}^{\mathrm{l,b}}-e_{i;n,m}^{\mathrm{l}}\right)\right]^2, 
\label{eqn:msb}
\end{equation}
such that the total number of images used to train the network is $n_{\mathrm{gal}}\times n_{\mathrm{real}}$.
The MSB loss function is used to mitigate the noise bias \citep[][]{kacprzak_noiseBias_2012,refregier_noiseBias_2012} which arises if the standard mean-square-error (MSE) is used as the objective function. This bias occurs because ellipticity, $e_i$, is not a linear function of the pixel intensities.
Although the simulated galaxies in the training sets are not explicitly sheared, their target ellipticities are defined to represent the post-lensing (observed) values. This ensures that the training procedure remains consistent with the quantities predicted for the sheared galaxies in the test sets.

Each CNN$_{e_i}$ model in a committee is trained independently using a unique set of simulated images. Extending the hyperparameter study based on noise-free images in \citet[][see their Figure 6]{voigt2024}, we adopt values optimised with networks trained on noisy images. These values, summarised in Table~\ref{tab:cnn_parameters}, provide a practical balance between performance and training efficiency. Each training set contains $n_{\mathrm{gal}} = 5\times 10^{4}$ unique galaxy images, with $n_{\mathrm{real}} = 300$ noisy realisations per image, and each training batch consists of all realisations of a single galaxy.

\begin{table}
\centering
\caption{
Hyperparameters used to train the CNNs for shear (CNN$_{e_i}$) and signal-to-noise (CNN$_\mathrm{snr}$) estimation. Table~\subref{tab:cnn_shared} lists parameters with shared values, and Table~\subref{tab:cnn_specific} those with CNN-specific values. For shear estimation, predictions are averaged over a committee of 31 trained CNN$_{e_i}$ models; for $S/N$ estimation, we use a single CNN.
}
\label{tab:cnn_parameters}
\begin{subtable}{\linewidth}
\centering
\caption{Shared hyperparameters}
\label{tab:cnn_shared}
\begin{tabular}{l c}
\hline
Hyperparameter & Value \\
\hline
Number of filters ($n_{\mathrm{fil}}$) & 40 \\
Filter size (pixels) & $3 \times 3$ \\
Stride (pixels) & 1 \\
Epochs & 150 \\
Batch size  & $n_{\mathrm{real}}$ \\
Learning rate & $10^{-3}$ \\
\hline
\end{tabular}
\end{subtable}

\vspace{0.5cm} 

\begin{subtable}{\linewidth}
\centering
\caption{CNN-specific hyperparameters}
\label{tab:cnn_specific}
\begin{tabular}{l c c}
\hline
Hyperparameter & CNN$_{e_i}$ & CNN$_\mathrm{snr}$ \\
\hline
Training set size ($n_{\mathrm{gal}}$) & $5 \times 10^4$ & $10^5$ \\
Noise realisations per galaxy ($n_{\mathrm{real}}$) & 300 & 1 \\
\hline
\end{tabular}
\end{subtable}
\end{table}

We exclude faint galaxies from the training sets and simulate BSGs with properties described in Section~\ref{sect:pops}. We do not apply any shear or size truncation to galaxies in the training sets and include all ellipticity magnitudes up to $|e|<1$.

\subsection{Shear bias estimation procedure}
\label{subsec:shear_est_procedure}
Shear biases are assessed by applying the CNN committees for $e_1$ and $e_2$ to test sets, each comprising independent, noisy, sheared galaxies with a different known constant input shear, $\gamma_i$. The predicted shear, $\gamma_i^{\mathrm{est,b}}$, is obtained by averaging the CNN committee predictions over all galaxies in each test set (see Equations~\ref{eqn:mean_ellipticity} and~\ref{eqn:shear_estimate} below), and then compared with the true input values. Multiplicative and additive biases are defined by assuming a linear relation between the estimated and true shears (Equation~\ref{eqn:shear_est}).
To reduce shape noise, we adopt the standard approach of simulating galaxy pairs, where each pair consists of identical sources rotated by $90^\circ$ \citep{step2_2007}. Each test set contains $n_{\mathrm{test}}$ such pairs.

Using training and test sets provides a useful framework for quantifying different sources of shear bias. For example, \citet{voigt2024} examined biases arising from mismatches in PSF or galaxy populations between training and test data. In this paper, we focus on biases introduced by faint galaxies that are present in the test images but excluded from the training sets. This approach quantifies the bias that would occur in our shear measurements if we assumed no faint galaxies in the data, while in reality such faint galaxies—matching the population in the test sets—are present. In Section~\ref{sec:biases_noSat}, we demonstrate that biases are consistent with requirements when faint galaxies are absent from the test sets.

As described above, we compute the predicted $i^{\mathrm{th}}$ component of the ellipticity for the $j^\mathrm{th}$ galaxy in a test set by averaging the predictions over a committee of $n_{\mathrm{cnn}}$ independently trained CNN models:
\begin{equation}
\left< e_{i;j}^{\mathrm{l,b}} \right> = \frac{1}{n_{\mathrm{cnn}}}\sum_{k=1}^{n_{\mathrm{cnn}}} e_{i;j,k}^{\mathrm{l,b}},
\label{eqn:mean_ellipticity}
\end{equation}
where $e_{i;j,k}^{\mathrm{l,b}}$ is the $e_i$ estimate obtained from the $k^\mathrm{th}$ CNN$_{e_i}$ model.

The shear estimate from the full test set is then computed by averaging over all $2n_{\mathrm{test}}$ galaxies i.e., all pairs:
\begin{equation}
\gamma_i^{\mathrm{est,b}} = \frac{1}{2n_{\mathrm{test}}} \sum_{j=1}^{2n_{\mathrm{test}}} \left< e_{i;j}^{\mathrm{l,b}} \right>.
\label{eqn:shear_estimate}
\end{equation}

To estimate shear biases, we generate 25 test sets corresponding to all combinations of five equally spaced shear values per component, $\gamma_i = \{-0.05, -0.025, 0, 0.025, 0.05\}$. Each test set has a distinct pair of $(\gamma_1, \gamma_2)$ values. The resulting shear estimates from the CNN$_{e_i}$ committees are then fit with a linear regression model to determine multiplicative and additive biases \citep[see also][]{voigt2024}.

\section{Baseline shear biases and $S/N$ estimation}
\label{sec:biases_noSat}
In this section, we present a new method for estimating $S/N$ directly from noisy galaxy images and establish a baseline measurement of shear biases in the absence of faint-galaxy contamination. These baseline biases provide the reference against which the impact of undetected galaxies is later assessed, and we confirm that they lie within the \emph{Euclid} requirements.

Even without faint galaxy contamination, shear measurement methods are affected by systematic biases. 
Since this paper focuses on biases introduced by nearby undetected galaxies, we control or eliminate other sources of bias, as outlined below \citep[see][for a review of weak lensing systematics]{Mandelbaum2018review}. 

While noise bias is commonly calibrated, for the CNN method applied here it is already reduced below the required thresholds (as demonstrated in \citealt{voigt2024}).
Model-fitting biases are avoided by adopting identical galaxy profiles in both the training and test sets. In addition, galaxies are sampled from the same population distributions, with the exception that the training sets allow larger values of $r_{\mathrm{e}}$ (effectively unbounded) and ellipticity (up to the physical limit $|e|<1$) to account for shearing applied in the test sets. 

Biases from PSF mis-modelling are eliminated by using the same, known PSF in both training and test sets (see Section~\ref{sec:gal_psf_models} for the PSF model). Detection biases are also absent: our simulations include all galaxies drawn from the distributions in Section~\ref{sect:pops}, removing any bias associated with selection at detection.

We note that biases from mismatches between simulated and observed galaxy intensity profiles and population distributions, as well as between PSF models, are discussed in \citet{voigt2024}.

Another potential source of bias in weak lensing pipelines is selection cuts,
with analyses typically removing objects with PSF-convolved galaxy to PSF size ratios $< 1.25$ or $S/N < 10$, or both \citep{laureijs_2017}. 
In this study, we do not impose a minimum size cut, but a cut on signal-to-noise is applied with $S/N \geq 10$. 

In benchmark shear measurement studies—for example the GREAT08 challenge \citep{great08_results}—$S/N$ is often calculated from noise-free images. While suitable for controlled validation, this approach does not account for potential biases arising when $S/N$ must be estimated directly from noisy data. Typically, $S/N$ estimates for survey data are obtained
using tools such as \texttt{SExtractor} \citep{sextractor1996}. 

Here, we introduce a CNN-based method for estimating $S/N$ from noisy galaxy images. The network, which we refer to as CNN$_\mathrm{snr}$, is trained to predict $S/N$ from individual noisy postage stamps. Its architecture mirrors that used for shear estimation (see Table~\ref{tab:arch}), except that the dense layer uses a linear activation function.\footnote{We also tested a sigmoid activation to constrain the normalized $S/N$ between 0 and 1, but found the linear function yielded better performance.} The model is trained on $10^5$ noisy galaxy images, with target $S/N$ values computed from the corresponding noise-free stamps. The target $S/N$ is defined as the ratio of the galaxy’s total flux, integrated over the noise-free postage stamp, to the per-pixel Gaussian noise level $\sigma_{\mathrm{n}}$ (see Equation~\ref{eqn:snr}).
We found no benefit from using multiple noise realisations ($n_{\mathrm{real}} > 1$) or from employing a CNN committee. The CNN$_\mathrm{snr}$ hyperparameters are listed in Table~\ref{tab:cnn_parameters}.

Figure~\ref{fig:snr_hist_res} shows the ``true'' (from noise-free stamps) and predicted (from CNN$_\mathrm{snr}$) $S/N$ distributions for an independent set of $10^4$ galaxies, along with residuals (predicted minus true) for a random subset of $10^3$ galaxies. The mean residuals deviate significantly from zero in the lowest two bins ($0 < S/N \le 10$ and $10 < S/N \le 20$) and marginally in the bin at $50 < S/N \le 60$. 
However, even the largest mean residual (in the lowest $S/N$ bin) is only $\sim 0.6$ in $S/N$ units
and we find
that using the predicted $S/N$ instead of the true $S/N$ has negligible impact on the shear biases obtained in this work.

\begin{figure}
\centering
\includegraphics[width=0.85\linewidth]{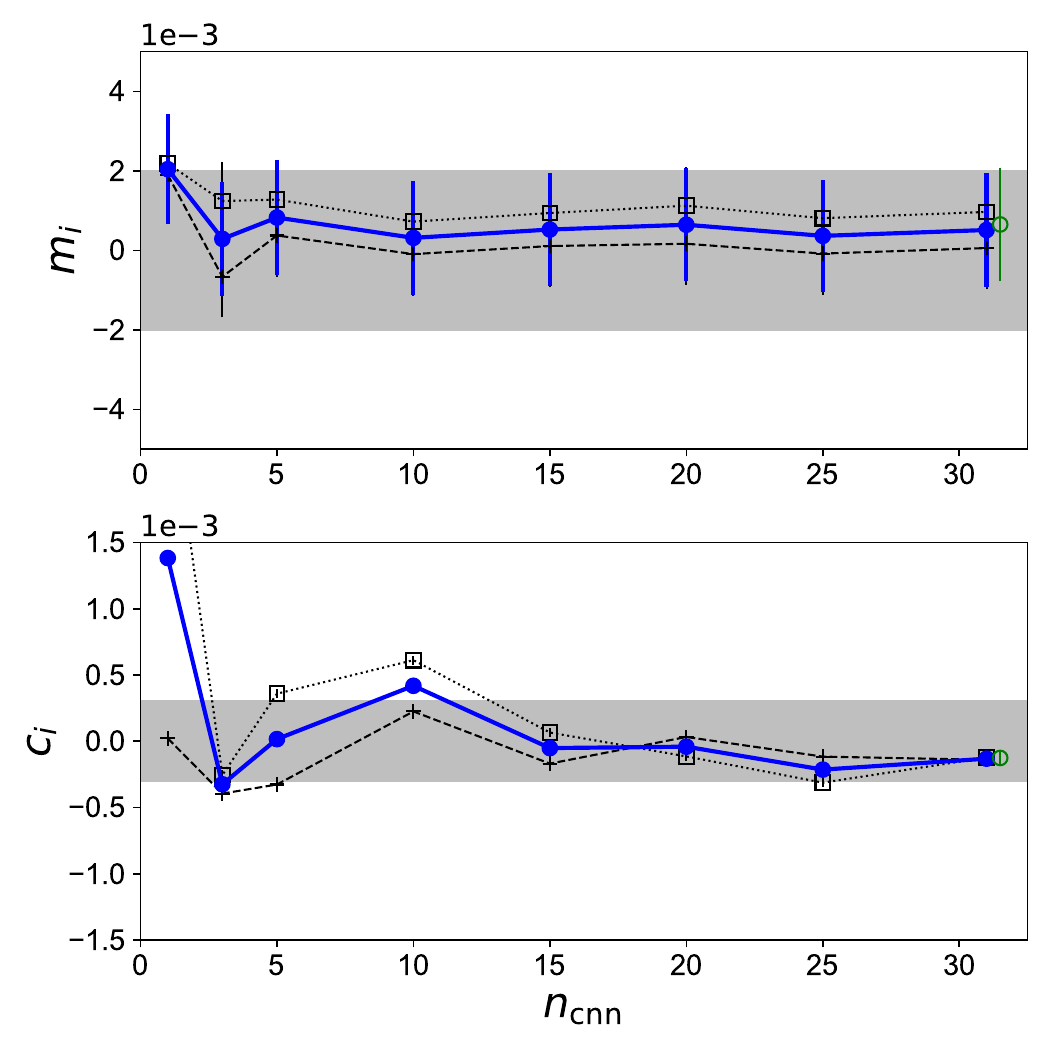}
\caption{
Multiplicative (top) and additive (bottom) biases adopting committees comprising $n_{\mathrm{cnn}}$ subsets of 31 trained CNN$_{e_i}$ models. 
A selection cut of $S/N \geq 10$, using the CNN$_{\mathrm{snr}}$ predictions, is applied.
Black crosses (squares) correspond to $i=1$ ($i=2$), while blue filled circles showing the mean across the two components. Green open circles indicate biases obtained using the ``true" $S/N$ with $n_{\mathrm{cnn}}=31$ (offset for clarity). Shaded regions show the top-level \emph{Euclid} bias requirements (see Section~\ref{sec:ellip_shear}).
}
\label{fig:mc_noSatellites}
\end{figure}

Figure~\ref{fig:mc_noSatellites} demonstrates that the multiplicative and additive biases obtained—excluding faint galaxies and applying a selection cut with $S/N\geq10$, estimated using CNN$_{\mathrm{snr}}$—are consistent with the \emph{Euclid} requirements. Biases are shown as a function of the committee size, $n_{\mathrm{cnn}}$. 
We find that multiplicative biases stabilise for $n_{\mathrm{cnn}} \gtrsim 5$, while additive biases stabilise for $n_{\mathrm{cnn}} \gtrsim 15$. In subsequent analyses we use all 31 trained CNN$_{e_i}$ models.

\section{Shear biases from faint galaxy contamination}
\label{sec:faint}

In this section, we assess the impact of faint, undetected galaxies (i.e., with magnitude $> 24.5$) on the accuracy of shear measurements made using CNN$_{e_i}$ committees trained on isolated bright galaxies (see Section~\ref{sec:shear_measurement_method}). Shear estimates ($\gamma_i^{\mathrm{est,b}}$) are obtained from postage stamps that include these faint contaminants, and the resulting shear biases are measured following the procedure outlined in Section~\ref{subsec:shear_est_procedure}, using 25 test sets. Galaxies with $S/N < 10$—estimated using CNN$_{\mathrm{snr}}$ (see Section~\ref{sec:biases_noSat})—are excluded from the sample. 
We find that the additive biases are negligible for all faint galaxy parameters explored. As such, we present only multiplicative biases for the remainder of the paper.

In the fiducial set-up, we simulate a random distribution of faint galaxies within each postage stamp, using the field densities provided in Table~\ref{tab:Nsat} and with an apparent magnitude distribution that matches the overall field distribution (see Section~\ref{sect:pops} and Table~\ref{tab:slopes}). Faint galaxy morphologies and ellipticities are drawn from the same distributions as the BSGs, described in Sections~\ref{sec:gal_psf_models} and~\ref{sect:pops}. 
We apply the same shear to the faint galaxies that is applied to the BSGs. All galaxies up to a limiting magnitude of 29 and within 3\,arcsec of the BSG centre are included in the simulations. This fiducial configuration provides the baseline for our tests; in subsequent sections we explore how the results change under different assumptions about the faint population. 

Assuming a random spatial distribution, the expected number of faint galaxies within a  circular region of radius $\theta_{r}$\,arcsec
and up to limiting magnitude $m_{\mathrm{lim}}$, is given by
\begin{equation}
\left<N_{\theta_{r}}\right> = \left[N(m_{\mathrm{lim}}) - N(24.5)\right] \times \pi \left(\frac{\theta_{r}}{60}\right)^2,
\end{equation}
where $\langle N(m_{\mathrm{AB}})\rangle$ is the cumulative projected number density per arcmin$^2$ defined in Equation~\ref{eqn:num_cum}; here it is evaluated at $m_{\mathrm{lim}}$ and at $m_{\mathrm{AB}}=24.5$. Table~\ref{tab:Nsat} lists $\left<N_{\theta_{r}}\right>$ for $\theta_{r} = 3$\,arcsec. For instance, the mean number of faint galaxies within a 3\,arcsec radius
for $m_{\mathrm{lim}} = 29$ is 1.47.

Within the halo model \citep{Cooray2002}, the number of satellite galaxies within a dark matter halo follows a Poisson distribution \citep[][]{Zheng_2005_HOD}, with a mean (the halo occupation number) that scales with halo mass (see Appendix~\ref{app:bm}). Although not all BSGs are central galaxies, we similarly adopt a Poisson model for the number of faint galaxies within $\theta_{r}$\,arcsec of a BSG. This assumption is justified because the various contributions to the faint galaxy population around BSGs—comprising foreground, background, and halo member galaxies—can be treated as independent Poisson processes, whose sum is also Poisson-distributed. As such, for a given BSG, the number of faint galaxies within $\theta_{r}$\,arcsec is drawn from:
\begin{equation}
N_{\theta_{r}} \sim \mathrm{Po}\left(\left<N_{\theta_{r}}\right>\right),
\end{equation}
where $\mathrm{Po}(\lambda)$ denotes a Poisson distribution with mean $\lambda$. In this Section, we assume $\left<N_{\theta_{r}}\right>$ is constant across all BSGs, i.e., independent of BSG properties.

\begin{figure*}
\centering
\includegraphics[width=0.9\linewidth]{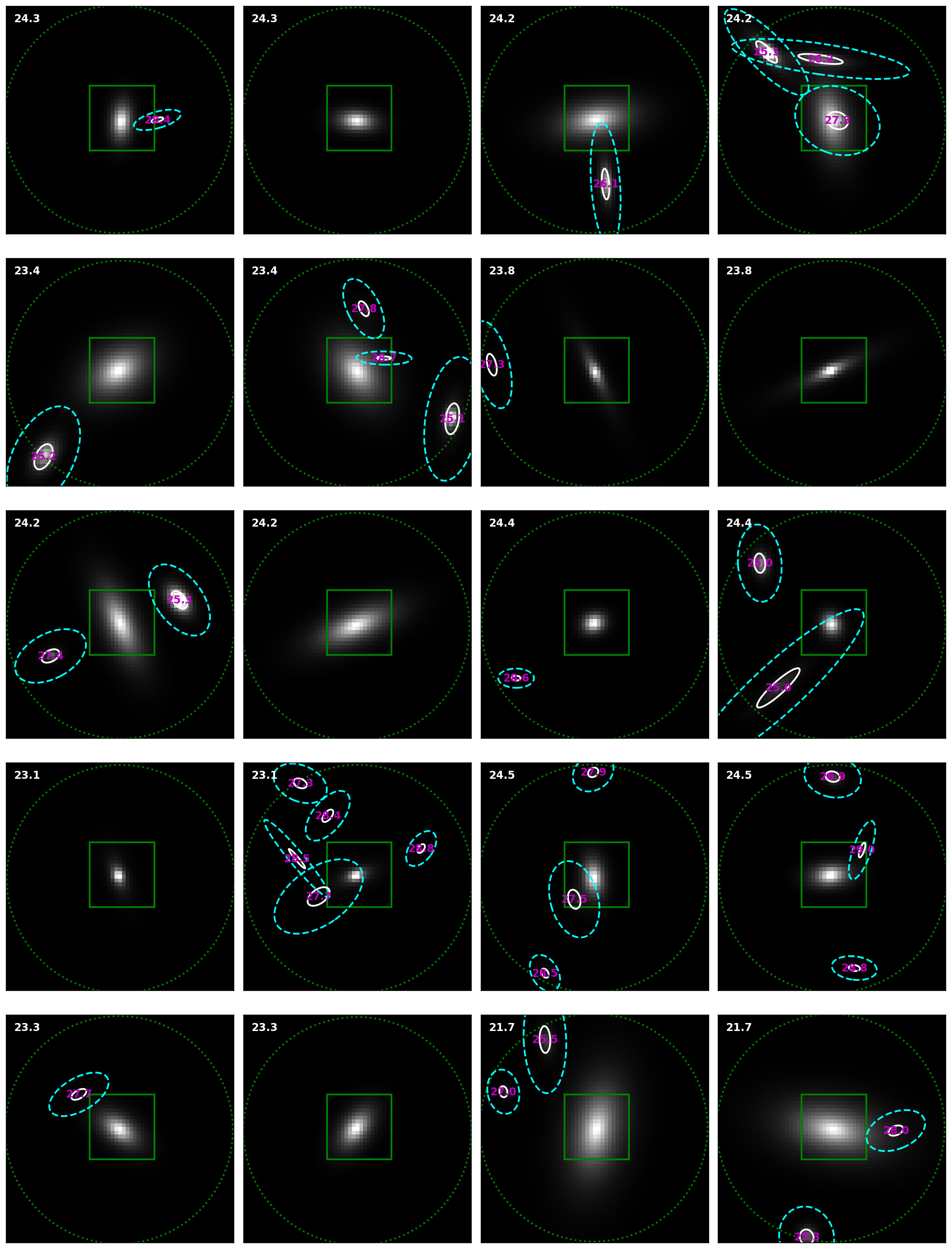}
\caption{Images showing 6 by 6\,arcsec$^2$ regions of sky centered on the BSG. Each panel displays the PSF-convolved image with zero applied shear, sampled at 0.1\,arcsec pixel scale, and scaled so that the central pixel has unit intensity. The images are shown without noise for 10 galaxy pairs (where each galaxy in a pair is the 90$^{\circ}$ rotated version of the other one). Faint neighbour galaxies are illustrated using ellipses that show their intrinsic (pre-PSF convolved) shapes with semi-major and semi-minor axes $a$ and $b$ (solid white; see Section~\ref{sec:gal_psf_models}), together with the truncation boundary at $4a$ and $4b$ (dashed cyan). Apparent magnitudes are indicated for the BSG (white; top-left corner of each panel) and for each faint neighbour (magenta; at ellipse centres). The green solid square denotes the 1.7 by 1.7\,arcsec cut-out region used for shear measurement. Only faint galaxies whose centre lies within the green dotted circle (with radius $\theta_{r}$\,arcsec, centred on the centre of the BSG), are simulated in the test sets. Shown for the field number density of faint galaxies to a limiting magnitude $m_{\mathrm{lim}}=29$ (see Table~\ref{tab:Nsat}).}
\label{fig:images_stamps}
\end{figure*}

Figure~\ref{fig:images_stamps} shows a random selection of BSGs and their associated faint neighbours on 6 by 6\,arcsec$^2$ grids. For each selected galaxy, we also show the $90^\circ$ rotated counterpart included in the test sets (see Section~\ref{subsec:shear_est_procedure}), each with independent faint galaxy realisations. The locations, sizes, ellipticities and orientations of faint neighbours are indicated using overlaid ellipses. Also shown are the faint galaxy circular inclusion regions (radius 3\,arcsec) and postage stamp cut-outs used for shear estimation. Intensities are displayed on a linear scale, so faint galaxies are only visible when their magnitudes are comparable to that of the BSG. Apparent magnitudes of both BSGs and faint neighbours are annotated.

In the following subsections, we examine deviations from the fiducial setup 
that may significantly influence the resulting shear biases—and thus require careful treatment in calibration simulations.

\subsection{Limiting apparent magnitude and clustering radius}
\label{subsec:mlim}

\begin{figure*}
\centering
\includegraphics[width=0.33\linewidth]{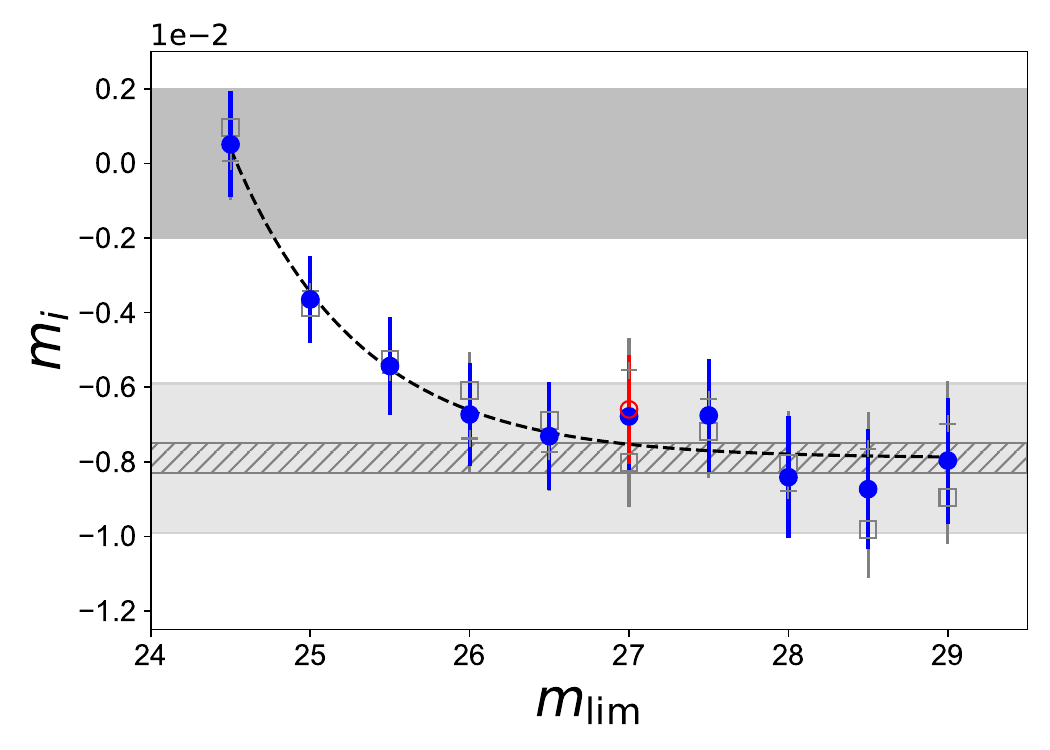}
\includegraphics[width=0.33\linewidth]{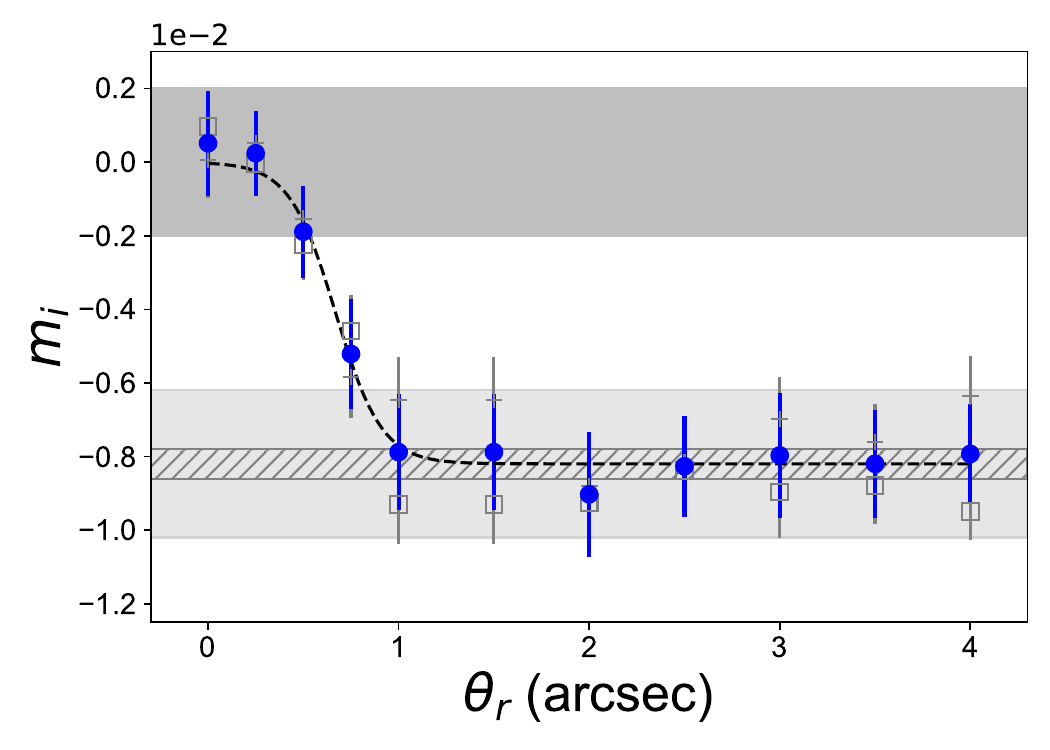} 
\includegraphics[width=0.33\linewidth]{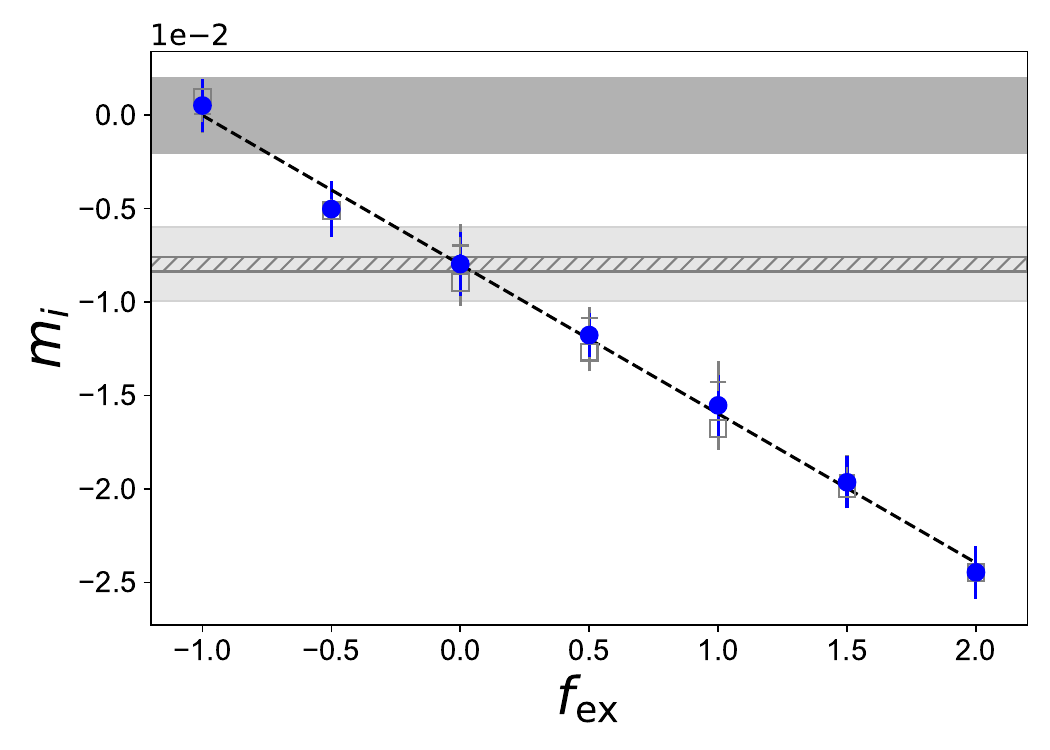} 
\caption{Multiplicative biases as a function of: the faintest galaxy magnitudes included in the postage stamps ($m_{\mathrm{lim}}$; left); the maximum radial distance between the BSG and faint galaxy centres ($\theta_{r}$; middle) and the excess faint galaxy density over the field density ($f_{\mathrm{ex}}$; right). 
$m_1$ ($m_2$) using the full model for the apparent magnitude distribution are shown as grey crosses (open squares), with blue solid circles indicating the mean bias across the two components. The red open circle (left) shows the corresponding mean bias value for the linear approximation model to the apparent magnitude distribution for $m_{\mathrm{lim}} \leq 27$. 
The dark grey shaded regions indicate the \emph{Euclid} bias requirement ($|m_i| < 2 \times 10^{-3}$). Black dashed lines show the best-fitting exponential (left), sigmoid (middle) and linear (right) regression models.
The lighter (hashed) regions show values within $\pm 2 \times 10^{-3}$ ($\pm 4 \times 10^{-4}$) of the asymptotic bias (left and middle) and the mean bias predicted by the model at $f_{\mathrm{ex}}=0$ (right).
}
\label{fig:mlim}
\end{figure*}

Figure~\ref{fig:mlim} shows how the multiplicative biases vary when changing either the faint galaxy limiting magnitude ($m_{\mathrm{lim}}$) or the maximum radial distance from the BSG centre ($\theta_{r}$). 
Faint galaxies are sampled from the apparent magnitude distribution given by Equation~\ref{eqn:num_cum}, and excluded from the postage stamps if their apparent magnitude is greater than the limiting value 
or their radial distance from the BSG centre exceeds $\theta_{r}$ (see Section~\ref{sec:sims}).
We find that faint contaminants introduce a multiplicative bias that becomes increasingly negative with $m_{\mathrm{lim}}$ up to $\sim 27$, and with separation up to $\theta_r\sim1\mathrm{arcsec}$. Similar trends have been reported in previous studies (e.g. H17, M19), although the absolute bias values, as well as the point at which they cease decreasing, depend on the implementation details, including the shape measurement method as well as the assumed properties of the faint population.

To characterise the trends we find more precisely and to obtain confidence intervals (CIs), we fit simple parametric models using error-weighted non-linear least squares regression. 
68\% CIs are estimated from 2000 Monte Carlo (MC) realizations, obtained by resampling each data point from a Gaussian with width given by its measurement error. 
In both cases, the asymptotic bias parameter ($a_{\ell}$ or $a_{\theta}$) describes the limiting value reached at large $m_{\mathrm{lim}}$ or $\theta_r$. 
We define the point at which the bias effectively flattens, $x^*$, as the minimum value of the independent variable where the model bias lies within $\Delta m = 4\times10^{-4}$ of its asymptotic value (i.e., within a factor of 5 of the top-level \emph{Euclid} requirement).  

For the dependence on limiting magnitude, we adopt an exponential form
\begin{equation}
m(m_{\mathrm{lim}}) = a_{\ell} + b_{\ell} \, \exp\left[-k_{\ell} \, (m_{\mathrm{lim}}-24.5)\right],    
\end{equation}
where $b_{\ell}$ sets the amplitude of the exponential term and $k_{\ell}$ controls the rate of flattening. 
The best-fit asymptotic bias is $a_{\ell}=-7.9\times10^{-3}$, corresponding to a flattening point $m_{\mathrm{lim}}^*=27.0$. 
From the MC realizations, we obtain $a_{\ell} \in [-9.4,-7.2]\times10^{-3}$ and $m_{\mathrm{lim}}^* \in [26.1,29.1]$, with medians $a_{\ell}=-8.1\times10^{-3}$ and $m_{\mathrm{lim}}^*=26.9$. 
Thus faint contaminants at least as faint as $m_{\mathrm{AB}}\sim26$ and potentially as faint as $m_{\mathrm{AB}}\sim29$ will need to be included in simulations.

For the dependence on clustering radius, we use a flipped sigmoid (3-parameter logistic) form
\begin{equation}
m(\theta_r) = \frac{a_\theta}{1 + \exp[-k_\theta (\theta_r - \theta_0)]} \,,
\end{equation}
where $k_\theta$ controls the steepness of the transition and $\theta_0$ is the midpoint.  
The best-fit asymptotic bias is $a_\theta=-8.2\times10^{-3}$, corresponding to a flattening radius $\theta_r^*=1.03\,\mathrm{arcsec}$. 
From the MC realizations, we obtain $a_\theta \in [-8.9,-7.6]\times10^{-3}$ and $\theta_r^* \in [0.79,1.25]$\,arcsec, with medians $a_\theta=-8.3\times10^{-3}$ and $\theta_r^*=0.99$\,arcsec. 
We verify that the choice of faint galaxy truncation has negligible effect: increasing the cut radius from the fiducial value of $R_{\rm cut}=4$ to $R_{\rm cut}=10$ does not significantly change the bias dependence on $\theta_r$.
The minimum required radius we find here is somewhat smaller than the $2.5$--$3$\,arcsec reported by M19, which may be due to the smaller postage stamps used in this study. 

In this section, we have adopted simple parametric forms that capture the observed trends with minimal parameters. Although the precise values of $x^*$ depend on the assumed functional form, the resulting CIs provide a useful indication of the depths to which calibration simulations may need to extend and the radii around BSGs within which faint galaxies must be included.

\subsection{Faint galaxy excess}
\label{subsec:excess}
We quantify the dependence of the biases on the mean number of faint galaxies around BSGs in terms of an excess relative to the mean across the field, denoted by $f_{\mathrm{ex}}$. The mean local number of faint galaxies within $\theta_{r}$ arcsec of a BSG is given by
\begin{equation}
\left<N_{\mathrm{\theta_{r};loc}}\right> = \left<N_{\mathrm{\theta_{r}}}\right> (1 + f_{\mathrm{ex}}),
\label{eqn:excess}
\end{equation}
where $f_{\mathrm{ex}}=0$ (the fiducial case) corresponds to the field density
(see also M19). In this section, we assume the excess is the same for all BSGs. In Section~\ref{sec:clustering}, we examine the impact of an excess that correlates with the apparent magnitude of the bright galaxy.

In Figure~\ref{fig:mlim}, we observe a strong dependence of the multiplicative biases on $f_{\mathrm{ex}}$. We fit a linear regression to the mean of $m_1$ and $m_2$, obtaining a slope of $-(8.0\pm0.2)\times 10^{-3}$ and an intercept of $-(8.0\pm0.3)\times 10^{-3}$, where the intercept represents the bias for a field-density level of faint galaxies. 
The slope implies that, for accurate calibration for \emph{Euclid}, the mean faint galaxy density 
close to BSGs must be known to within $\pm 0.25$, or to within $\pm 0.05$ to be a factor of 5 below the \emph{Euclid} requirement.

\subsection{Maximum Halo Occupancy}
As discussed in Section~\ref{sec:faint}, we sample faint galaxies from a Poisson distribution with mean $\left<N_{\mathrm{\theta_{r}}}\right>$. Since this distribution allows a non-zero probability of unrealistically high occupancies, we explore the impact of imposing an upper limit on the number of faint galaxies per bright galaxy, denoted $k_{\mathrm{max}}$, while keeping the mean number of faint galaxies across the BSG sample fixed at the field value.

For a given BSG, the number of nearby faint galaxies is drawn from a truncated Poisson distribution, whose probability mass function (PMF) is defined as:
\begin{equation}
 P(N=k)=   
 \begin{cases}
     e^{-\lambda}\lambda^k/k! & k=0,1,\dots,k_{\mathrm{max}} \\
     0 & k > k_{\mathrm{max}}
 \end{cases}
\end{equation}
where $\lambda$ is computed numerically so that the expected number of faint neighbours, given by $\mathrm{E}[N_{\mathrm{\theta_{r}}}]=\sum_{k=0}^{k_{\mathrm{max}}} kP(N=k)$, is equal to $\left<N_{\mathrm{\theta_{r}}}\right>$, for all $k_{\mathrm{max}}$.

\begin{figure*}
\centering
\includegraphics[width=0.33\linewidth]{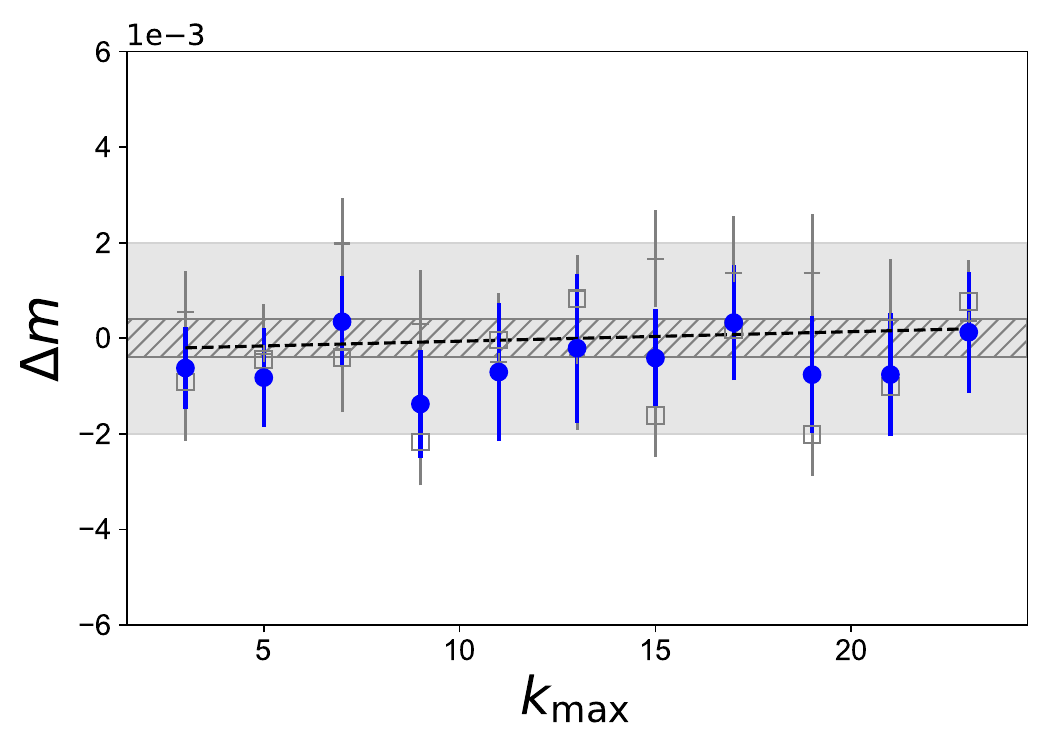} 
\includegraphics[width=0.33\linewidth]{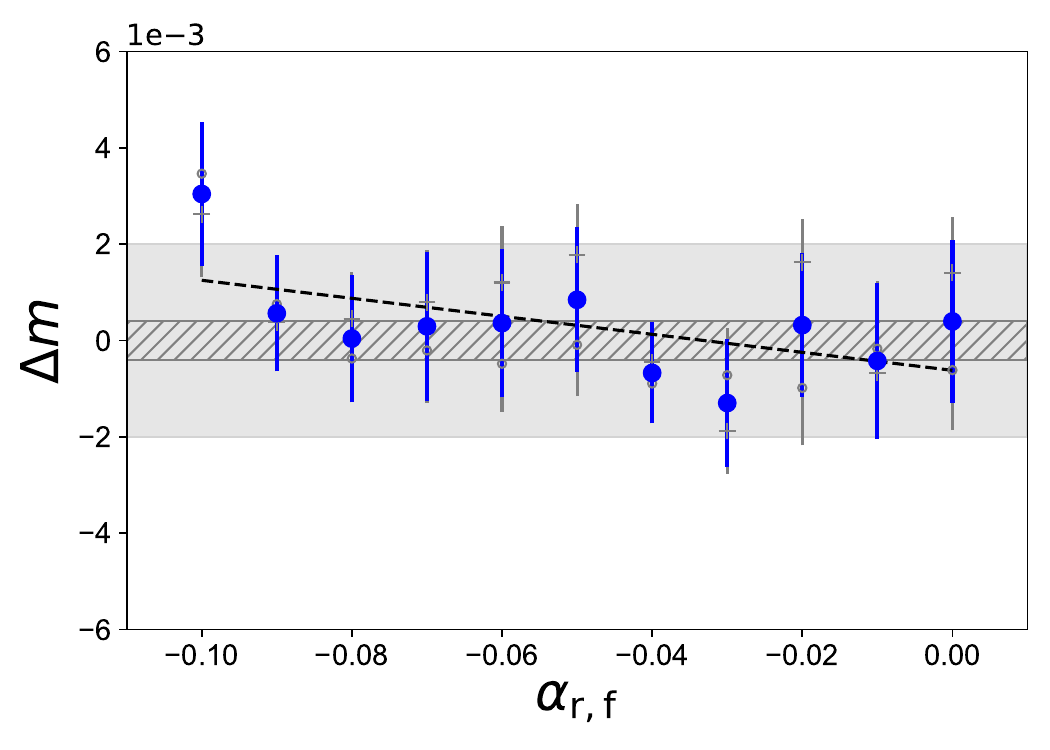} 
\includegraphics[width=0.33\linewidth]{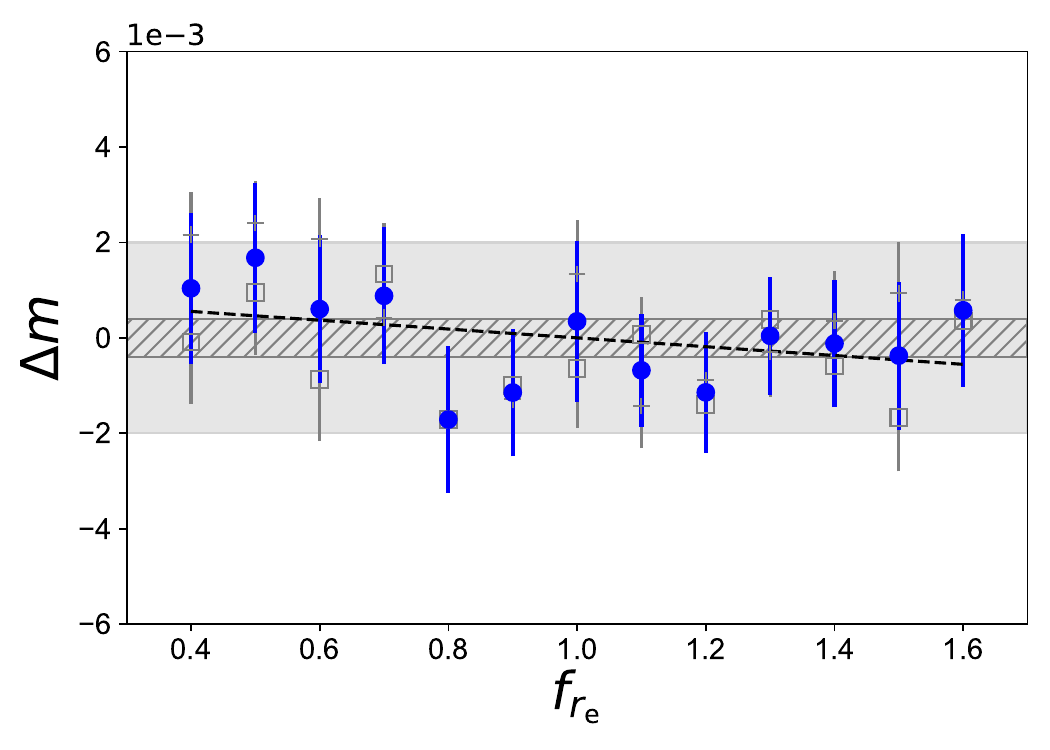} 
\caption{Residual multiplicative biases ($\Delta m$) as a function of the maximum halo occupancy ($k_{\mathrm{max}}$; left) and the slope ($\alpha_{\mathrm{r,f}}$; middle) and normalisation ($f_{r_{\mathrm{e}}}$; right) of the faint galaxy effective radius–apparent magnitude relation. $\Delta m_1$ ($\Delta m_2$) values are shown as grey crosses (open squares), with blue filled circles indicating the mean bias across the two components. Black dashed lines show the best-fit regression lines to the blue data points. 
For $k_{\mathrm{max}}$, biases are shown relative to the average over the mean biases (i.e., measured mean bias minus average mean bias). For $\alpha_{\mathrm{r,f}}$ and $f_{r_{\mathrm{e}}}$, biases are shown relative to those predicted by the best-fit regression models at the fiducial parameter values (i.e., measured mean bias minus predicted fiducial mean bias).
Results use the fiducial setup with the full model for the faint galaxy apparent magnitude distribution.
The light (hashed) region shows values within $\pm 2 \times 10^{-3}$ ($\pm 4 \times 10^{-4}$) of zero residual bias.}
\label{fig:kmax_alpharf}
\end{figure*}

Figure~\ref{fig:kmax_alpharf} shows the multiplicative biases as a function of $k_{\mathrm{max}}$ for the fiducial set-up. 
Fitting a linear regression model, we find that the biases are insensitive to the truncation threshold, with no evidence for a slope different from zero ($p$-value = 0.5).
This suggests that although a true Poisson distribution permits unphysical high-occupancy outliers, these rare cases have negligible impact on the overall shear bias, provided the mean number of faint galaxies is accurate. It is therefore acceptable to use an untruncated Poisson distribution in calibration simulations.

\subsection{Size--magnitude relation}
\label{subsec:alphar}
\begin{figure*}
\centering
\includegraphics[width=0.33\linewidth]{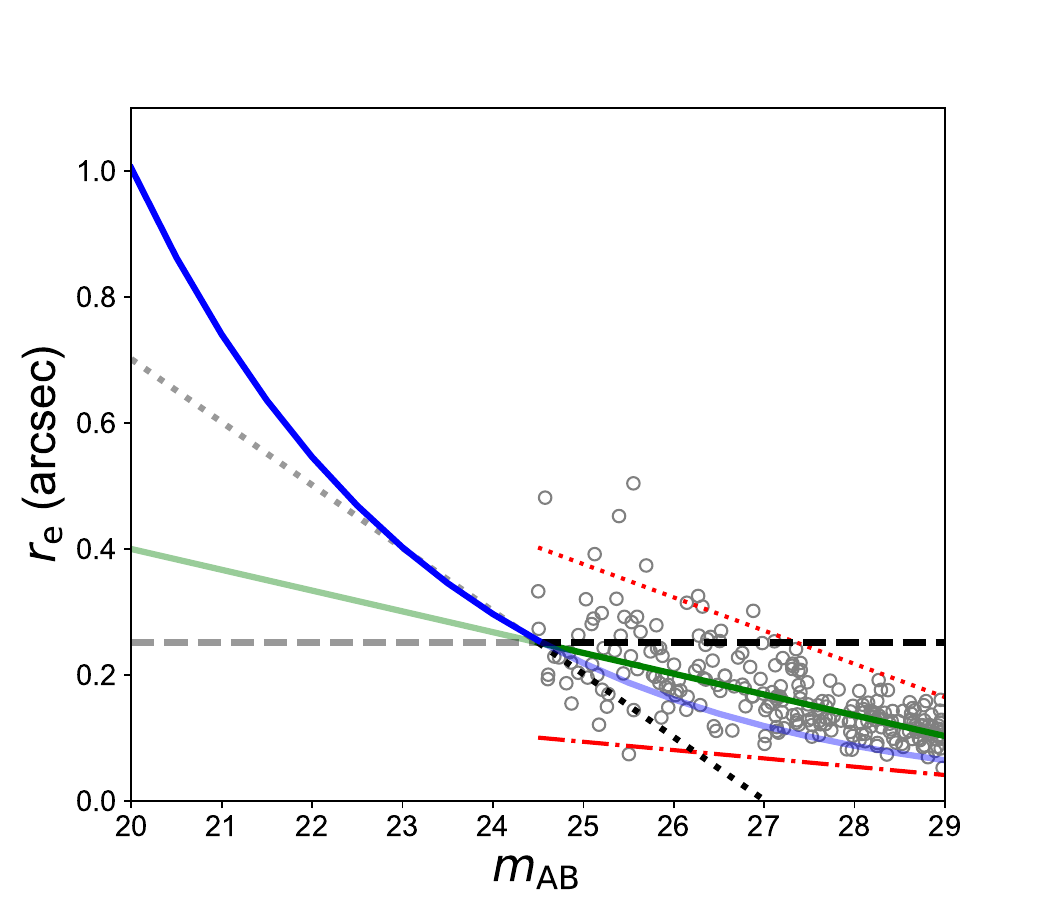} 
\includegraphics[width=0.33\linewidth]{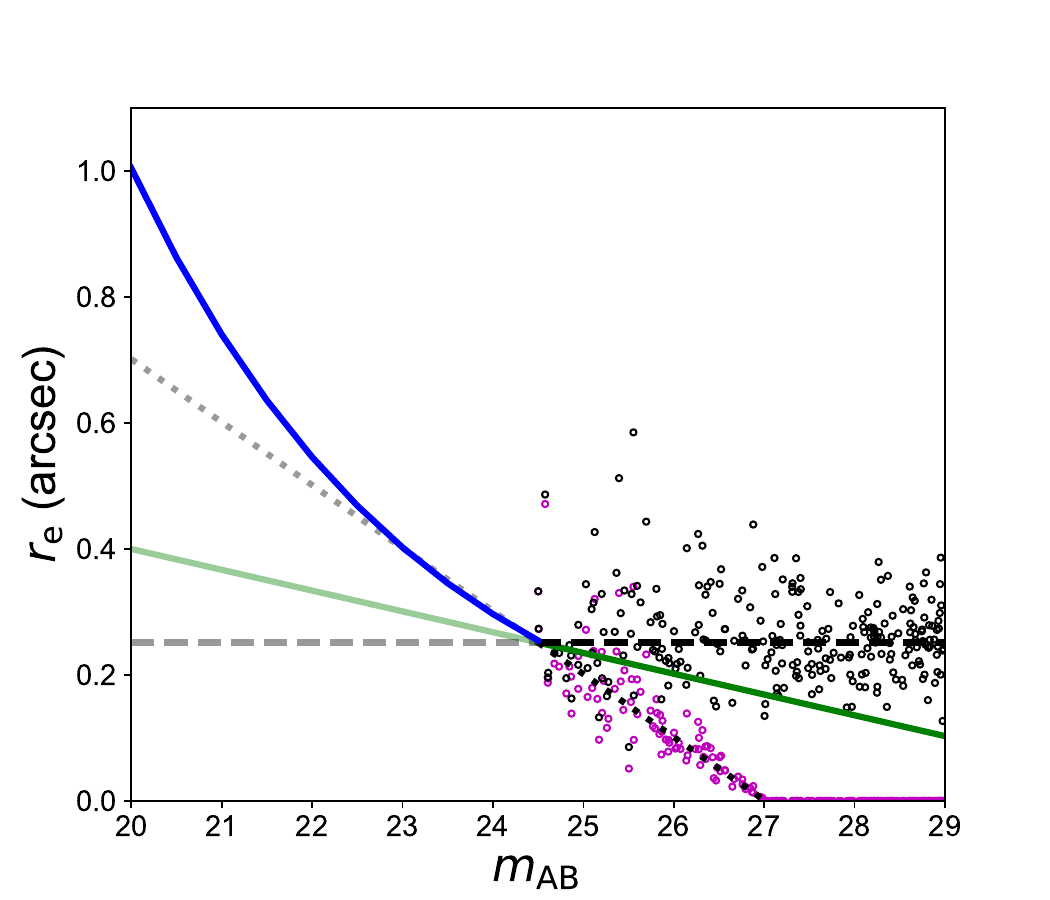} 
\includegraphics[width=0.33\linewidth]{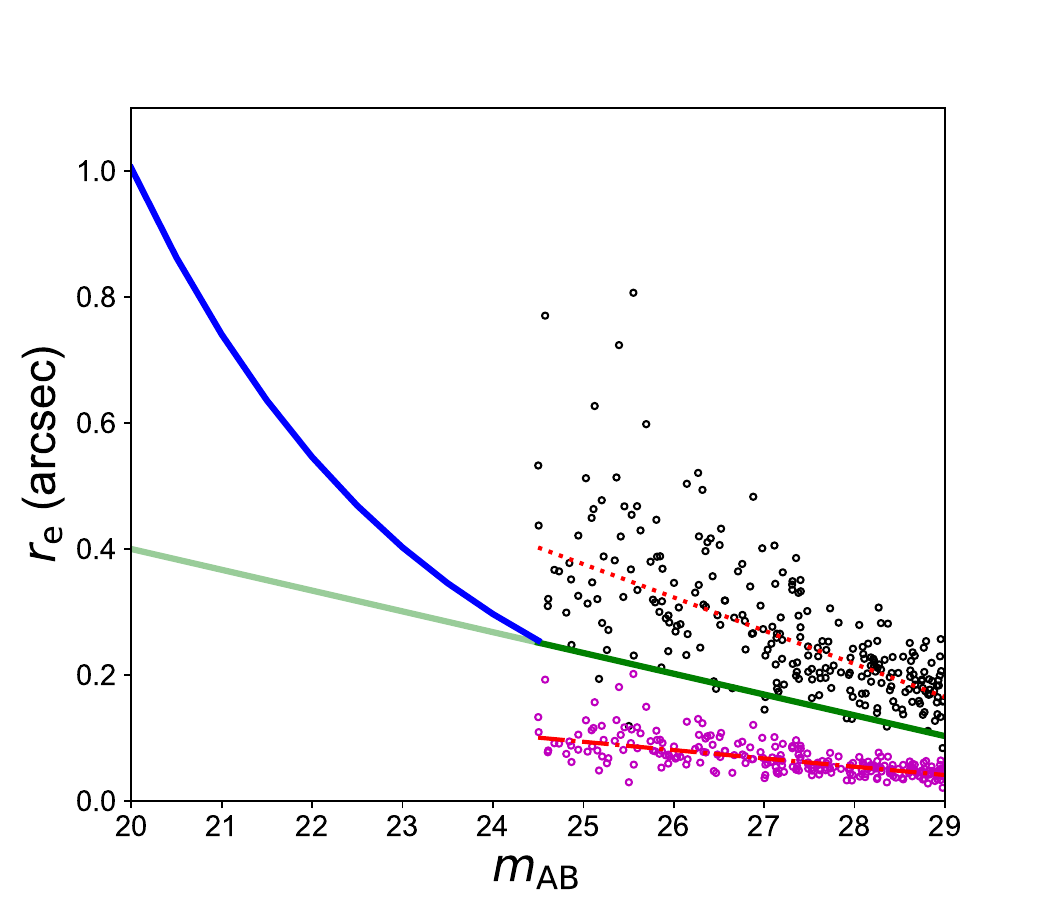} 
\caption{Relation between effective radius, $r_{\mathrm{e}}$, and apparent magnitude, $m_{\mathrm{AB}}$, used in this study. The blue solid line shows the relation for BSGs, and the green solid line for the fiducial faint population. Also shown are: the steep ($\alpha_{\mathrm{r,f}} = -0.1$; black dotted) and shallow ($\alpha_{\mathrm{r,f}} = 0$; black dashed) slopes for the faint population, and the upper ($f_{r_{\mathrm{e}}} = 1.6$; red dotted) and lower ($f_{r_{\mathrm{e}}} = 0.4$; red dash-dotted) size scalings, explored in Section~\ref{subsec:alphar}. The size–magnitude relation adopted for the bright sample in M19 corresponds to the black dotted line. Faint curves represent extrapolations beyond the region where the relation is applied. Open grey circles (left panel) show faint galaxies sampled from the fiducial population. Magenta and black small open circles show faint galaxies sampled at the minimum and maximum investigated values of $\alpha_{\mathrm{r,f}}$ (middle panel) and $f_{r_{\mathrm{e}}}$ (right panel), respectively.}
\label{fig:size-mag-relation}
\end{figure*}

We investigate how variations in the size--magnitude relation of faint galaxies (see Equations~\ref{eqn:size-mag_dispersion} and ~\ref{eqn:size-mag_faint}) affect the biases. Our fiducial slope matches that adopted for faint galaxies in M19. Here, we assess the sensitivity of the biases to changes in this slope — that is, to how galaxy size scales with magnitude. A steeper slope implies that faint galaxies are more compact, with their flux distributed over fewer pixels, while a shallower slope leads to more extended profiles. Figure~\ref{fig:size-mag-relation} illustrates the range of slopes we explore, from flat (no dependence of size on magnitude) to one consistent with the bright-end relation in M19 ($\alpha_{\mathrm{r,f}} = -0.1$).

The results, shown in Figure~\ref{fig:kmax_alpharf}, indicate that the slope of the size–magnitude relation has minimal impact on the biases. A linear regression yields a slope consistent with zero ($p = 0.07$). There is a slight suggestion that the absolute value of the bias decreases for the steepest slope, but this trend is likely not physical. Instead, it arises from the way our model treats faint galaxies with very small sizes: whenever a draw gives $r_{\mathrm{e}} < 0$, we set the effective radius to zero rather than re-sampling. This choice avoids distorting the magnitude distribution, but it means that very faint galaxies are effectively excluded from the simulations. As a result, when $\alpha_{\mathrm{r,f}} = -0.1$, the number of galaxies with $m_{\mathrm{lim}} > 27$ drops (see Figure~\ref{fig:size-mag-relation}), leading to a reduction in the contribution of the faintest sources to the measured bias.

We also examine how the biases respond to an overall scaling of faint galaxy sizes. This shifts the size–magnitude relation vertically and slightly alters its slope (see Figure~\ref{fig:size-mag-relation}). Results are shown in Figure~\ref{fig:mlim} for scaling factors between 0.4 and 1.6. For the parameter range explored, we find a minimal impact on the biases ($p$-value = 0.2).

\subsection{Faint galaxy shear}
\label{subsec:faint-shear}
In the fiducial set-up, faint galaxies are sheared by the same shear as the BSG, implicitly assuming that they are physically close and thus subject to the same lensing distortion. In reality, however, only a subset of faint neighbours projected on the sky lie at similar redshifts to the BSG; background and foreground galaxies experience different lensing due to matter distributions along their respective lines of sight, only partially correlated with those affecting the BSG.

To test the sensitivity of our results to this assumption, we vary the coherence between the shear applied to the BSG and that applied to the faint galaxies. Specifically, we model the shear of each faint galaxy as a linear combination of the BSG shear and a random shear component:
\begin{equation}
\gamma_{i,\mathrm{faint}} = \rho_{\mathrm{\gamma}}\,\gamma_{i,\mathrm{BSG}} + \sqrt{1 - \rho_{\mathrm{\gamma}}^2}\,\gamma_{i,\mathrm{rand}},
\label{eqn:gamma_faint_rho}
\end{equation}
where $\rho_{\mathrm{\gamma}} \in [0, 1]$ controls the level of shear coherence. Each random component $\gamma_{i,\mathrm{rand}}$ is drawn independently from a Gaussian distribution, $\gamma_{i,\mathrm{rand}} \sim \mathcal{N}(0, \sigma^2_{\gamma_i})$, with $\sigma_{\gamma_i} = 0.02$, representative of the cosmic shear variance on arcminute scales. 
Although this model does not capture the full statistical properties of the cosmic shear field—particularly non-Gaussianity on small scales and correlations between shear components—it provides a straightforward test of robustness to variations in the shear applied to faint galaxies.

This test smoothly interpolates between two limiting cases: $\rho_{\mathrm{\gamma}} = 1$, where faint galaxies experience identical shear to the BSG (the fiducial set-up), and $\rho_{\mathrm{\gamma}} = 0$, where the faint galaxy shears are entirely uncorrelated with the BSG shears. The results are shown in Figure~\ref{fig:align_asym_shear}, which plots the resulting shear bias as a function of $\rho_{\mathrm{\gamma}}$. Across the full range of shear coherence, 
we find no significant change in the measured bias, with no evidence for a slope different from zero ($p$-value = 0.25), indicating that our results are robust to assumptions about the lensing relationship between the BSG and faint neighbours.

\subsection{Faint galaxy alignments relative to the BSG}
\label{subsec:align}
In this section we investigate how alignments between faint galaxies and the BSG affect shear biases. We consider three effects: the orientation of the faint galaxy relative to the BSG centre, the relative orientations of the faint and bright galaxies, and the location of the faint galaxy with respect to the BSG major axis. These are examined in the following subsubsections. 
We also briefly discuss the physical mechanisms that give rise to them.

\subsubsection{Faint galaxy orientation}
We investigate the impact of alignments between faint galaxy orientations and the position and orientation of the BSG. Specifically, we consider:
(i) radial alignment, where the faint galaxy’s major axis lies along the line connecting the centres of the faint galaxy and the BSG ($0^{\circ}$ offset);
(ii) tangential alignment, where the major axis is perpendicular to this line ($90^{\circ}$ offset); and
(iii) parallel alignment, where the faint galaxy’s major axis is aligned with that of the BSG.
In all cases, the degree of alignment is quantified prior to lensing, and both the BSG and faint galaxies are subsequently lensed by the same shear. We discuss the validity of this approach further below.

Radial and parallel alignments arise when faint galaxies are physically close to the BSG. Radial alignments can occur when the BSG is a Bright Central Galaxy (BCG) and the faint galaxy is one of its satellites, while parallel alignments may be observed when both the BSG and the faint galaxy are satellites within the same halo. These alignments are attributed to tidal gravitational interactions and are examples of intrinsic alignments \citep[e.g.,][and references therein]{Mandelbaum2018review}. 
Tangential alignments, by contrast, are expected when faint galaxies projected close to the BSG lie at higher redshift, such that they are lensed by the BSG host halo. In this case, the tangential orientation of the faint galaxy represents the shear induced by the BSG halo itself.  
In all three cases, the BSG and faint galaxies are then subject to a similar foreground shear from matter between the BSG and the observer. This justifies our procedure of applying the same shear to both the BSG and faint galaxies after imposing the initial alignment.  

The faint galaxy orientation, $\phi_{\mathrm{f}}$, is drawn from a von Mises distribution\footnote{The von Mises distribution is the circular analogue of the normal distribution.} as follows:
\begin{equation*}
\phi_{\mathrm{f}} \sim \mathrm{vM}(\psi,\kappa_{\mathrm{vM}}),
\label{eqn:phi}
\end{equation*}
where $\psi = \phi$ for alignment with the BSG major axis. For alignment with the BSG position, $\psi = \theta_{\mathrm{p}} + \delta$, where $\theta_{\mathrm{p}}$ is the angle to the line joining the bright and faint galaxy centres, given by:
\begin{equation*}
\theta_{\mathrm{p}} = \arctan\left(\frac{y_{0,\mathrm{f}} - y_{0,\mathrm{b}}}{x_{0,\mathrm{f}} - x_{0,\mathrm{b}}}\right),
\label{eqn:theta}
\end{equation*}
with $\delta = 0^{\circ}$ ($90^{\circ}$) corresponding to radial (tangential) alignment. 
We note that faint galaxies are placed at random positions around the BSG and are not translated under shear. In reality, lensing also induces small positional shifts, which would require a full multi-plane ray-tracing treatment to model accurately. Modelling such shifts is beyond the scope of this work.

The von Mises concentration parameter, $\kappa_{\mathrm{vM}}$, controls the degree of alignment: $\kappa_{\mathrm{vM}} = 0$ corresponds to random orientations,\footnote{For $\kappa_{\mathrm{vM}} = 0$, the von Mises distribution reduces to a uniform distribution over the range $[0, 2\pi)$.} while $\kappa_{\mathrm{vM}} \gtrsim 100$ yields orientations tightly clustered around $\psi$, approximating perfect alignment. All position angles $\phi_{\mathrm{f}}$ drawn from the von Mises distribution are mapped to the range $[0^\circ, 180^\circ)$.

We vary $\kappa_{\mathrm{vM}}$ and quantify the degree of alignment using the statistic:
\begin{equation*}
A_{\psi} \equiv \left<\cos^{2}(\phi_{\mathrm{f}} - \psi)\right>,
\end{equation*}
where the average is taken over all faint galaxies in an unlensed BSG sample. For the range of $\kappa_{\mathrm{vM}}$ values we explore, $A_\psi$ varies from 0.5 (random orientations) to 1 (perfect alignment).
We note that the alignments we impose are quantified prior to lensing, whereas observationally alignments are measured post-lensing and would therefore appear weaker. This does not reduce the relevance of our approach: the pre-lensing alignments we quantify are the physically relevant quantities for shear bias, and our procedure captures their impact accurately.

In Figure~\ref{fig:align_asym_shear}, we show the biases associated with each type of alignment, together with linear regression fits. The impact is significant for radial ($p$-value = 0.001) and tangential ($p$-value = 0.002) cases, but not for parallel alignment ($p$-value = 0.2). Radial alignment increases the magnitude of the bias (regression slope $-(6.4 \pm 1.3)\times10^{-3}$), while tangential alignment reduces it (slope $(6.7 \pm 1.5)\times10^{-3}$). Overall, these results suggest that a realistic treatment of faint galaxy–BSG alignments is important for calibration simulations.

\subsubsection{Faint galaxy spatial distribution}
Satellite galaxies are known to exhibit anisotropic spatial distributions around the BCG (referred to in the literature as the `host' galaxy), typically aligning along the host's major axis\footnote{Some studies have reported minor-axis alignments—known as the Holmberg effect \citep[e.g.][]{Holmberg1969}—but this is generally observed for satellites at large projected separations. Since we are concerned with satellites close to the central, this effect is not expected to be significant here.} \citep[e.g.][]{Brainerd_2005,Wang_2018,liu2024originlopsidedsatellitegalaxy}. The strength of this alignment depends on both galaxy colour and morphology: red centrals with red satellites exhibit the strongest anisotropy, while systems with blue centrals typically show nearly isotropic satellite distributions \citep[][]{2008MNRAS.390.1133B}. Of particular relevance to this study is that alignment strength increases with decreasing projected separation from the central and may be more pronounced when the satellite is significantly fainter than its host \citep[][]{Yang_2006}.

These observational trends are broadly supported by structure formation simulations \citep[e.g.][]{2007MNRAS.378.1531K,Agustsson2006}, and they challenge the assumption in our baseline model that faint galaxies are isotropically distributed around BSGs. Although not all BSGs correspond to BCGs with nearby faint satellites, we nevertheless examine the impact of anisotropic faint galaxy distributions on shear bias estimates under this extreme scenario, in order to quantify a plausible "worst-case" bias.

\begin{figure}
\centering
\includegraphics[width=1\linewidth]{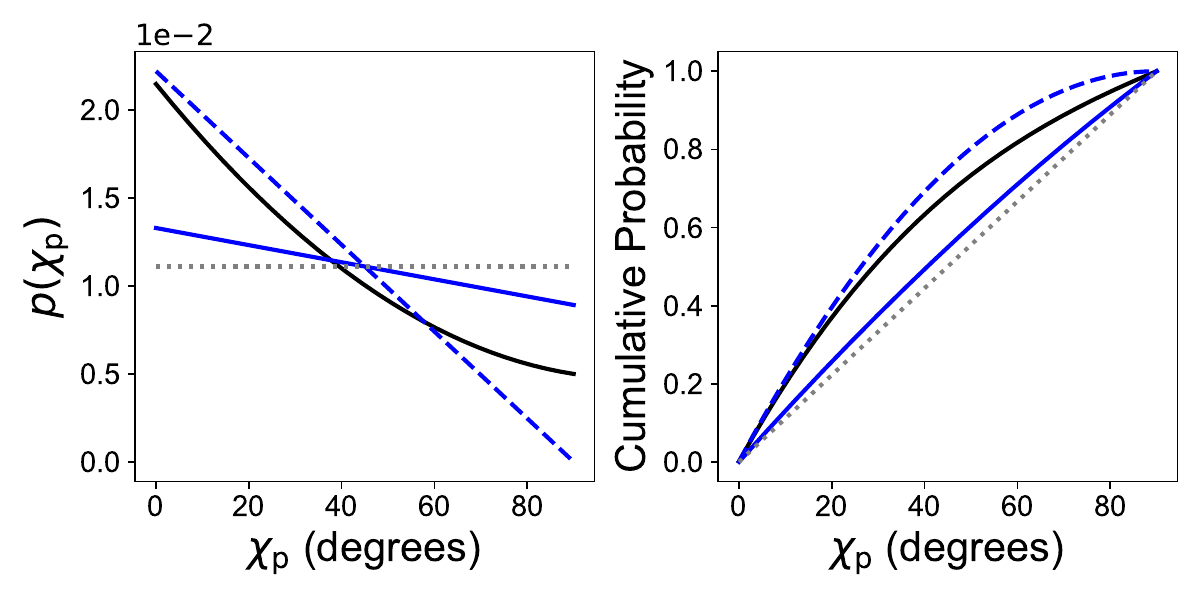}
\caption{
Probability density function (left; $p(\chi_{\mathrm{p}})$) and cumulative density function (right) for the angle $\chi_{\mathrm{p}}$ between the major axis of the BSG and the position of the faint galaxy. The fiducial model assumes a uniform distribution (grey dotted; $\mathbb{E}[\chi_{\mathrm{p}}] = 45^{\circ}$). Linear models are shown in blue: the solid line corresponds to the relation in \citet{Brainerd_2005} ($\mathbb{E}[\chi_{\mathrm{p}}] \approx 42^{\circ}$), while the dashed line represents a more extreme case with $\mathbb{E}[\chi_{\mathrm{p}}] \approx 30^{\circ}$. The black solid curve shows a quadratic model approximating the relation found in \citet{Agustsson2006}, with $\mathbb{E}[\chi_{\mathrm{p}}] \approx 34^{\circ}$.
}
\label{fig:anisotropy-pdf}
\end{figure}

We adopt two different forms for the probability density function (PDF) of the position angle, $\chi_{\mathrm{p}}$, of the faint galaxy relative to the major axis of the BSG. The first is a linear form, motivated by the results of \citet[][]{Brainerd_2005} based on isolated host galaxies in the Sloan Digital Sky Survey \citep[SDSS;][]{2000AJ....120.1579Y}:
\begin{equation}
p(\chi_{\mathrm{p}}) = \alpha_{\chi} \chi_{\mathrm{p}} + \beta_{\chi},
\end{equation}
and the second is a quadratic form, representing the results from $\Lambda$CDM simulations presented in \citet[][]{Agustsson2006}:
\begin{equation*}
p(\chi_{\mathrm{p}}) = \alpha_{\chi} + \beta_{\chi} \chi_{\mathrm{p}} + \gamma_{\chi} \chi_{\mathrm{p}}^2.
\end{equation*}
The parameter values are estimated from the plots in the respective papers and chosen to ensure proper normalisation of the PDFs over the range $[0^\circ, 90^\circ]$. These curves are shown in Figure~\ref{fig:anisotropy-pdf}, alongside the uniform distribution used as our fiducial model and an extreme linear case for comparison.

To construct the full two-dimensional angular distribution of faint galaxies while preserving alignment with the BSG’s major axis, we draw angles $\chi_{\mathrm{p}}$ from $p(\chi_{\mathrm{p}})$ over $[0^\circ, 90^\circ]$, and extend this to the full circle via the transformation:
\[
\chi_{\mathrm{p}}' = \epsilon_\chi \chi_{\mathrm{p}} + \delta_\chi,
\]
where $\epsilon_\chi \in \{-1, +1\}$ and $\delta_\chi \in \{0^\circ, 180^\circ\}$ are chosen randomly with equal probability. This construction yields a distribution that is symmetric about the major axis while retaining the alignment preference encoded in $p(\chi_{\mathrm{p}})$.

\begin{figure*}
\centering
\includegraphics[width=0.33\linewidth]{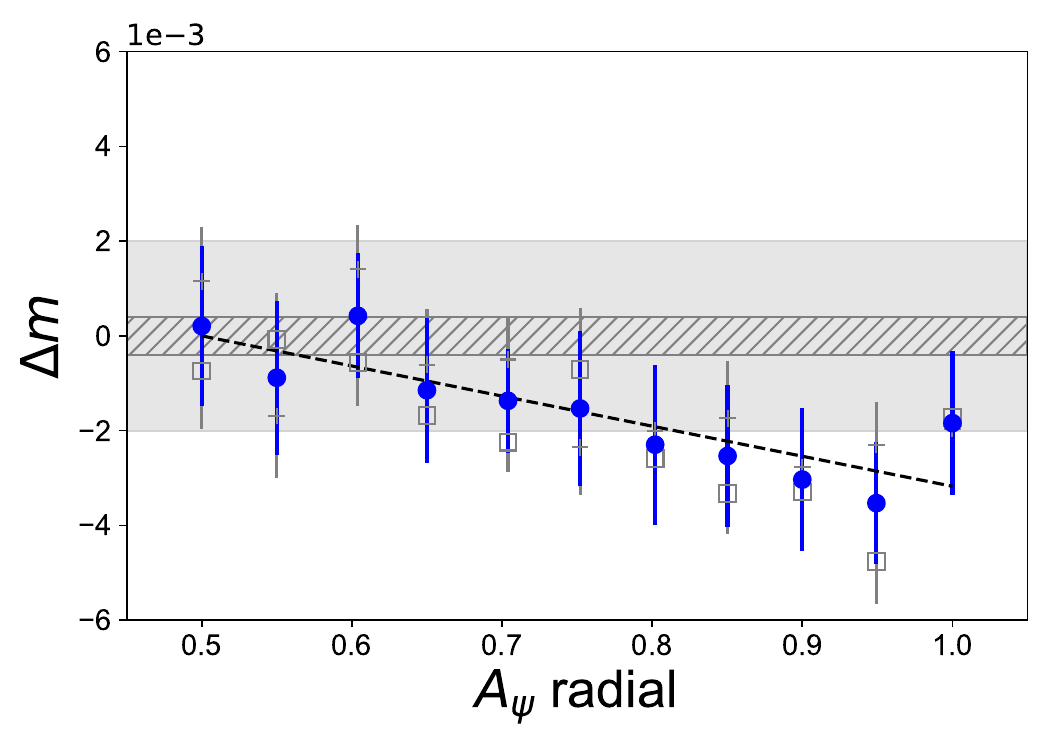} 
\includegraphics[width=0.33\linewidth]{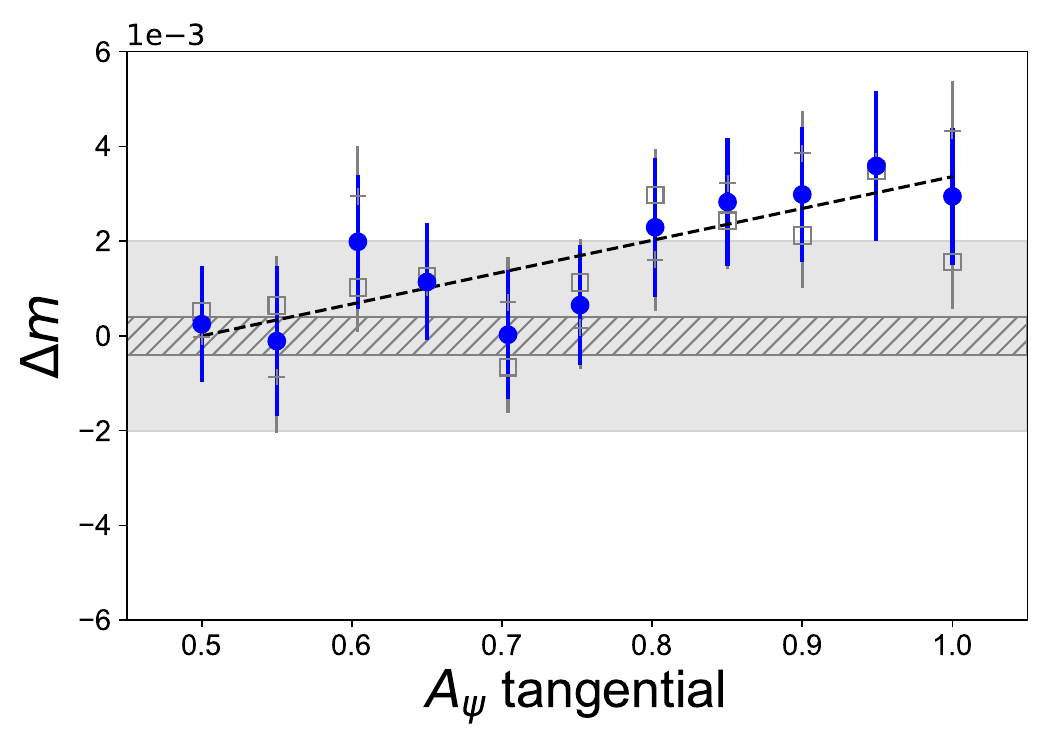} 
\includegraphics[width=0.33\linewidth]{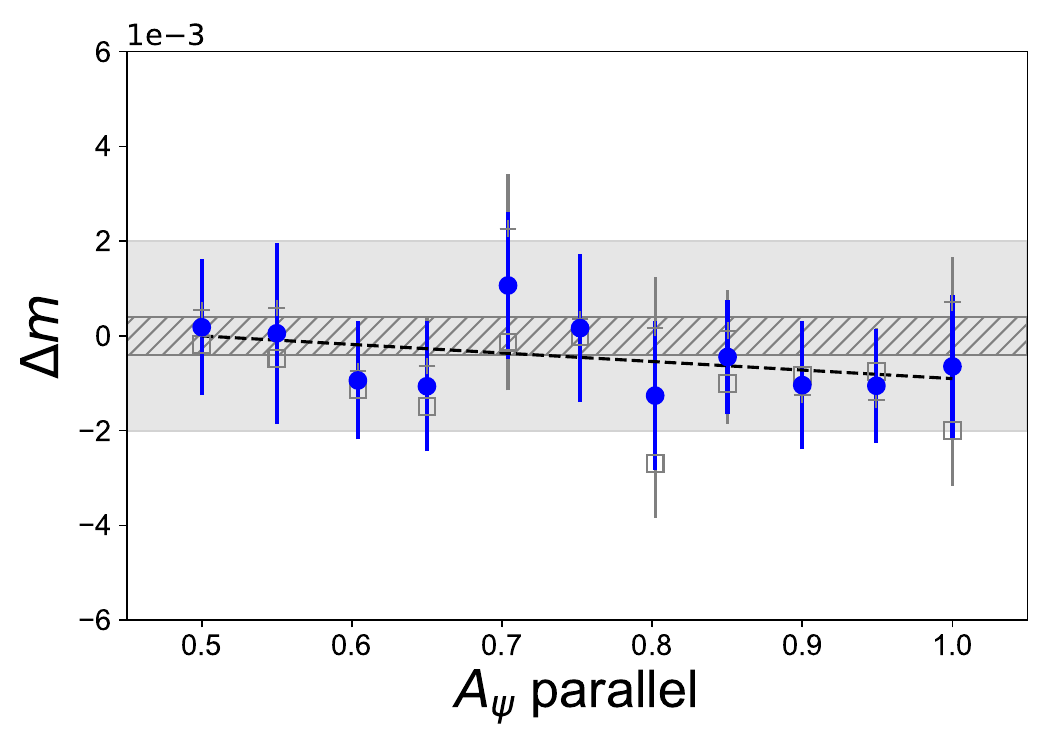} 
\includegraphics[width=0.33\linewidth]{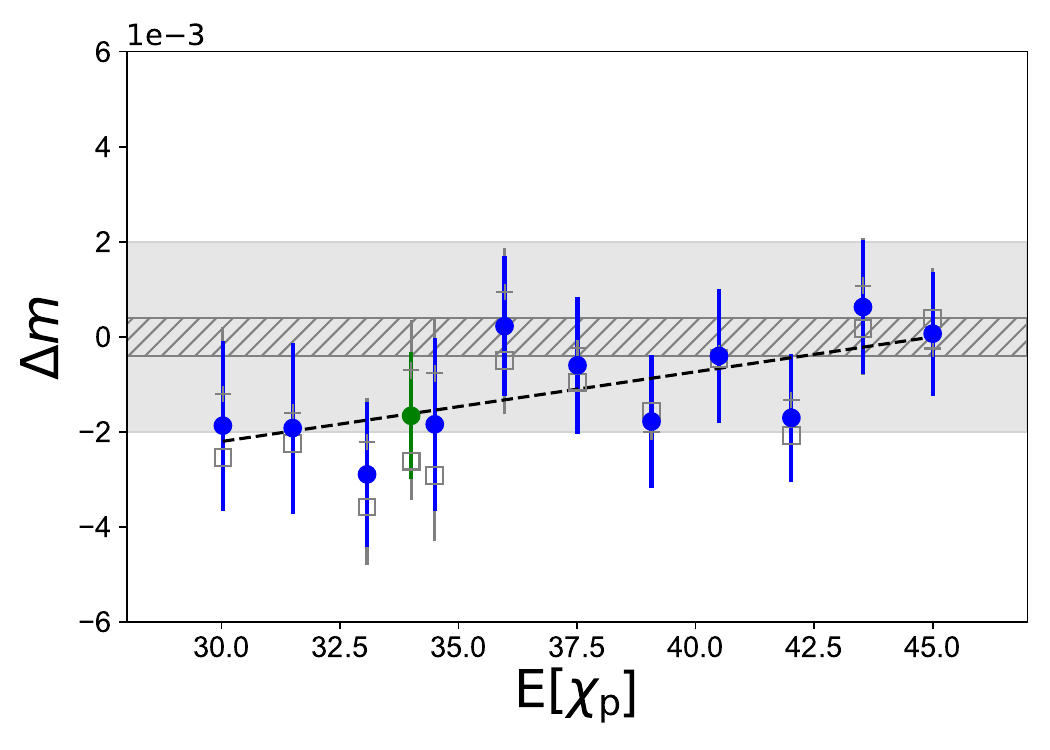} 
\includegraphics[width=0.33\linewidth]{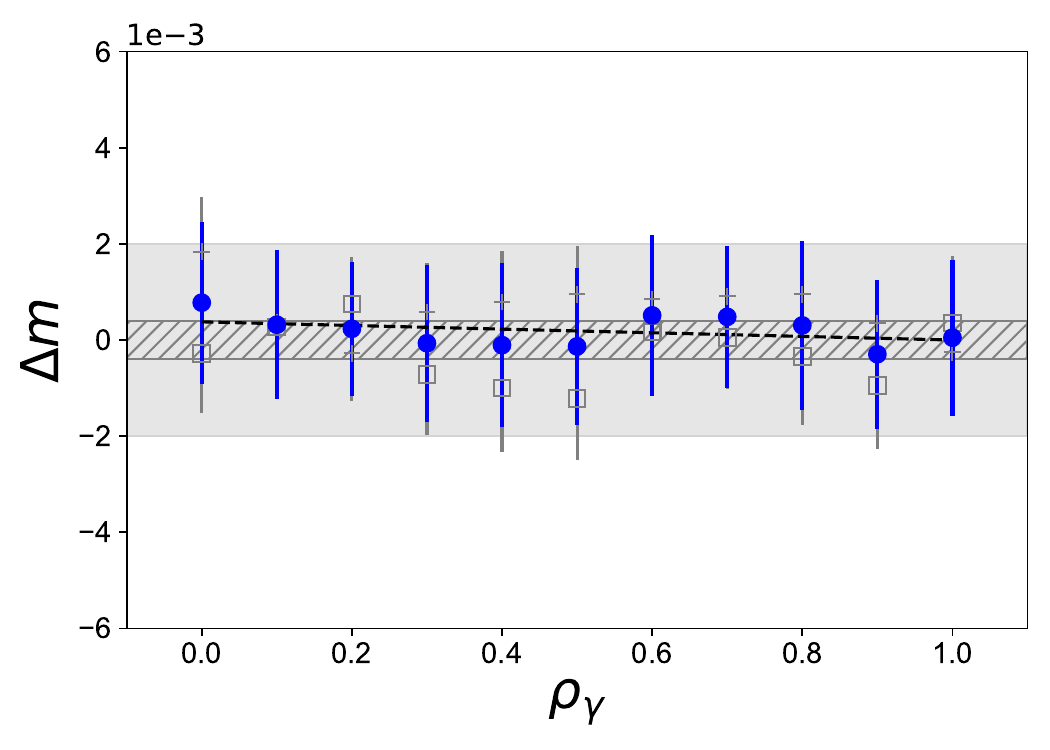} 
\caption{
Impact of various alignment and shear coherence scenarios for faint galaxies on multiplicative shear bias.
Top row, left to right: residual bias as a function of the alignment parameter \( A_\psi \), which quantifies 
(i) radial and (ii) tangential alignment of faint galaxy orientations with respect to the line connecting the faint galaxy and BSG centres, and 
(iii) the parallel alignment of faint and BSG orientations. \( A_\psi = 1 \) corresponds to perfect alignment, and \( A_\psi = 0.5 \) to random orientations.
Bottom left: residual bias as a function of the spatial alignment of faint galaxies with the BSG’s major axis, quantified using the anisotropy metric \( E(\chi_\mathrm{p}) \), where \( \chi_\mathrm{p} \) is the angle between the faint galaxy position and the BSG’s major axis; \( E(\chi_\mathrm{p}) = 45^\circ \) corresponds to random spatial positions. Blue points correspond to the linear form of $p(\chi_{\mathrm{p}})$; the single green point corresponds to the quadratic form.
In all cases, the specified alignment is applied prior to shearing.
Bottom right: residual bias as a function of the shear coherence parameter \( \rho_{\gamma} \) (Equation~\eqref{eqn:gamma_faint_rho}), which quantifies the correlation between the applied shear on faint galaxies and that of the BSG. \( \rho_{\gamma} = 1 \) corresponds to identical shears, while \( \rho_{\gamma} = 0 \) represents completely uncorrelated shears.
Grey and blue points, black dashed lines, and shaded regions are as in Figure~\ref{fig:kmax_alpharf}. For the anisotropy panel, the green point is omitted from the regression. For all panels biases are shown relative to those predicted by the linear model fits at the fiducial parameter value.
}
\label{fig:align_asym_shear}
\end{figure*}

We quantify the degree of anisotropy using the expected value of the position angle:
\begin{equation}
\mathrm{E}[\chi_{\mathrm{p}}] = \int_{0^\circ}^{90^\circ} \chi_{\mathrm{p}}\, p(\chi_{\mathrm{p}})\, \mathrm{d}\chi_{\mathrm{p}}.
\end{equation}
In practice, we estimate $\mathrm{E}[\chi_{\mathrm{p}}]$ by computing the mean $\chi_{\mathrm{p}}$ across an unlensed simulated population of BSGs and faint galaxies.
Figure~\ref{fig:align_asym_shear} shows the resulting shear biases as a function of $\mathrm{E}[\chi_{\mathrm{p}}]$, which ranges from $45^\circ$ (isotropic distribution) to $30^\circ$. The distributions based on \citet{Brainerd_2005} and \citet{Agustsson2006} correspond to $\mathrm{E}[\chi_{\mathrm{p}}] \approx 42^\circ$ and $34^\circ$, respectively.

Fitting a linear regression model to the results based on the linear form of $p(\chi_{\mathrm{p}})$, we find a statistically significant slope ($p$-value = 0.03) with gradient $(1.5 \pm 0.6)\times10^{-4}$. This implies that the anisotropy of faint galaxies around BSGs, expressed through $\mathrm{E}[\chi_{\mathrm{p}}]$, must be constrained to within $\pm2.7^\circ$ in order for biases to remain at least a factor of five below the top-level \emph{Euclid} requirement.

\subsection{Faint galaxy apparent magnitude slope}
\label{subsec:alphaf}
The apparent magnitude distribution of faint galaxies surrounding BSGs may differ from that of the general faint galaxy population across the field. For example, brighter satellites may preferentially reside closer to the halo centre \citep[][]{Tal_2012}. Additionally, observational studies suggest a correlation between the absolute magnitude of the BCG and that of its brightest satellite. We investigate this latter effect further in Section~\ref{sec:clustering}.

Here, we quantify the effect on the biases when varying the slope of the faint galaxy apparent magnitude distribution for the entire BSG sample. Since the biases flatten for $m_{\mathrm{lim}} \gtrsim 27$ (see Figure~\ref{fig:mlim}), we simplify the analysis by including only faint galaxies up to this magnitude limit. 
In this range, the cumulative distribution in Equation~\ref{eqn:num_cum} is well approximated by an exponential in magnitude (equivalently a power law in flux) with $\beta_{\mathrm{m}}=1$, so that $\log_{10}(N)$ varies linearly with $m_{\mathrm{AB}}$. We refer to this as the linear approximation to the full model. This simplified model allows us to vary $\alpha_{\mathrm{m,f}}$ while keeping the total number density of faint galaxies fixed, thereby isolating the impact of the slope alone.

\begin{figure*}
\centering
\includegraphics[width=0.49\linewidth]{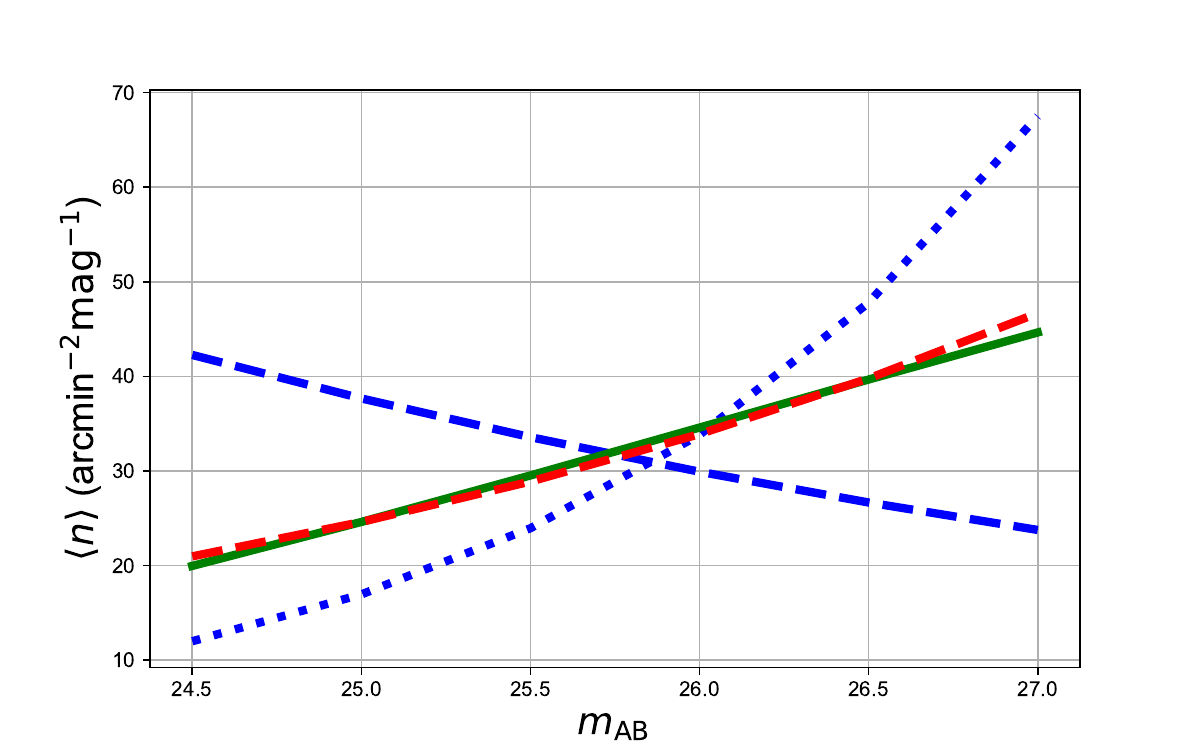}
\includegraphics[width=0.49\linewidth]{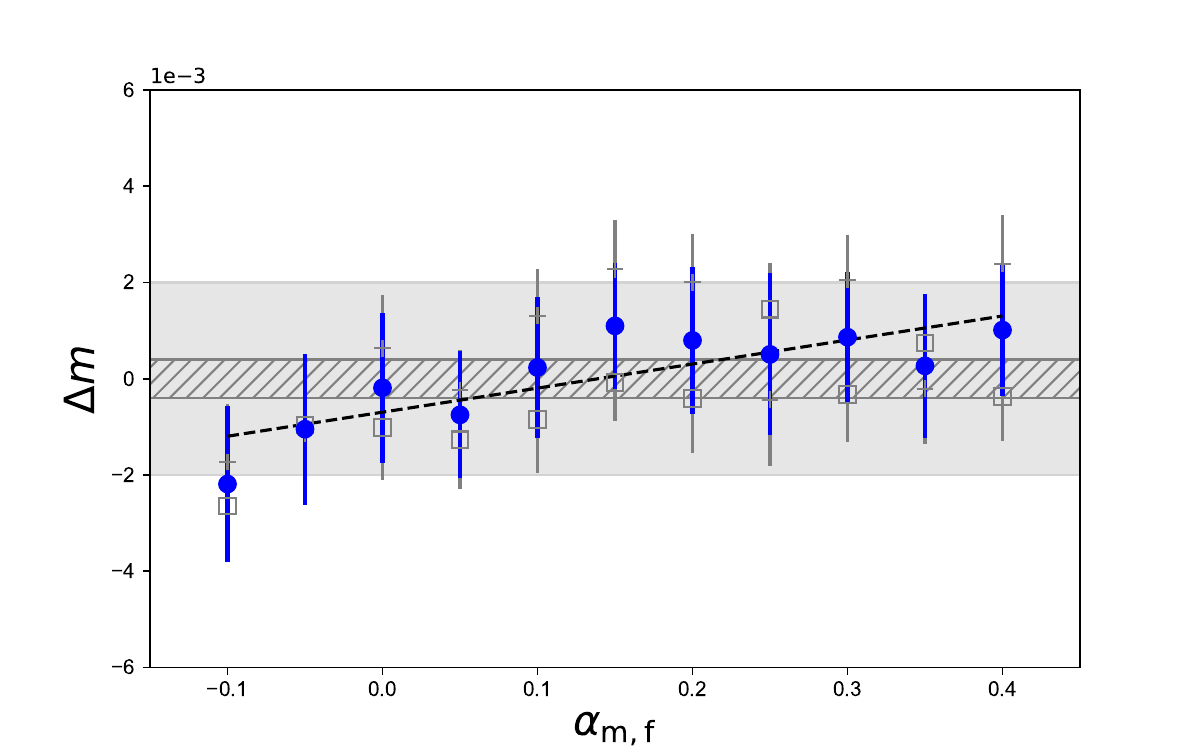} 
\caption{Left: mean faint galaxy number density $\left<n(m_{\mathrm{AB}})\right>$ for $\alpha_{\mathrm{m,f}}=-0.1$ (blue dashed), $0.4$ (blue dotted), and the fiducial value $\alpha_{\mathrm{m,f}}=0.139$ (red dashed) using the ``linear" model. 
The area under each curve in the range $24.5 < m_{\mathrm{AB}} < 27$ is fixed to $N_{\mathrm{fid;27}}=80.34$ galaxies per arcmin$^2$. 
The green solid curve shows $\left<n(m_{\mathrm{AB}})\right>$ for the ``full'' model (see Table~\ref{tab:slopes}). 
Right: residual multiplicative biases as a function of $\alpha_{\mathrm{m,f}}$ using the linear model. 
Grey and blue points, black dashed line, and shaded regions are as in Figure~\ref{fig:kmax_alpharf}. Biases are shown relative to the regression model prediction at the fiducial parameter value.
}
\label{fig:bias_alphamf}
\end{figure*}

Differentiating Equation~\ref{eqn:num_cum}, 
we obtain the following expression for the mean number density (per arcmin$^2$) of faint galaxies in the magnitude range $m_{\mathrm{AB}}$ to $m_{\mathrm{AB}}+dm_{\mathrm{AB}}$:
\begin{equation}
n_{\mathrm{f}}(m_{\mathrm{AB}})\,dm_{\mathrm{AB}} = A_{\mathrm{m,f}}\, 10^{\alpha_{\mathrm{m,f}} m_{\mathrm{AB}}}\, dm_{\mathrm{AB}},
\label{eqn:nm}
\end{equation}
where we have set $\beta_{\mathrm{m}}=1$\footnote{More generally, for $\beta_{\mathrm{m}} \ne 1$, 
\[
n_{\mathrm{f}}(m_{\mathrm{AB}})\,dm_{\mathrm{AB}} = A_{\mathrm{m,f}}\, \beta_{\mathrm{m}}\, m_{\mathrm{AB}}^{\beta_{\mathrm{m}} - 1}\, 10^{\alpha_{\mathrm{m,f}} m_{\mathrm{AB}}}\, dm_{\mathrm{AB}}.
\]
This form is used in Figure~\ref{fig:bias_alphamf} where we compare the full model and linear approximations.} and the subscript $\mathrm{f}$ refers to the faint galaxy population.

Fiducial values for the parameters $A_{\mathrm{m,f}}$ and $\alpha_{\mathrm{m,f}}$ in the linear approximation model are obtained by imposing two conditions. First, the faint galaxy number density integrated over the range $24.5 < m_{\mathrm{AB}} < 27$ must equal the fiducial value for the full model, such that:
\begin{equation}
\int_{24.5}^{27}  n_{\mathrm{f}}(m_{\mathrm{AB}})\,dm_{\mathrm{AB}} = N_{\mathrm{fid;27}},
\label{eqn:int_nf}
\end{equation}
where $N_{\mathrm{fid;27}}=80.34$ galaxies per arcmin$^2$ (see Table~\ref{tab:Nsat}). Second, we require that $n_{\mathrm{f}}(m_{\mathrm{AB}})$ at $m_{\mathrm{AB}} = 25$ is the same in both the full and linear approximation models; this somewhat arbitrary choice
ensures a reasonable match between the two models within the relevant magnitude range (see Figure~\ref{fig:bias_alphamf}). The resulting fiducial values of $A_{\mathrm{m,f}}$ and $\alpha_{\mathrm{m,f}}$ for the linear approximation are listed in Table~\ref{tab:slopes}.
In Figure~\ref{fig:mlim}, we plot the bias obtained when we adopt the linear approximation with $m_{\mathrm{lim}}=27$ and find that it is consistent with the bias obtained using the ``full" model.

As $\alpha_{\mathrm{m,f}}$ is varied, the mean number density is held fixed by adjusting the normalization constant $A_{\mathrm{m,f}}$ according to:
\begin{equation}
A_{\mathrm{m,f}} =    
\begin{cases}
\frac{2}{5}N_{\mathrm{fid;27}} & \alpha_{\mathrm{m,f}} = 0,\\
\\
\frac{\alpha_{\mathrm{m,f}}\ln(10) N_{\mathrm{fid;27}}}{10^{27\alpha_{\mathrm{m,f}}}-10^{24.5\alpha_{\mathrm{m,f}}}} & \mathrm{otherwise}.
\end{cases}
\label{eqn:Amf} 
\end{equation}
For reference, we plot the linear density profiles for $\alpha_{\mathrm{m,f}}=-0.1$ and 0.4 in Figure~\ref{fig:bias_alphamf}.
We note that the number density distribution is discontinuous at $m_{\mathrm{AB}} = 24.5$ when $\alpha_{\mathrm{m,f}}$ is varied away from the fiducial value, as may be expected if BSGs and faint galaxies are drawn from separate populations.

We plot the biases as a function of $\alpha_{\mathrm{m,f}}$ in Figure~\ref{fig:bias_alphamf}. The fiducial slope is $\alpha_{\mathrm{m,f}} =0.139$.
As expected, the absolute value of the bias increases for flatter or negative slopes (corresponding to a higher proportion of brighter faint neighbours) and decreases for steeper positive slopes. 
Fitting a linear regression model, we find a statistically significant slope ($p$-value = 0.002) with gradient $(5.0 \pm 1.2) \times 10^{-3}$.
This implies that, for calibration simulations, $\alpha_{\mathrm{m,f}}$ must be determined to within $\pm0.4$ of its true value to satisfy the Euclid top-level bias requirement, or to within $\pm0.08$ to remain within a factor of five of the requirement.

\section{Shear biases from correlations between faint-galaxy properties and BSG magnitude}
\label{sec:clustering}

The local mean density of faint galaxies around a BSG is likely to vary, depending on its environment. For instance, when the BSG is the BCG of a massive halo, the excess can reach factors of order 5 (e.g. M19). In contrast, isolated galaxies, as well as those in 
low-mass groups, will tend to be surrounded by an under-density of faint neighbours relative to the average across all BSGs. This suggests a positive correlation between halo mass and faint-galaxy excess. If BSG apparent magnitude is taken as a proxy for mass, then a corresponding dependence on BSG magnitude is also expected.

A second effect arises from the magnitude gap: halos hosting brighter BCGs tend to have larger gaps between the first and second brightest members \citep{2014A&A...566A.140G}. This is thought to reflect the halo’s formation history\footnote{Large magnitude gaps may also be associated with isolated groups located away from the dense nodes of the cosmic web \citep{2023A&A...676A.133Z}.}  \citep{2010MNRAS.405.1873D,vitorelli2018mass} and implies that the faint galaxy apparent magnitude distribution depends on the BCG’s luminosity. Again, if apparent magnitude is used as a proxy for absolute magnitude, this translates to a correlation between the faint-galaxy distribution and the BSG apparent magnitude.

We investigate these two effects separately in the following sections. We adopt the linear form for the faint galaxy apparent magnitude distribution (see Table~\ref{tab:slopes} and Section~\ref{subsec:alphaf}) and include neighbouring galaxies up to a limiting magnitude of $m_{\mathrm{lim}}=27$.

\subsection{Including a dependence of faint galaxy excess on BSG magnitude}
We consider a model in which the mean number of faint galaxies within $\theta_{r}$ arcsec of a BSG is conditional on the apparent magnitude of the BSG as follows:
\begin{equation}
\left<N_{\theta_{r}}(m_{\mathrm{AB,b}})\right> = B_{\mathrm{m}}10^{-b_{\mathrm{m}}m_{\mathrm{AB,b}}},
\label{eqn:bm}
\end{equation}
where $m_{\mathrm{AB,b}}$ is the apparent magnitude of the BSG, $b_{\mathrm{m}}=0$ for a constant mean density for all BSGs and $B_{\mathrm{m}}$ is given by
\begin{equation}
B_{\mathrm{m}} = \frac{N_{\mathrm{fid;27}}\pi(\theta_{r}/60)^2(\alpha_{\mathrm{m,b}}-b_{\mathrm{m}})\left(10^{24.5\alpha_{\mathrm{m,b}}}-10^{20\alpha_{\mathrm{m,b}}}\right)}{\alpha_{\mathrm{m,b}}\left(10^{24.5(\alpha_{\mathrm{m,b}}-b_{\mathrm{m}})}-10^{20(\alpha_{\mathrm{m,b}}-b_{\mathrm{m}})}\right)}
\label{eqn:Bm}
\end{equation}
so that the mean excess across all BSGs is held equal to the fiducial excess (see derivation in Appendix~\ref{app:Bm_derivation}). We include all faint galaxies within $\theta_{r}=3$ of the BSG and up to a limiting magnitude $m_{\mathrm{lim}}=27$. The number density of faint galaxies per arcmin$^{2}$ is $N_{\mathrm{fid;27}}$ (see Section~\ref{subsec:alphaf}) and the slope of the BSG apparent magnitude distribution is $\alpha_{\mathrm{m,b}}=0.36$ (see Table~\ref{tab:slopes}). 

\begin{figure*}
\centering
\includegraphics[width=1\linewidth]{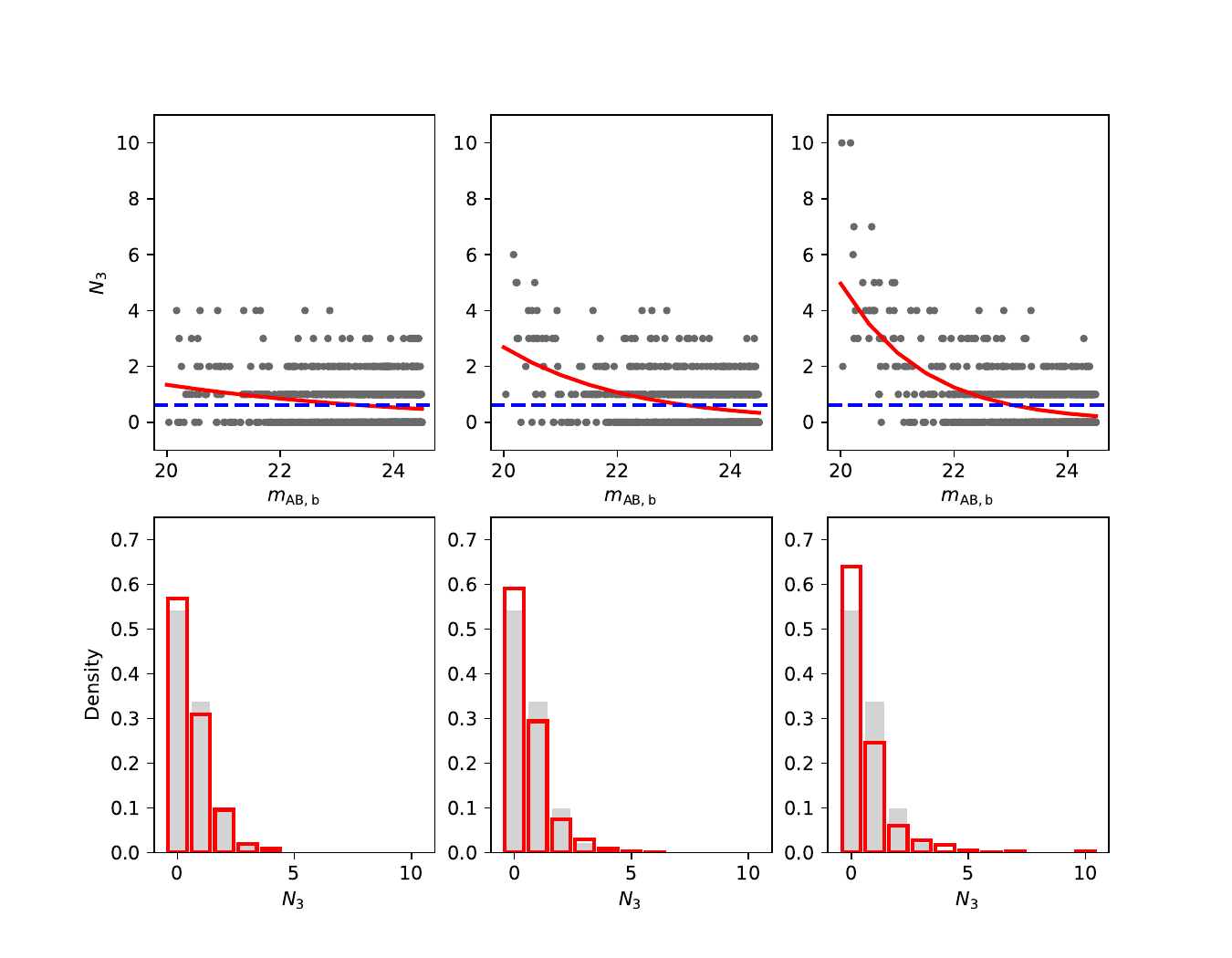}
\caption{
Top panels: Number of faint galaxies within 3\,arcsec of each BSG (\( N_3 \)) as a function of the BSG apparent magnitude, for three values of the correlation parameter: \( b_{\mathrm{m}} = 0.1 \) (left), \(0.2\) (middle), and \(0.3\) (right). The red lines show \( \left<N_3\right> \) as a function of BSG magnitude, while grey dots represent \(N_3\) for individual BSGs. The horizontal blue dashed lines indicate \( \left<N_3\right> \) for \( b_{\mathrm{m}} = 0 \).  
Bottom panels: Distribution of \( N_3 \) across all BSGs. Grey bars show the distribution for the fiducial setup with \( b_{\mathrm{m}} = 0 \), while red-outlined bars show the distributions for the corresponding \( b_{\mathrm{m}} > 0 \) cases. The overall mean excess is fixed to zero.
}
\label{fig:N3_mb}
\end{figure*}

We justify the use of this empirical relation in Appendix~\ref{app:bm}, finding an approximate value for $b_{\mathrm{m}}\sim0.3$.
In practice, its value will depend on the survey parameters, in particular the survey depth and filters, and will need to be determined from observational data, such as the \emph{Euclid} deep field.
For a positive $b_{\mathrm{m}}$ the number density of faint galaxies is greater for a brighter BSG, as is supported in the literature \citep[e.g.,][]{2023AAS...24146011S,Zarattini2021}. In Figure~\ref{fig:N3_mb}, we plot the relationship between the number of faint galaxies within 3\,arcsec of a BSG ($N_3$) and the apparent magnitude of the BSG for a range of $b_{\mathrm{m}}$ values, together with barplots showing the frequency density of $N_3$ across the sample.  

\begin{figure*}
\centering
\includegraphics[width=0.8\linewidth]{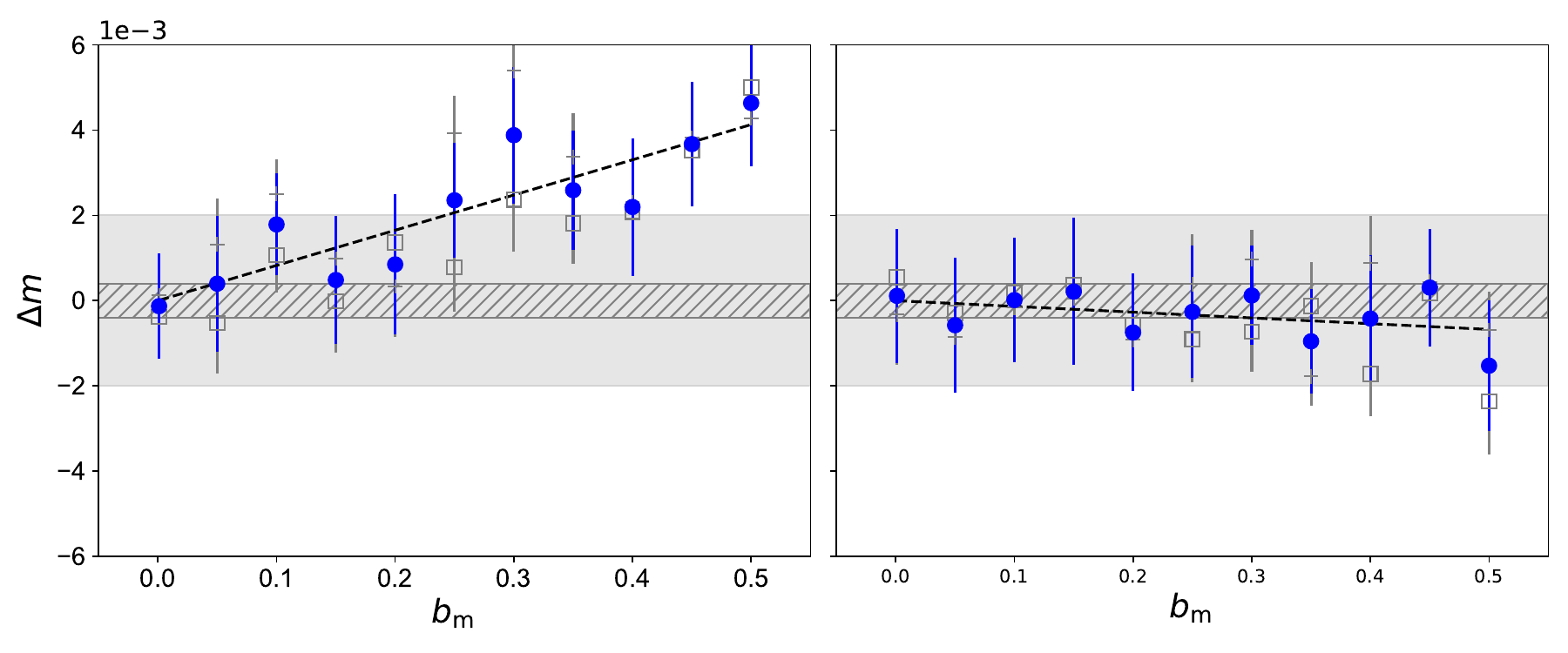}
\caption{Residual multiplicative biases as a function of $b_{\mathrm{m}}$, the parameter defining the correlation between the mean faint galaxy excess and BSG apparent magnitude (left-hand plot; see Equation~\ref{eqn:bm}). Also shown are the biases obtained using the same distribution of excesses among BSGs, but with zero correlation with the BSG magnitude (right-hand plot). 
Grey and blue points, black dashed lines, and shaded regions are as in Figure~\ref{fig:kmax_alpharf}. Biases are shown relative to that predicted by the regression model fit at the fiducial parameter value (i.e., at $b_{\mathrm{m}}=0$).
}
\label{fig:bias_bm}
\end{figure*}

The multiplicative biases are shown in Figure~\ref{fig:bias_bm} for values of $b_{\mathrm{m}}$ between 0 and 0.5. Also shown are the biases when we implement the same distribution of excesses provided by Equation~\ref{eqn:bm}, but with zero correlation between $\left<N_{\theta_{r}}(m_{\mathrm{AB,b}})\right>$ and the BSG apparent magnitude. This is achieved in practice by drawing a different random magnitude for the BSG than that used to obtain $\left<N_{\theta_{r}}(m_{\mathrm{AB,b}})\right>$.

Assuming a linear relation between the multiplicative bias and \( b_{\mathrm{m}} \), we find a statistically significant slope ($p$-value = 0.0004) when fitting a linear regression model to the mean of \( m_1 \) and \( m_2 \) as a function of \( b_{\mathrm{m}} \), with \(\mathrm{d}\Delta m/\mathrm{d}b_{\mathrm{m}} = (8.3 \pm 1.5)\times 10^{-3}\). In contrast, when we preserve the same distribution of faint galaxy excesses across BSGs but remove the correlation with BSG magnitude, the slope is consistent with zero ($p$-value = 0.24). This latter result is expected given the linear response of the multiplicative bias to a uniform change in faint galaxy excess across the sample, demonstrated in Section~\ref{subsec:excess} and Figure~\ref{fig:mlim}. Nevertheless, the zero-correlation test provides a useful check that confirms the correlation between the mean faint galaxy density and BSG magnitude is the factor driving the residual biases, rather than differences in the overall \(N_3\) distribution relative to the fiducial case.  
To ensure that residual biases remain at least a factor of five below the top-level \emph{Euclid} requirement, calibration simulations must therefore model this correlation accurately, constraining the parameter \( b_{\mathrm{m}} \) to within \( \pm 0.05 \) of its true value.

\subsection{Including a dependence of the faint galaxy magnitude distribution on BSG magnitude}
In this section, we explore the effect of a correlation between BSG apparent magnitude and the magnitude distribution of the surrounding faint galaxies. 
We adopt a simple model in which the conditional slope of the faint galaxy apparent magnitude distribution, $\alpha_{\mathrm{m,f;c}}$, varies linearly with the BSG apparent magnitude $m_{\mathrm{AB,b}}$:
\begin{equation}
\alpha_{\mathrm{m,f;c}} = \alpha_{\mathrm{m,f;fid}} - B_{\mathrm{c}}(m_{\mathrm{AB,b}} - m_{\mathrm{AB,b;p}}),
\label{eqn:alphac}
\end{equation}
where $\alpha_{\mathrm{m,f;fid}}=0.139$ is the fiducial slope (see Table~\ref{tab:slopes}), $m_{\mathrm{AB,b;p}}$ is the ``pivot" magnitude (see below) and $B_{\mathrm{c}}$ controls the strength of the correlation. When $B_{\mathrm{c}} = 0$, the slope is independent of the BSG magnitude. We fix $m_{\mathrm{AB,b;p}} = 23.5$ (chosen close to the median BSG magnitude\footnote{The median BSG magnitude is $\approx23.4$ and the mean $\approx23.7$.}) so that approximately half the galaxy sample has an associated faint galaxy magnitude slope below the fiducial value, and half above. 

For each BSG, the mean number of faint galaxies is held constant at the field value (i.e., with $f_{\mathrm{ex}}=0$) by re-normalizing the distribution using Equation~\ref{eqn:Amf} with $\alpha_{\mathrm{m,f}} = \alpha_{\mathrm{m,f;c}}$. Thus, while the distribution of faint magnitudes varies with BSG brightness, the mean number of faint galaxies is fixed for each BSG at the fiducial value.  

\begin{figure*}
\centering
\includegraphics[width=0.8\linewidth]{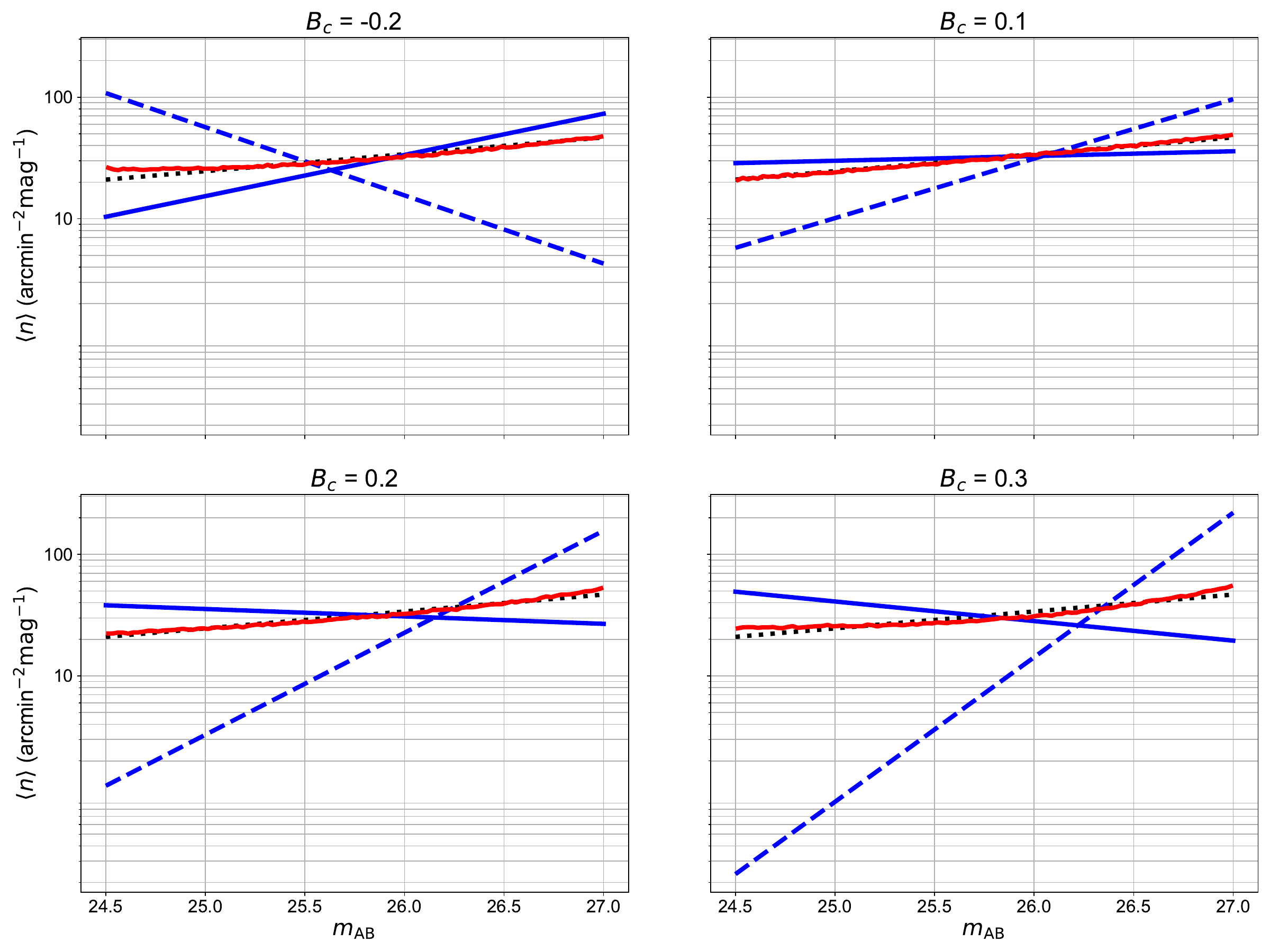}
\caption
{Faint galaxy apparent magnitude distribution for $B_{\mathrm{c}}=-0.2$ (top left), 0.1 (top right), 0.2 (bottom left) and 0.3 (bottom right) for BSG apparent magnitudes $m_{\mathrm{AB,b}}=20$ (blue dashed) and 24.5 (blue solid). 
The red solid line shows the faint galaxy distribution across all BSGs when using Equation~\ref{eqn:alphac} with $m_{\mathrm{AB,b;p}}=23.5$. The fiducial case ($B_{\mathrm{c}}=0$) is shown with a black dotted line.}
\label{fig:n_mb_Bc}
\end{figure*}

However, because the BSG number density itself varies with magnitude, the overall faint galaxy apparent magnitude distribution across the BSG sample will differ from the fiducial (uncorrelated, $B_{\mathrm{c}} = 0$) case. Figure~\ref{fig:n_mb_Bc} compares the overall faint galaxy distribution with the fiducial case for different $B_{\mathrm{c}}$ values, along with distributions associated with the brightest and faintest BSGs. Across the range tested, the total distribution is very close to fiducial. Nonetheless, because even small changes in the faint distribution can affect shear biases (see Section~\ref{subsec:alphaf} and Figure~\ref{fig:bias_alphamf}), we perform a sensitivity test designed to isolate the impact of the correlation itself from that of changes in the overall faint magnitude distribution. Specifically, we compute biases using the same distribution as that obtained when $B_{\mathrm{c}} \neq 0$, but remove the correlation with BSG magnitude by randomly reassigning BSG magnitudes when generating the faint population. This ensures that any difference in biases between the two cases arises solely from the correlation, rather than from a shift in the global faint galaxy magnitude distribution.

\begin{figure}
\centering
\includegraphics[width=0.9\linewidth]{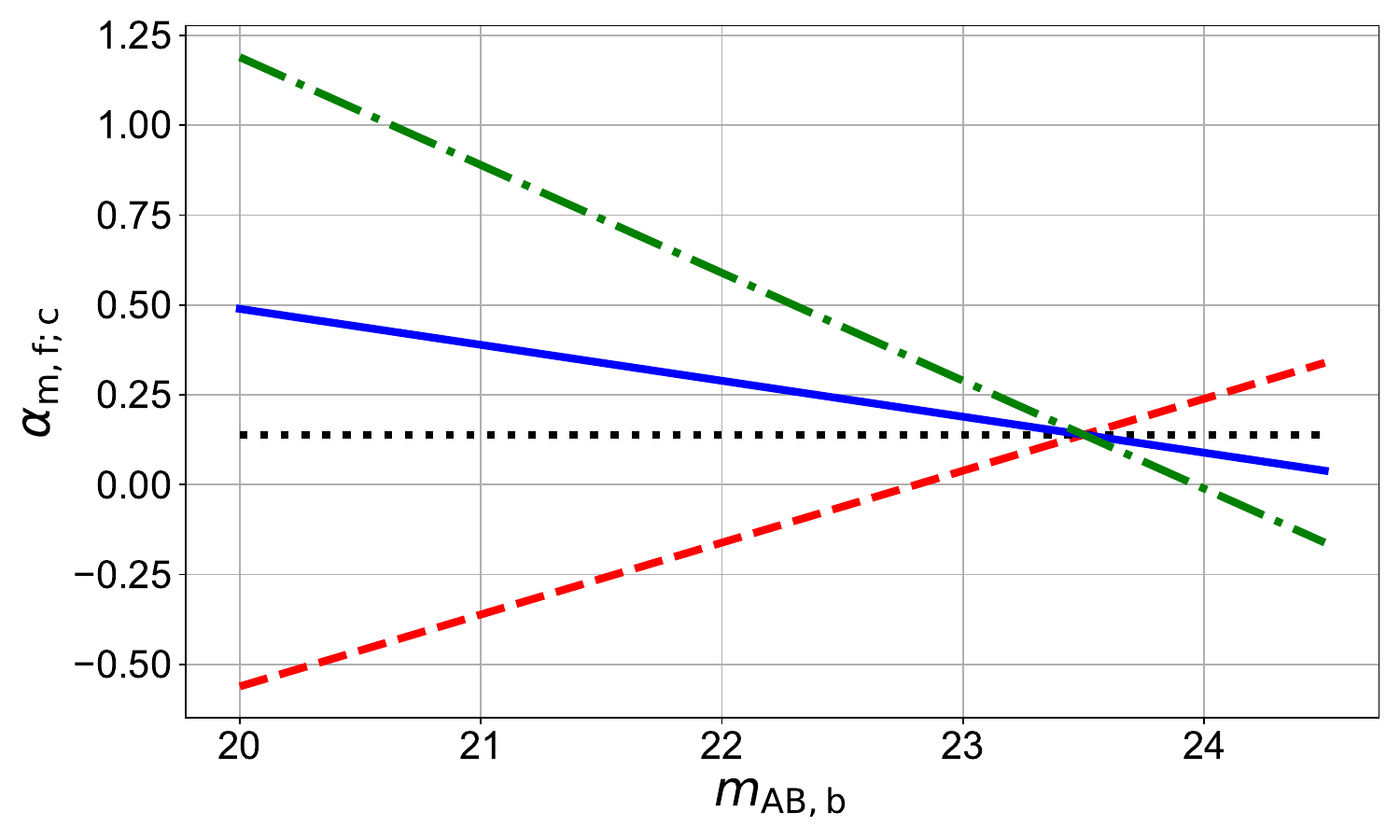}
\caption{Dependence of the faint galaxy apparent magnitude slope on BSG magnitude for $B_{\mathrm{c}}=-0.2$ (red dashed), 0 (black dotted), 0.1 (blue solid) and 0.3 (green dash-dotted) with $m_{\mathrm{AB,b;p}}=23.5$ (see Equation~\ref{eqn:alphac}). For $B_{\mathrm{c}}=0$, the slope is fixed at the fiducial value ($\alpha_{\mathrm{m,f}}=0.139$; see ``Fiducial Linear'' model in Table~\ref{tab:slopes}).}
\label{fig:alphaf_mb_Bc}
\end{figure}

\begin{figure}
\centering
\includegraphics[width=0.95\linewidth]{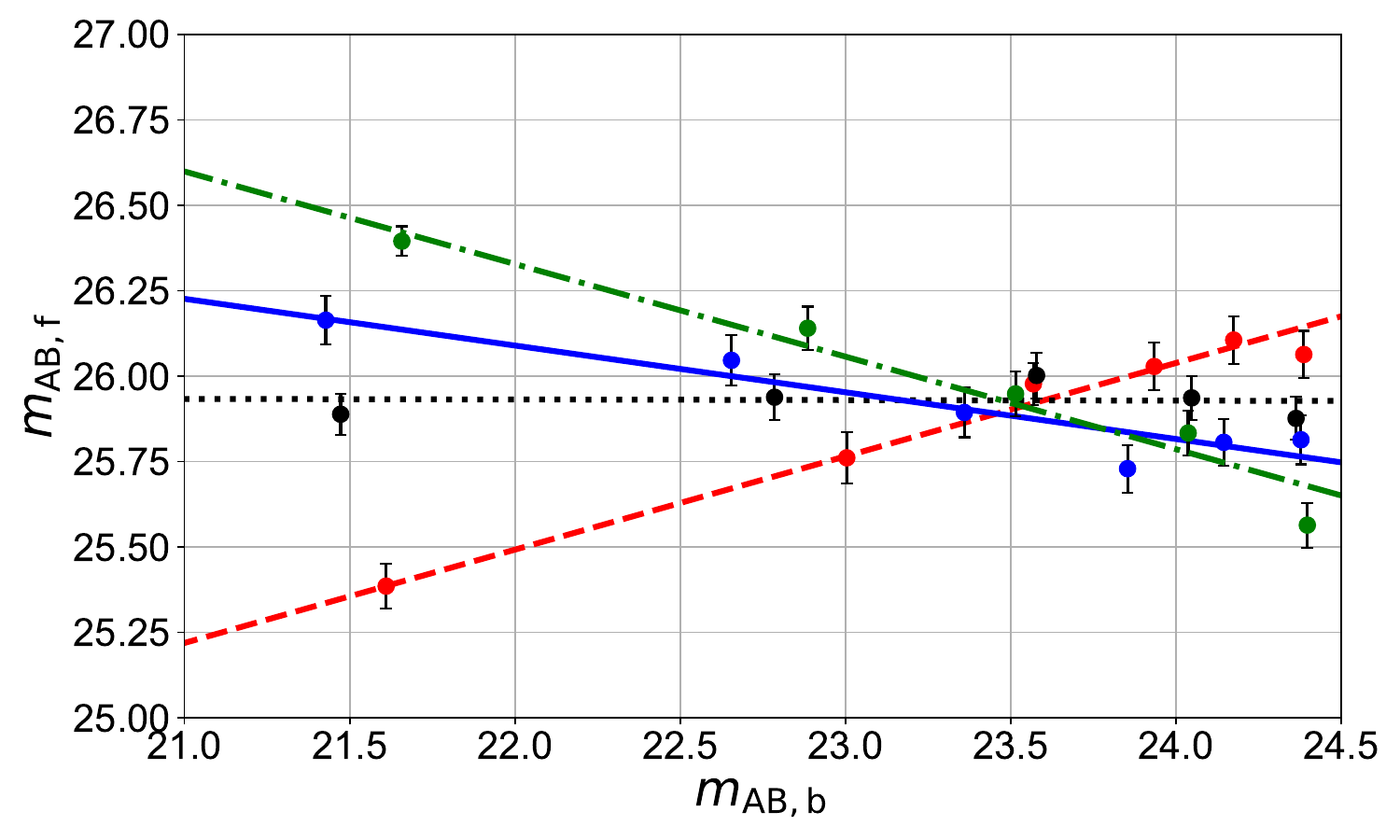}
\caption{Relationship between faint galaxy magnitudes and BSG magnitudes for different $B_{\mathrm{c}}$ values (colours and linestyles as in Figure~\ref{fig:alphaf_mb_Bc}). All faint galaxies within 3\,arcsec of a BSG are included. Results are shown with $\sim$100 BSGs per bin. Best-fit regression parameters and correlation coefficients are given in Table~\ref{tab:Bc_corr_mf_mb}.}
\label{fig:Bc_corr_mf_mb}
\end{figure}

We explore correlation strengths in the range $B_{\mathrm{c}} \in [-0.2, 0.35]$. 
Figure~\ref{fig:alphaf_mb_Bc} illustrates how the conditional slope $\alpha_{\mathrm{m,f;c}}$ varies with BSG magnitude for representative $B_{\mathrm{c}}$ values. 
To connect $B_{\mathrm{c}}$ with an observable quantity, Figure~\ref{fig:Bc_corr_mf_mb} shows, for each $B_{\mathrm{c}}$, the mean faint galaxy magnitude as a function of BSG magnitude, binned in quantiles of $\sim100$ BSGs each.
The relationship is approximately linear\footnote{
The distribution of magnitude gaps between the brightest and second-brightest galaxies in a halo is commonly studied \citep[e.g.,][]{More_2012,Zarattini2021}. 
However, since our analysis excludes neighbours brighter than 24.5, a direct comparison is not possible. 
Nevertheless, the linear relation observed between BCG absolute magnitude and magnitude gap in previous work \citep[e.g.,][]{2014A&A...566A.140G} motivates the form of Equation~\ref{eqn:alphac}.}; 
Table~\ref{tab:Bc_corr_mf_mb} provides the regression slope, intercept, and Pearson's correlation coefficient for each $B_{\mathrm{c}}$.

\begin{table}
\centering
\begin{tabular}{c|ccc}
$B_{\mathrm{c}}$ &  Slope & Intercept & $r$\\
\hline
-0.2   & 0.27 & 19.48 & 0.37 \\
0 & 0.00 & 25.97 & 0.00 \\
0.1 & -0.14 & 29.10 & -0.19\\
0.3 & -0.27& 32.3 & -0.38\\
\end{tabular}
\caption{Best-fit linear regression parameters for the relation between faint galaxy magnitudes, $m_{\mathrm{AB,f}}$, and BSG magnitudes, $m_{\mathrm{AB,b}}$, for different $B_{\mathrm{c}}$ values (see Figure~\ref{fig:Bc_corr_mf_mb}). Also shown is Pearson's correlation coefficient $r$.}
\label{tab:Bc_corr_mf_mb}
\end{table}

\begin{figure*}
\centering
\includegraphics[width=0.8\linewidth]{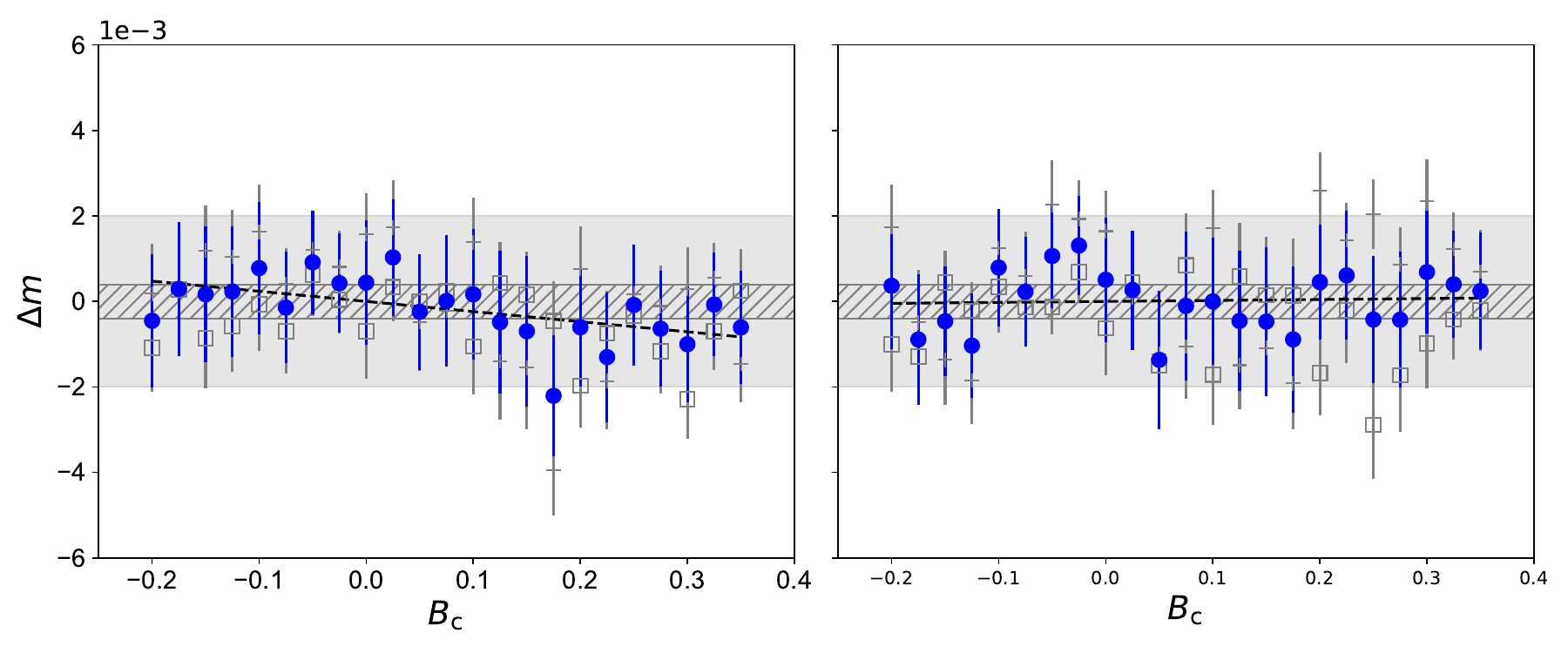}
\caption{Residual multiplicative biases as a function of $B_{\mathrm{c}}$, both with (left) and without (right) the correlation between the faint-end slope and BSG magnitude included. 
At a given $B_{\mathrm{c}}$, the overall distribution of faint galaxy magnitudes across all BSGs is the same in both cases.
Grey and blue points, black dashed lines, and shaded regions are as in Figure~\ref{fig:kmax_alpharf}. Biases are shown relative to that predicted by the regression model fit at the fiducial parameter value (i.e., at $B_{\mathrm{c}}=0$).
}
\label{fig:bias_Bc}
\end{figure*}

Figure~\ref{fig:bias_Bc} shows the multiplicative biases versus $B_{\mathrm{c}}$, both with ($r\neq0$) and without ($r\approx 0$) the correlation between $\alpha_{\mathrm{m,f}}$ and BSG magnitude included in the simulations.
We find that with $r\approx 0$ there is no evidence for a linear regression slope different from zero ($p$-value = 0.8). This demonstrates that any differences in the overall faint galaxy apparent magnitude distribution (i.e., across all BSGs) from the fiducial case has a negligible impact on the biases. However, when we include the correlation between the faint-end magnitude slope and the BSG magnitude there is a significant effect ($p$-value = 0.007) with slope $-(2.0\pm0.8)\times10^{-3}$.
Assuming the relation shown in Figure~\ref{fig:Bc_corr_mf_mb}, our results suggest that $B_{\mathrm{c}}$ must be constrained to within $\pm0.2$ to ensure biases are a factor of five below the \emph{Euclid} requirement.

\section{Discussion}
\label{sec:discussion}

We have investigated the impact of undetected galaxies on shear calibration, focusing on how various properties of the faint population influence multiplicative biases. Using independent image simulations\footnote{We generate PSF-convolved galaxy images using an independent image-generation pipeline, providing a useful verification of results obtained by H17, M19, and H21, who use the \texttt{GalSim} galaxy simulation toolkit \citep[][]{2015A&C....10..121R_galsim}.} and a noise-bias-free machine learning shape measurement code \citep[][]{voigt2024}, we confirm previous findings that failing to account for faint blends leads to unacceptably large biases. In our fiducial simulations, this bias reaches $m_i \sim -8 \times 10^{-3}$—well above the \emph{Euclid} top-level requirement of $|m_i| < 2 \times 10^{-3}$. 

Consistent with previous studies (e.g. M19), we find that calibration simulations must include faint galaxies to a limiting magnitude \(m_{\rm AB,f} \sim 27.0^{+2.1}_{-0.9}\) in order to capture the dominant contributions to blending-induced bias.\footnote{This agrees with the finding in M19 that the required limiting magnitude is largely insensitive to the choice of shear measurement method.} However, we find that galaxies need only be included out to projected separations \(\theta_{r}\sim1.03^{+0.22}_{-0.24}\)\,arcsec (approximately 10 pixels for \emph{Euclid}) from each BSG, compared to the $\sim$2.5\,arcsec inclusion radius found by M19. This difference may arise from the smaller postage stamps used in our setup (M19 use 64 by 64 pixels), and suggests that calibration simulations should be adapted to the specifics of the shear measurement pipeline.

Our study builds on previous work by systematically varying key properties of the faint galaxy population that have not yet been explicitly explored in this context. Specifically, we examine the size–magnitude relation, the slope of the apparent magnitude distribution,\footnote{H17 also examine the slope of the apparent magnitude distribution, but here we fix the mean galaxy number density to isolate the effect of the distribution shape from changes in projected number density.} and correlations between faint galaxy properties and those of the BSG, including its position, orientation, and brightness. Crucially, where possible, we quantify biases in terms of parameters that can in principle be measured directly from survey data.

In Table~\ref{tab:results_linreg}, we summarise which properties of the faint population significantly impact multiplicative biases over the range of values explored. We refer to these as the “critical’’ parameters and provide constraints on how precisely each one must be determined to suppress residual biases to at least a factor of five below \emph{Euclid}'s top-level requirement. We note that even stricter thresholds may be required to accommodate multiple systematics in shear measurement.

\begin{table*}
\centering
\caption{
Summary of the linear regression model fits used to quantify the dependence of
multiplicative biases on faint galaxy parameters. For each parameter, we fit a simple
linear regression model and test the null hypothesis $H_0:\beta=0$ (no
dependence of bias on the parameter) against the alternative $H_A:\beta\neq0$,
where $\beta$ is the slope of the best--fit regression line. We report the
two-sided $p$-value from this test.
The parameters are grouped into two subtables: those with $p<0.05$, referred to
as \emph{critical} because they show a statistically significant dependence of
shear bias on the parameter, and those with $p\ge0.05$, for which no significant
dependence is detected over the range explored. For all parameters we list the
fiducial value adopted in the simulations and the range of values included in the regression fit. For the
critical parameters we additionally report the constraint required to keep the
residual multiplicative bias below $4\times10^{-4}$,
corresponding to a factor of five more stringent than the top-level \emph{Euclid}
requirement.
}
\label{tab:results_linreg}

\begin{subtable}{\linewidth}
\centering
\caption{Critical parameters ($p<0.05$)}
\label{tab:results_critical}
\begin{tabular}{lcccc}
\hline
Parameter & $p$-value & Constraint & Fiducial & Range explored \\
\hline
$f_{\mathrm{ex}}$              & $4\times10^{-7}$ & $\pm0.05$        & 0     & $[-1,2]$ \\
$A_\psi$ radial                & 0.001            & $\pm0.06$        & 0.5   & $[0.5,1]$ \\
$A_\psi$ tangential            & 0.002            & $\pm0.06$        & 0.5   & $[0.5,1]$ \\
$\mathrm{E}[\chi_{\mathrm{p}}]$ & 0.03             & $\pm2.7^\circ$   & $45^\circ$ & $[30^\circ,45^\circ]$ \\
$\alpha_{\mathrm{m,f}}$        & 0.002            & $\pm0.08$        & 0.139 & $[-0.1,0.4]$ \\
$b_{\mathrm{m}}$               & 0.0004           & $\pm0.05$        & 0     & $[0,0.5]$ \\
$B_{\mathrm{c}}$               & 0.007            & $\pm0.2$         & 0     & $[-0.2,0.35]$ \\
\hline
\end{tabular}
\end{subtable}

\vspace{0.5cm}

\begin{subtable}{\linewidth}
\centering
\caption{Non-critical parameters ($p\ge0.05$)}
\label{tab:results_noncritical}
\begin{tabular}{lccc}
\hline
Parameter & $p$-value & Fiducial & Range explored \\
\hline
$k_{\mathrm{max}}$       & 0.5   & $\infty$ & $[3,23]$ \\
$\alpha_{\mathrm{r,f}}$  & 0.07  & -0.033   & $[-0.1,0]$ \\
$f_{r_{\mathrm{e}}}$     & 0.2   & 1        & $[0.4,1.6]$ \\
$A_\psi$ parallel        & 0.2   & 0.5      & $[0.5,1]$ \\
$\rho_{\gamma}$          & 0.25  & 1        & $[0,1]$ \\
\hline
\end{tabular}
\end{subtable}

\end{table*}

As expected, blending biases depend sensitively on the mean faint galaxy density in the vicinity of BSGs, which we define as the excess over the mean density of faint galaxies across the field. We find that this excess must be constrained to within $\pm0.05$ to reduce residual biases sufficiently. Furthermore, a correlation between the local faint galaxy density and the BSG apparent magnitude can induce large residual biases if unaccounted for. We model this relation (Appendix~\ref{app:bm}), described by the parameter $b_{\mathrm{m}}$ (Equation~\ref{eqn:bm}), and find that it must be known to within at least $\pm0.05$ of its true value. 

In addition, the slope of the faint galaxy apparent magnitude distribution—measured across all BSGs and with the mean faint galaxy number density held constant—has a pronounced impact.
Shallower slopes (i.e., a higher fraction of relatively bright faint galaxies) increase the absolute magnitude of the bias, while steeper slopes reduce it. Our results indicate that the slope must be known to within $\pm0.08$ to limit residual biases. We also examine the impact of a linear relation between the slope and BSG apparent magnitude. This correlation has a statistically significant effect on shear biases and requires the parameter $B_{\mathrm{c}}$ (Equation~\ref{eqn:alphac}) to be determined to within $\pm0.2$ to
keep residual biases under control.

Faint galaxy orientations and positions must also be taken into account. Radial and tangential alignments of faint galaxies with respect to the BSG centre, as well as anisotropy in their spatial distribution relative to the BSG major axis, all substantially alter the biases. By contrast, correlations between the shears of faint galaxies and the BSG, as well as parallel alignments of their intrinsic orientations, do not have a measurable impact. Across the parameter ranges explored, variations in the slope or normalisation of the size–magnitude relation also do not substantially affect the biases, indicating robustness to moderate uncertainties in size–magnitude modelling.

Any shear estimator that does not explicitly account for flux contamination from unresolved neighbours is likely to exhibit similar sensitivities to faint blends as those described above.
Our results thus inform the design of calibration simulations used by shape-measurement pipelines, identifying the critical faint-galaxy properties that must be included in, for example, the Euclid Flagship Simulation \citep[][]{potter2016pkdgrav3trillionparticlecosmological,euclidcollaboration2024euclidvflagshipgalaxy}, which plays a central role in modelling detection, selection, and shape-measurement systematics.

Furthermore, our work highlights the need to measure faint-galaxy properties directly from deep-field data in order to achieve the required precision on these critical parameters (see Table~\ref{tab:results_critical}).
While theoretical models and $N$-body simulations provide valuable insights into large-scale galaxy distributions \citep[e.g.][]{Vogelsberger_2014}, they are less reliable for capturing the small-scale distributions of faint-galaxy positions, magnitudes, and orientations that matter for blending. These challenges stem from uncertainties in the galaxy–halo connection and the impact of baryonic physics on galaxy formation and clustering.

To constrain biases from contaminants as faint as $m_{\rm AB} \sim 27$, deep fields must reliably detect galaxies to at least this limit. The \emph{Euclid} Deep Survey, although covering most of the faint range ($m_{\rm AB} \sim 24.5$–$26.5$), may fall short of capturing the full contribution from the faintest blends. This highlights the importance of complementary ultra-deep datasets such as the Hubble Ultra Deep Field (HUDF) or Hubble eXtreme Deep Field, which reach $i' \sim 29$. Previous simulation-based studies have drawn on HUDF data to model faint-galaxy clustering (M19), but our results emphasise the need for tighter quantitative constraints on parameters such as clustering excess, magnitude slope, and local correlations with BSG properties.

\section*{Acknowledgements}
The authors are grateful to the DiRAC High Performance Computing (HPC) Facility for the allocation of seedcorn time and to Stuart Newman for assistance using the CERES HPC Facility at The University of Essex.

\section*{Data Availability}
The data underlying this article will be shared on reasonable request to the corresponding author.



\bibliographystyle{mnras}
\bibliography{example} 



\appendix
\section{Deriving the faint-galaxy excess -- BSG brightness correlation normalisation parameter}
\label{app:Bm_derivation}

Here we derive the parameter $B_{\mathrm{m}}$, which quantifies the excess number of faint galaxies around a BSG as a function of its apparent magnitude.

The mean total number of faint galaxies within a radius $\theta_r$ (in arcsec) of a BSG, integrated over all BSGs (per arcmin$^2$), is
\begin{equation}
\left<N_{\theta_r,\mathrm{tot}} \right>= \int_{20}^{24.5} \left<N_{\theta_r}(m_{\mathrm{AB,b}})\right> n_{\mathrm{b}}(m_{\mathrm{AB,b}}) \, dm_{\mathrm{AB,b}},
\label{eqn:N_thetar_tot_app}
\end{equation}
where $\left<N_{\theta_r}(m_{\mathrm{AB,b}})\right>$ is defined in Equation~\ref{eqn:bm} and $n_{\mathrm{b}}(m_{\mathrm{AB,b}})$ is the BSG number density per arcmin$^2$ per unit magnitude.

Substituting $\left<N_{\theta_r}(m_{\mathrm{AB,b}})\right>$ in terms of $B_{\mathrm{m}}$ and $b_{\mathrm{m}}$ into Equation~\ref{eqn:N_thetar_tot_app} gives
\begin{equation}
\left<N_{\theta_r,\mathrm{tot}} \right> = \frac{B_{\mathrm{m}} A_{\mathrm{m,b}}}{(\alpha_{\mathrm{m,b}}-b_{\mathrm{m}})\ln 10} 
\left(10^{24.5(\alpha_{\mathrm{m,b}}-b_{\mathrm{m}})} - 10^{20(\alpha_{\mathrm{m,b}}-b_{\mathrm{m}})} \right).
\end{equation}

Alternatively, expressing $\left<N_{\theta_r,\mathrm{tot}} \right>$ in terms of the fiducial mean number of faint galaxies within $\theta_r$,
\begin{equation}
\left<N_{\theta_r,\mathrm{tot}} \right> = \frac{\pi \theta_r^2 N_{\mathrm{fid};27}}{60^2} \int_{20}^{24.5} n_{\mathrm{b}}(m_{\mathrm{AB,b}}) \, dm_{\mathrm{AB,b}},
\end{equation}
with
\begin{equation}
\int_{20}^{24.5} n_{\mathrm{b}}(m_{\mathrm{AB,b}}) \, dm_{\mathrm{AB,b}} = \frac{A_{\mathrm{m,b}} \left(10^{24.5\alpha_{\mathrm{m,b}}} - 10^{20\alpha_{\mathrm{m,b}}} \right)}{\alpha_{\mathrm{m,b}} \ln 10}.
\end{equation}

Equating the two expressions for $\left<N_{\theta_r,\mathrm{tot}}\right>$ then yields Equation~\ref{eqn:Bm}, giving the required form of $B_{\mathrm{m}}$.

\section{Relationship between number of satellites in a fixed aperture and halo mass}
\label{app:bm}

In this Appendix, we present an approximate derivation of the relationship between the number of satellites within a fixed circular aperture of radius $R$ around a BCG and the host halo mass. This modelling does not enter into the main analysis, but is used to justify the form of the relation provided in Equation~\ref{eqn:bm} and to obtain an approximate value for $b_{\mathrm{m}}$.

Within the halo occupation framework, the total number of satellites (above a fixed luminosity threshold) residing in a halo of mass $M_{200}$ scales approximately as
\begin{equation}
\left<N_{\mathrm{sat,tot}}\right> \propto \left(M_{200}\right)^{\alpha_{\mathrm{tot}}},
\end{equation}
with $\alpha_{\mathrm{tot}} \sim 0.9$–1 \citep{2006MNRAS.371.1173V}. Under the assumption that satellites follow the dark matter distribution, the number of satellites enclosed within a fixed aperture grows more slowly with halo mass, since a smaller fraction of satellites falls within the aperture at higher halo masses due to the increasing halo size.

Satellite galaxies are often assumed to follow the same spatial distribution as the dark matter \citep[e.g.][see also comments in the final paragraph of this Section]{2011ApJ...736...59Z}, typically modelled by a Navarro–Frenk–White (NFW) profile \citep{1997ApJ...490..493N},
with density at radius $r$ given by
\begin{equation}
  \rho(r) = \frac{\delta_\mathrm{c} \rho_\mathrm{c} }{(r/r_\mathrm{s})(1+r/r_\mathrm{s})^2},
\label{eqn:rho_dm}
\end{equation}
where $\rho_{\mathrm{c}}=3H(z)^2/(8\pi G)$ is the critical density of the Universe at redshift $z$, $H(z)$ is the Hubble parameter, $G$ is Newton's gravitational constant, $\delta_{\mathrm{c}}$ is the characteristic overdensity\footnote{Defined such that the mean overdensity within $r_{200}$ is 200 times the critical density.}, 
given by
\begin{equation}
\delta_{\mathrm{c}}=\frac{200}{3}\frac{c^3}{\ln{(1+c)}-c/(1+c)},
\label{eqn:deltac}
\end{equation}
and $r_{\mathrm{s}}$, the scale radius, is related to the virial radius $r_{200}$ and the dimensionless concentration parameter $c$ via $r_{200} = c r_{\mathrm{s}}$.

We adopt a relationship between the virial mass, $M_{200}$\footnote{$M_{200}$ is the mass enclosed within $r_{200}$.}, and concentration from \citet{2017ApJ...840..104S}, such that
\begin{equation}
c(M_{200}) = C_0 \left( \frac{M_{200}}{10^{12} M_{\odot}} \right)^{-\gamma_c},
\label{eqn:shan}
\end{equation}
with $C_0 = 6.61$ and $\gamma_c = 0.15$ for galaxies in the redshift range $0.4 < z < 0.6$. These parameter values were determined by \citet{2017ApJ...840..104S} based on the Canada-France-Hawaii Telescope (CFHT) Stripe 82 Survey, covering halos in the mass range $5 \times 10^{12}$–$2 \times 10^{14} M_{\odot}$. We allow for an evolution with redshift using the relation from \citet[][]{2001MNRAS.321..559B}, given by
\begin{equation}
c(M_{200})\propto (1+z)^{-1}.     
\label{eqn:conc_z}
\end{equation}

Integrating Equation~\ref{eqn:rho_dm} and assuming spherical symmetry, the three-dimensional halo mass enclosed within a fixed radius $R$ is
\begin{equation}
M(<R) = 4 \pi \delta_\mathrm{c} \rho_\mathrm{c} r_{\mathrm{s}}^3 \left[ \ln \left( 1 + \frac{R}{r_{\mathrm{s}}} \right) - \frac{R / r_{\mathrm{s}}}{1 + R / r_{\mathrm{s}}} \right].
\label{eqn:mass_dm}
\end{equation}
In our fiducial setup, we use an aperture with radius 3\,arcsec, corresponding to $R \approx 24$\,kpc at $z=1$ (the median redshift for \emph{Euclid}).

\begin{figure}
\centering
\includegraphics[width=0.95\linewidth]{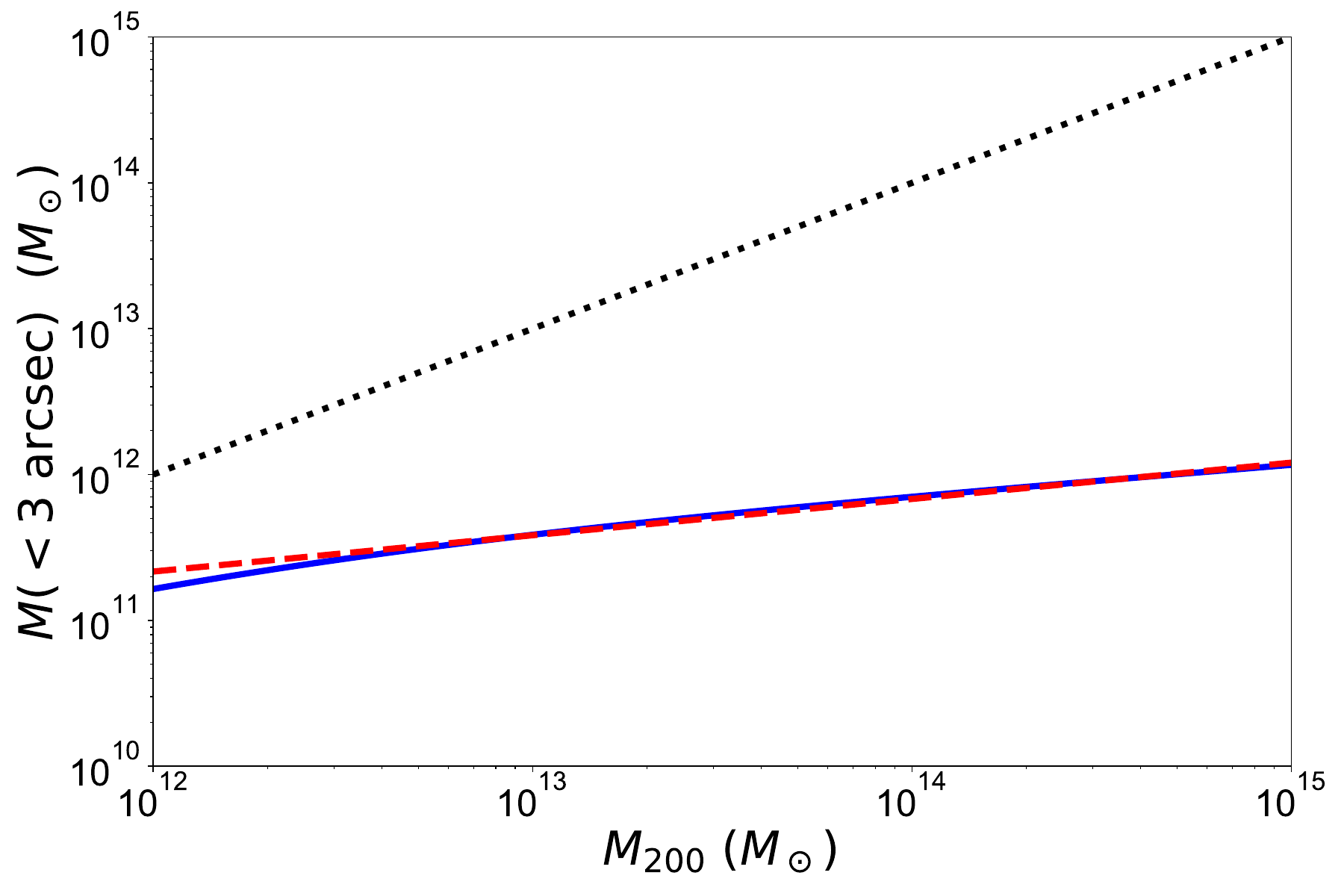}
\caption{Three-dimensional halo mass enclosed within a 3\,arcsec radius at $z=1$ (blue solid) using the mass--concentration relation from \citet[][]{2017ApJ...840..104S} (Equation~\ref{eqn:shan}). The red dashed curve shows the best-fit power law: $M(<3~\mathrm{arcsec})\propto M_{200}^{\alpha_{\mathrm{R}}}$, with $\alpha_{R}=0.25$. For reference, the dotted black line shows the one-to-one relation, $M(<3~\mathrm{arcsec}) = M_{200}$. At low halo masses, $r_{200}$ falls within the aperture, so the enclosed mass approaches the total halo mass.}
\label{fig:Mproj_M200}
\end{figure}

Using Equations~\ref{eqn:deltac}, ~\ref{eqn:shan}, ~\ref{eqn:conc_z} and ~\ref{eqn:mass_dm},
we plot in Fig.~\ref{fig:Mproj_M200} the three-dimensional mass enclosed within 3\,arcsec of the halo centre as a function of virial mass\footnote{The halo mass is truncated at $r_{200}$, so that $M(<3\,\mathrm{arcsec})$ tends to $M_{200}$ as the halo mass decreases and $r_{200} < R$.}. We find that a power law provides a reasonable approximation to the distribution, such that
\begin{equation}
M(<R) \propto \left(M_{200}\right)^{\alpha_{R}},
\end{equation}
with $\alpha_{R} \sim 0.25$ for $R=3\,\mathrm{arcsec}$ and $z=1$. Thus, we can write the mean number of satellites within radius $R$ as
\begin{equation}
\left<N_{\mathrm{sat}}(<R)\right>\propto \left(M_{200}\right)^{\alpha_{R}\alpha_{\mathrm{tot}}}.
\end{equation}

We now relate $\left<N_{\mathrm{sat}}(<R)\right>$ to the BCG apparent magnitude. Assuming all galaxies lie at the same redshift, the BCG luminosity is related to its apparent magnitude by
\begin{equation}
L_{\mathrm{BCG}} \propto 10^{-0.4m_{\mathrm{AB}}}.
\end{equation}
We find that this relation is only marginally shallower if we instead use a realistic redshift distribution. Adopting a sub-linear relation between BCG luminosity and halo mass,
\begin{equation}
L_{\mathrm{BCG}}\propto \left(M_{200}\right)^{\beta_{\mathrm{L}}},
\end{equation}
with $\beta_{\mathrm{L}}\sim0.3$ \citep[][and references therein]{2008MNRAS.385L.103B}, we obtain
\begin{equation}
\left<N_{\mathrm{sat}}(<R)\right>\propto 10^{-0.4\alpha_{R}\alpha_{\mathrm{tot}}m_{\mathrm{AB}}/\beta_{\mathrm{L}}}.
\end{equation}
Using representative values of $\alpha_{R}=0.25$, $\alpha_{\mathrm{tot}}=0.9$ and $\beta_{\mathrm{L}}=0.3$, and assuming the number of satellites within $R$ corresponds to the projected number, we find $b_{\mathrm{m}} \sim 0.3$ (see Equation~\ref{eqn:bm}).

We emphasise that the above derivation provides only a rough estimate and the actual relation will likely differ from the one provided here. 
For simplicity, we have shown the enclosed three-dimensional mass, whereas observations correspond to satellite counts in projected apertures.
Other factors may also affect the satellite number density–magnitude relation. For example, we assume
 satellites follow the dark matter distribution, but simulations suggest they are less centrally concentrated \citep[e.g.][]{2004MNRAS.355..819G}, although the picture is complex \citep[e.g.][and references therein]{2006ApJ...647...86C,sales_lambas_2005MNRAS.356.1045S,Chen08}. In addition, the inner density profiles of dark matter halos may be more cuspy than the NFW profile \citep[e.g.][]{2014MNRAS.441.3359D}, which would increase the enclosed mass at small radii and modify the halo mass dependence. Studies also suggest a more complex redshift dependence of the concentration parameter than the simple $(1+z)^{-1}$ scaling adopted here \citep[e.g.][]{2011MNRAS.411..584M}.
Finally, we note that only a proportion of BSGs will be BCGs, so the comparison to Equation~\ref{eqn:bm} is approximate. It will therefore be important to determine $b_{\mathrm{m}}$ empirically from deep field surveys.


\bsp	
\label{lastpage}
\end{document}